\documentclass[twocolumn,trackchanges]{aastex62}
\usepackage[section]{placeins}
\usepackage{longtable}
\usepackage{hyperref}
\usepackage{url}
\usepackage{amsmath}
\graphicspath{{./}{figures/}}

\usepackage{enumitem}
\usepackage{comment}
\newcommand{\mosfit}{{\tt MOSFiT} }
\newcommand{\Ni}{$^{56}$Ni }
\usepackage{float}

\usepackage[hang,flushmargin]{footmisc} 
\usepackage{url}

\newcommand{\STScI}{\affiliation{Space Telescope Science Institute, 3700 San Martin Dr, Baltimore, MD 21218, USA}}

\newcommand{\CfA}{\affiliation{Center for Astrophysics \textbar{} Harvard \& Smithsonian, 60 Garden Street, Cambridge, MA 02138-1516, USA}}

\newcommand{\Birmingham}{\affiliation{Birmingham Institute for Gravitational Wave Astronomy and School of Physics and Astronomy, University of Birmingham, Birmingham B15 2TT, UK}}
\newcommand{\CIERA}{\affiliation{Center for Interdisciplinary Exploration and Research in Astrophysics and Department of Physics and Astronomy, \\Northwestern University, 1800 Sherman Ave., 8th Floor, Evanston, IL 60201, USA}}

\newcommand{\Steward}{\affiliation{Steward Observatory, University of Arizona, 933 North Cherry Avenue, Tucson, AZ 85721, USA}}

\shorttitle{LSNe}
\shortauthors{Gomez et al.}

\begin{document}

\title{Luminous Supernovae: Unveiling a Population Between Superluminous and Normal Core-collapse Supernovae}

\correspondingauthor{Sebastian Gomez}
\email{sgomez@stsci.edu}

\author[0000-0001-6395-6702]{Sebastian Gomez}
\STScI
\CfA

\author[0000-0002-9392-9681]{Edo Berger}
\CfA

\author[0000-0002-2555-3192]{Matt Nicholl}
\Birmingham

\author[0000-0003-0526-2248]{Peter K. Blanchard}
\CIERA

\author[0000-0002-0832-2974]{Griffin Hosseinzadeh}
\Steward

\begin{abstract}

Stripped-envelope core-collapse supernovae can be divided into two broad classes: the common Type Ib/c supernovae (SNe Ib/c), powered by the radioactive decay of $^{56}$Ni, and the rare superluminous supernovae (SLSNe), most likely powered by the spin-down of a magnetar central engine. Up to now, the intermediate regime between these two populations has remained mostly unexplored. Here, we present a comprehensive study of 40 \textit{luminous supernovae} (LSNe), SNe with peak magnitudes of $M_r = -19$ to $-20$ mag, bound by SLSNe on the bright end and by SNe Ib/c on the dim end. Spectroscopically, LSNe appear to form a continuum between Type Ic SNe and SLSNe. Given their intermediate nature, we model the light curves of all LSNe using a combined magnetar plus radioactive decay model and find that they are indeed intermediate, not only in terms of their peak luminosity and spectra, but also in their rise times, power sources, and physical parameters. We sub-classify LSNe into distinct groups that are either as fast-evolving as SNe Ib/c or as slow-evolving as SLSNe, and appear to be either radioactively or magnetar powered, respectively. Our findings indicate that LSNe are powered by either an over-abundant production of $^{56}$Ni or by weak magnetar engines, and may serve as the missing link between the two populations.

\end{abstract}

\keywords{supernovae: general -- methods: statistical -- surveys}

\section{Introduction}\label{sec:intro}

Stars more massive than $\sim 8$ M$_\odot$ end their lives in a core-collapse supernova (CCSN). Some of these massive stars lose their hydrogen and/or helium envelopes before explosion, either through winds or interaction with a binary companion (e.g., \citealt{Heger03, Smith14_ARAA}). These stripped stars can lead to either Type Ib SNe (lacking hydrogen), or Type Ic SNe (lacking hydrogen and helium) \citep{Woosley95,Filippenko97}. Based on their light curve evolution and spectral properties, we know SNe Ib/c are powered by the radioactive decay of \Ni synthesized during the explosion \citep{Arnett82, Gal-Yam17}. SNe Ib/c are relatively dim and fast-evolving, reaching typical peak magnitudes of $M_r= -17.7\pm 0.9$ within about $20\pm 10$ days after explosion \citep{Barbarino20}. Spectroscopically, SNe Ib/c exhibit strong suppression blueward of $\sim 4000$ \AA\ due to line blanketing from Fe-peak elements. In terms of their environments, SNe Ib/c tend to occur in galaxies with relatively high metallicities of \mbox{$12 + \log($O/H$) = 8.8\pm 0.3$} \citep{Modjaz20}.

More recently, a new class of stripped-envelope CCSN was discovered and designated Type-I superluminous supernovae (hereafter, SLSNe; \citealt{Chomiuk11,Quimby11}). SLSNe can be up to 100 times more luminous than SNe Ib/c and reach typical peak magnitudes of $M_r = -21.7\pm 0.7$ (e.g., \citealt{Gal-Yam19,Gomez20,Chen22}) with longer rise times of $\sim 20-80$ days \citep{Nicholl17_mosfit}. The spectra of SLSNe are marked by distinctive W-shaped \ion{O}{2} absorption features around $\sim 3500-5000$ \AA\ at early times \citep{Chomiuk11,Quimby11}; and while the spectra of SLSNe tend to be much bluer than those of SNe Ib/c before peak, as they evolve and cool they begin to more closely resemble normal SNe Ic (e.g., \citealt{Pastorello10, Quimby18, Blanchard19, Nicholl19_nebular}). Unlike SNe Ib/c, SLSNe generally occur in low-metallicity galaxies with typical values of \mbox{$12 + \log($O/H$) = 8.4\pm 0.3$} \citep{Lunnan14}. Their blue spectra, bright and slowly evolving light curves, and low metallicity environments all point to an energy source distinct from radioactive decay \citep{Angus16,Nicholl17_mosfit,Margalit18}, most likely the spin-down energy of a millisecond magnetar born in the explosion \citep{Kasen10,Woosley10}. Additionally, it has been shown that SLSNe progenitors are generally more massive before explosion ($\approx 3-40$ M$_\odot$; \citealt{Blanchard20}) than those of SNe Ib/c ($4.5\pm 0.8$ M$_\odot$; \citealt{Barbarino20}). SLSNe are also rare, representing $\lesssim 1\%$ of the SNe Ib/c volumetric rate \citep{Frohmaier2020}. 

Given the distinct energy sources of SNe Ib/c and SLSNe, we may expect SNe in the intermediate regime to exist. These intermediate SNe could either have weaker magnetar engines than those of normal SLSNe, or an over-abundant production of $^{56}$Ni compared to SNe Ib/c. Recent studies have begun exploring these intermediate SNe, suggesting they are powered by more than just radioactive decay (e.g., SN\,2012aa from \citealt{Roy16} and SNe 2019dwa, 2019cri, 2019hge, and 2019unb from \citealt{Prentice21}). Here, we report the first systematic search and analysis of such intermediate events, their properties, and power sources. We present a list of 40 SNe with intermediate luminosities between SLSNe and SNe Ib/c, compiled either through our own observational program, or publicly available transients from the literature. We define the sample of \textit{luminous supernovae} (LSNe) as SNe with spectra consistent with a stripped-envelope CCSN and a peak absolute $r$-band magnitude of $M_r = -19$ to $M_r = -20$ mag, bound by SLSNe on the bright end and by SNe Ib/c on the dim end. We caution that the LSN label does not necessarily imply a physical connection between these objects, but is rather a phenomenological grouping based solely on their peak luminosity. We use these selection criteria to explore their physical properties, connections to both SLSNe and SNe Ib/c, and sub-groupings within the LSNe sample.

This paper is structured as follows: In \S\ref{sec:sample} we present the samples of LSNe, SNe Ib/c and SLSNe used in our analysis, as well as the sources of their photometry and spectroscopy. In \S\ref{sec:modeling} we describe the light curve modeling, and in \S\ref{sec:results} we discuss the results of these models. In \S\ref{sec:grouping} we outline the possible sub-groupings for LSNe. In \S\ref{sec:properties} we discuss the observational features and rates of the LSNe population, and finally conclude in \S\ref{sec:conclusions}. Throughout this paper we assume a flat $\Lambda$CDM cosmology with \mbox{$H_{0} = 69.3$ km s$^{-1}$ Mpc$^{-1}$}, $\Omega_{m} = 0.286$, and $\Omega_{\Lambda} = 0.712$ \citep{Hinshaw13}.

\section{Sample of Luminous Supernovae}\label{sec:sample}

\begin{figure*}[t]
	\begin{center}
		\includegraphics[width=\textwidth]{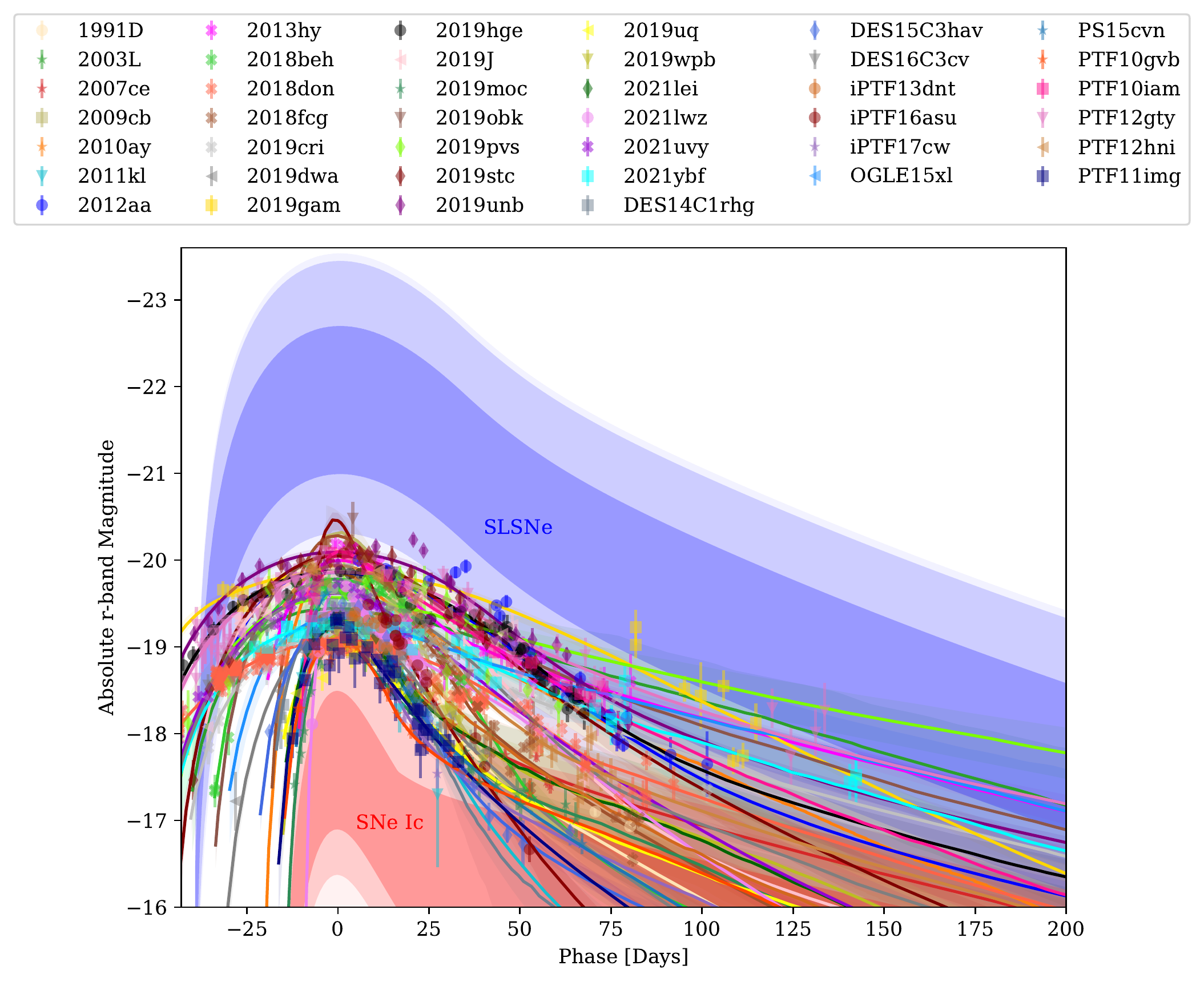}
		\caption{Light curves of all the Gold and Silver LSNe in our sample. The individual data points are $r$-band magnitudes of the SNe, and the lines are the corresponding best-fit models described in \S\ref{sec:modeling}. The shaded regions represent the 1, 2, and $3\sigma$ intervals for typical light curves of SLSNe (\textit{blue}) and SNe Ic (\textit{red}) obtained from averaging their light curve models. \label{fig:KLSNe_LCs}}
	\end{center}
\end{figure*}

We begin our analysis by gathering a sample of 315 stripped-envelope CCSNe, including a list of all known SLSNe in addition to SNe Ib/c and SNe Ic-BL with above-average luminosities, and a sample of SNe Ic/Ic-BL for comparison. The SNe in this master list were obtained either from the Open Supernova Catalog\footnote{\label{ref:osc}\url{https://sne.space/}; frontend now defunct, but backend available.} (OSC; \citealt{Guillochon17}), the Transient Name Server (TNS)\footnote{\label{ref:tns}\url{https://www.wis-tns.org/}}, the Weizmann Interactive Supernova Data Repository (WISeREP; \citealt{Yaron12})\footnote{\url{https://wiserep.weizmann.ac.il/}}, a literature search, or from our own FLEET transient follow-up program \citep{Gomez20_FLEET}. We include comparison samples of SNe Ic-BL from \cite{Taddia19_broadlined}, SNe Ic from \cite{Barbarino20}, and SLSNe from Gomez et al., in prep; where a full description of the SLSNe sample will be presented in a forthcoming paper. In total, we include 149 SLSNe and 61 SNe Ic/Ic-BL for our comparative analysis. The individual SNe used for this work are all listed in the Appendix.

We select the LSNe from our master sample by focusing on the objects that have a peak absolute magnitude of $M_r = -19$ to $-20$ mag. For SNe that either do not have $r$-band observations available, or that were not observed during peak, we estimate their peak absolute magnitude using the light curve models discussed in \S\ref{sec:modeling}. This range is motivated by the fact that SNe brighter than $M_r \approx -20$ mag tend to show relatively uniform spectroscopic and photometric features that allow us to classify them as SLSNe, and SNe dimmer than $M_r \approx -19$ mag can usually be confidently classified as SNe Ib/c.

Our final sample is made up of 59 LSNe, in addition to the 149 SLSNe and 61 SNe Ic/Ic-BL used for comparison. These SNe are not easily classified into either the SLSNe or SNe Ib/c categories, but lie somewhere in an intermediate regime in terms of both their light curves and spectra. Of these 59 LSNe, we designate 25 as ``Gold'' LSNe, when we have both enough photometry to be able to model their light curves, and spectroscopic observations from which we can verify their classification as stripped-envelope CCSNe. We designate 15 objects as ``Silver" when either their spectroscopic data is of poor quality but still consistent with stripped-envelope CCSNe, or they lack good photometric coverage before peak, but we are still able to constrain their peak using light curve models. Lastly, we label 19 objects as ``Bronze'' LSNe when either they do not have photometry available near peak, have less than four epochs of photometry, or have no public spectra available, any of which prevents us from producing trustworthy light curve models and/or confident spectroscopic classifications. We do not include the Bronze LSNe in our analysis. The full list of 59 LSNe, along with their individual data sources, notes, and peculiarities are presented in the Appendix. The final working sample of 40 Gold and Silver LSNe is listed in Table~\ref{tab:classes}.

\begin{deluxetable*}{cccccccccc}
    \tablecaption{Luminous Supernovae \label{tab:classes}}
    \tablehead{\colhead{Name} & \colhead{Redshift} & \colhead{Literature Class.} & \colhead{Spectral Group} & \colhead{Light curve} & \colhead{Quality} }
    \startdata
    DES14C1rhg  & 0.481 & SLSN-I        &   Superluminous   & Fast    & Gold     \\ % Fast LSN
    DES15C3hav  & 0.392 & SLSN-I        &   Superluminous   & Fast    & Gold     \\ % Fast LSN
    DES16C3cv   & 0.727 & SLSN-I        &   Normal          & Slow    & Gold     \\ % Slow Ic
    iPTF13dnt   & 0.137 & Ic-BL         &   Normal          & Fast    & Silver   \\ % Fast Ic
    iPTF16asu   & 0.187 & SLSN-I        &   Superluminous   & Fast    & Gold     \\ % Fast LSN
    iPTF17cw    & 0.093 & Ic-BL         &   Normal          & Fast    & Gold     \\ % Fast Ic
    OGLE15xl    & 0.198 & SLSN-I        &   Superluminous   & Slow    & Silver   \\ % Other
    PS15cvn     & 0.058 & Ic-BL         &   Normal          & Fast    & Gold     \\ % Fast Ic
    PTF10gvb    & 0.098 & Ic-BL         &   Normal          & Fast    & Gold     \\ % Fast Ic
    PTF10iam    & 0.109 & SLSN-I / Ic   &   Superluminous   & Medium  & Silver   \\ % Fast LSN
    PTF11img    & 0.158 & Ic-BL         &   Normal          & Fast    & Gold     \\ % Fast Ic
    PTF12gty    & 0.177 & SLSN-I / Ic   &   Superluminous   & Slow    & Gold     \\ % Slow LSN
    PTF12hni    & 0.106 & SLSN-I        &   Ambiguous       & Medium  & Gold     \\ % Other
    SN\,1991D   & 0.042 & Ib            &   Ambiguous       & Fast    & Silver   \\ % Other
    SN\,2003L   & 0.021 & Ib/c          &   Normal          & Slow    & Silver   \\ % Slow Ic
    SN\,2007ce  & 0.046 & Ic-BL         &   Normal          & Fast    & Silver   \\ % Fast Ic
    SN\,2009cb  & 0.187 & SLSN-I        &   Superluminous   & Fast    & Silver   \\ % Fast LSN
    SN\,2010ay  & 0.067 & Ic-BL         &   Normal          & Fast    & Silver   \\ % Fast Ic
    SN\,2011kl  & 0.677 & SLSN-I / GRB  &   Superluminous   & Fast    & Gold     \\ % Fast LSN
    SN\,2012aa  & 0.083 & SLSN-I / Ibc  &   Ambiguous       & Slow    & Gold     \\ % Other
    SN\,2013hy  & 0.663 & SLSN-I        &   Superluminous   & Medium  & Silver   \\ % Slow LSN
    SN\,2018beh & 0.060 & Ib/ Ic        &   Superluminous   & Slow    & Gold     \\ % Slow LSN
    SN\,2018don & 0.073 & SLSN-I / Ic   &   Superluminous   & Slow    & Silver   \\ % Other
    SN\,2018fcg & 0.101 & SLSN-I        &   Normal          & Fast    & Gold     \\ % Fast Ic
    SN\,2019cri & 0.050 & Ic            &   Normal          & Slow    & Gold     \\ % Slow Ic
    SN\,2019dwa & 0.082 & Ic            &   Ambiguous       & Medium  & Gold     \\ % Other
    SN\,2019gam & 0.124 & SLSN-Ib/IIb   &   Superluminous   & Slow    & Silver   \\ % Slow LSN
    SN\,2019hge & 0.086 & SLSN-Ib       &   Superluminous   & Slow    & Gold     \\ % Slow LSN
    SN\,2019J   & 0.120 & SLSN-I        &   Superluminous   & Slow    & Gold     \\ % Slow LSN
    SN\,2019moc & 0.056 & Ic            &   Normal          & Fast    & Gold     \\ % Fast Ic
    SN\,2019obk & 0.166 & SLSN-Ib       &   Superluminous   & Slow    & Silver   \\ % Slow LSN
    SN\,2019pvs & 0.167 & SLSN-I        &   Superluminous   & Slow    & Gold     \\ % Slow LSN
    SN\,2019stc & 0.117 & Ic            &   Normal          & Slow    & Gold     \\ % Slow Ic
    SN\,2019unb & 0.064 & SLSN-Ib       &   Superluminous   & Slow    & Gold     \\ % Slow LSN
    SN\,2019uq  & 0.100 & Ic            &   Normal          & Fast    & Silver   \\ % Fast Ic
    SN\,2019wpb & 0.068 & Ic            &   Normal          & Fast    & Silver   \\ % Fast Ic
    SN\,2021lei & 0.112 & Ic            &   Normal          & Fast    & Silver   \\ % Fast Ic
    SN\,2021lwz & 0.065 & SLSN-I        &   Superluminous   & Fast    & Gold     \\ % Fast LSN
    SN\,2021uvy & 0.095 & SLSN-I / Ib/c &   Normal          & Slow    & Gold     \\ % Slow Ic
    SN\,2021ybf & 0.130 & SLSN-I        &   Normal          & Slow    & Gold     \\ % Slow Ic
    \enddata
    \tablecomments{List of all the Gold and Silver LSNe used for this work, sorted alphabetically. We include classifications from the literature for each object from references listed in the Appendix, in addition to our own label based solely on their spectral features as either ``Superluminous" for SLSNe-like events or ``Normal" for SNe Ic/Ic-BL-like events. We add a light curve classification based on the duration of the light curve rise and whether it is fast like SNe Ic ($\lesssim 25$ days), slow like SLSNe ($\gtrsim 35$ days), or intermediate. Additional Bronze objects are listed in the Appendix but are otherwise excluded from this work.}
    \end{deluxetable*}

\subsection{Photometry}\label{sec:photometry}

We collect all available photometry for the 315 SNe in our sample. We obtain publicly available photometry from the OSC, TNS, and WISeREP for the SNe that have these data available. In addition, we include photometry from the Zwicky Transient Facility (ZTF; \citealt{Bellm19}) taken from the Automatic Learning for the Rapid Classification of Events (ALeRCE) broker \citep{Forster20}, the Asteroid Terrestrial-impact Last Alert System (ATLAS; \citealt{Tonry18}), the All Sky Automated Survey for SuperNovae (ASAS-SN; \citealt{Kochanek17}), the \textit{Gaia} Science Alerts (GSA; \citealt{Wyrzykowski16}), the Optical Gravitational Lensing Experiment (OGLE; \citealt{Wyrzykowski14}), the Catalina Real Time Transient Survey (CRTS; \citealt{Drake09}), and the Pan-STARRS Survey for Transients (PSST; \citealt{Huber15}). Photometry from ZTF, ATLAS, PSST, and ASAS-SN is reported from difference images and therefore already has the host flux subtracted. Photometry from the GSA, CRTS, and OGLE does not have the host contribution subtracted, so we subtract the corresponding host magnitude whenever necessary and possible.

\begin{figure*}
	\begin{center}
		\includegraphics[width=0.7\textwidth]{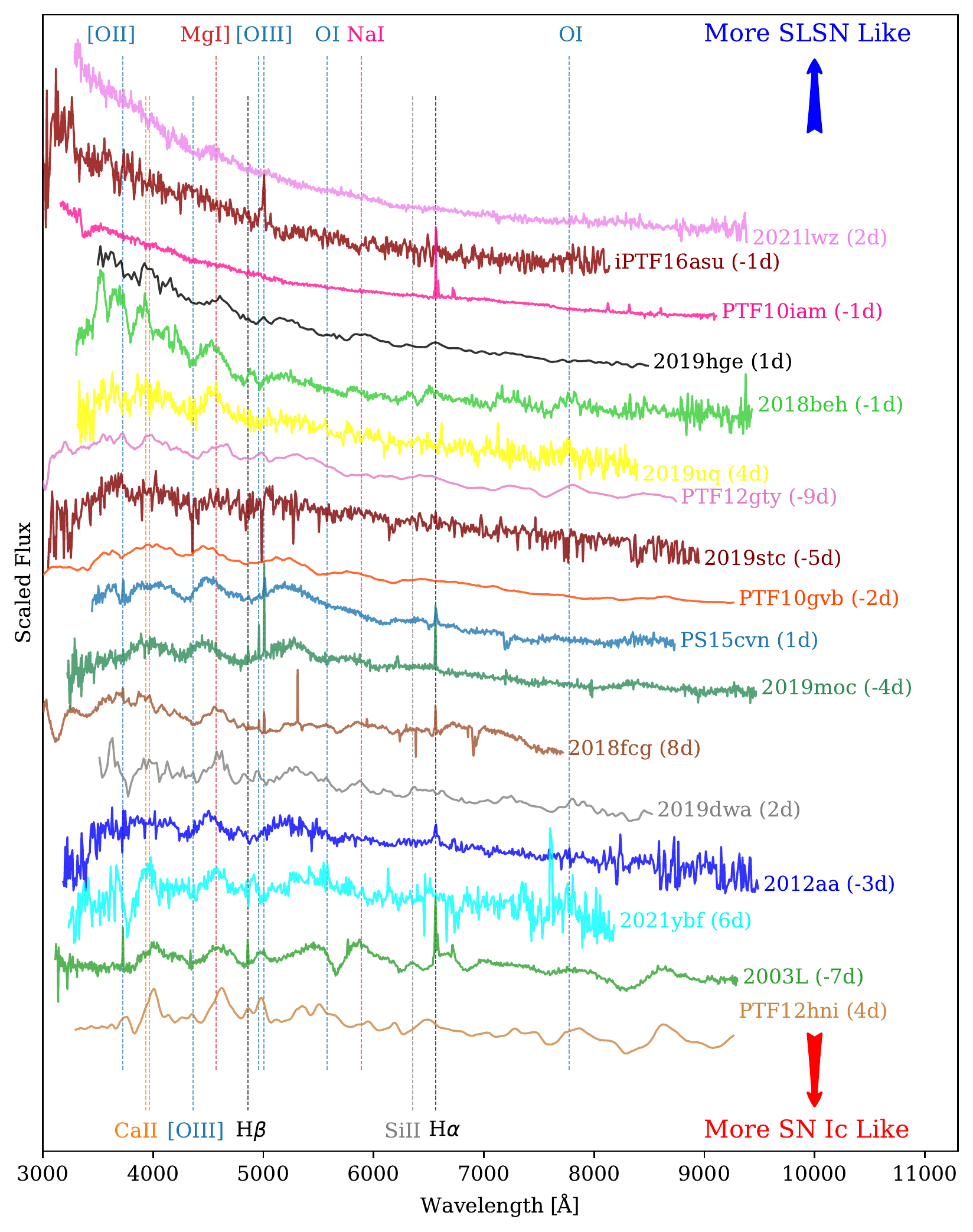}
		\caption{Representative spectra for all the Gold and Silver LSNe that have spectra available within $\pm10$ days of peak, sorted by color and spectral features. We find that LSNe appear to form a continuous distribution, from blue SLSN-like spectra to red SN Ic-like spectra. Individual references for each spectrum are listed in the Appendix. \label{fig:spectra}}
	\end{center}
\end{figure*}

In addition to publicly available photometry, we perform our own photometry on either public images or images from our own FLEET transient follow-up program \citep{Gomez20}. A few SNe observed by ZTF had sparsely sampled light curves reported by the automatic photometry pipeline. For these SNe, we download the raw ZTF images from the NASA/IPAC Infrared Science Archive\footnote{\url{https://irsa.ipac.caltech.edu/Missions/ztf.html}} to re-do the photometry and recover any sub-threshold detections that were missed by the automated pipeline. Additionally, we include $gri$ images of SNe that were observed by the Global Supernova Project (GSP) with the Las Cumbres Observatory Global Telescope Network (LCO; \citealt{Brown13}). Finally, we include images taken as part of FLEET with either KeplerCam on the 1.2-m telescope at the Fred Lawrence Whipple Observatory (FLWO), the Low Dispersion Survey Spectrograph (LDSS3c; \citealt{stevenson16}) or Inamori-Magellan Areal Camera and Spectrograph (IMACS; \citealt{dressler11}) both on the Magellan Clay 6.5-m telescopes at Las Campanas Observatory, or Binospec \citep{Fabricant19} on the MMT 6.5-m telescope.

We perform photometry on all images from ZTF, LCO, and FLEET in the same manner. Instrumental magnitudes are measured by modeling the point-spread function (PSF) of each image using field stars and subtracting the model PSF from the target. The magnitudes are then calibrated to AB magnitudes from the PS1/$3\pi$ catalog \citep{Chambers16}. For the majority of sources, we separate the flux of the SN from its host galaxy by performing difference imaging using a pre-explosion PS1/$3\pi$ template for comparison. We subtract the template from the science images using {\tt HOTPANTS} \citep{Becker15}. For sources where there is no host galaxy detected above the PS1/$3\pi$ detection limit of $\approx 23$ mag, we report PSF photometry taken directly from the science images without subtracting a template. The details for the data reduction of each SN are listed in the Appendix.

Finally, we verify that all the photometry is either already corrected for Galactic extinction, or correct it directly. We use the dust maps from \cite{Schlafly11} to obtain an estimate of $E(B-V)$ and the \cite{Barbary16} implementation of the \cite{Cardelli89} extinction law to calculate the corresponding extinction in each band.

A plot showing the $r$-band light curves of all 40 Gold and Silver LSNe is shown in Figure~\ref{fig:KLSNe_LCs}. The blue and red shaded regions in Figure~\ref{fig:KLSNe_LCs} represent the 1, 2, and 3$\sigma$ intervals for the light curves of SLSNe and SNe Ic, respectively. These regions were estimated by averaging the light curve models of all the SLSNe and SNe Ic/Ic-BL used for our comparative analysis, the full list of SNe is listed in \S\ref{sec:compare} of the Appendix. 

\subsection{Spectra}\label{sec:spectra}

In this work, we focus on the photometric properties of LSNe, but for comparison and to verify their classification as stripped-envelope CCSNe, we make use of either publicly available optical spectra, or our own newly collected spectra. The public spectra were obtained from WISeREP, the OSC, or the TNS. Some spectra were also obtained from published papers, either from journal databases or private communication with the authors. We use these public spectra to verify the classification given to each SN. Special care is taken for objects that have only been classified in an Astronomer's Telegram or TNS report, but not in a refereed publication. We then verify the redshift of each SN and update it if a better estimate was found from newer or higher quality spectra. Finally, we correct these spectra for Galactic extinction using the extinction maps from \cite{Schlafly11} and the \cite{Barbary16} implementation of the \cite{Cardelli89} extinction law. The individual data sources and any notes regarding the spectra of each SN are listed in the Appendix.

The newly collected spectra presented here are part of our FLEET observational program. These spectra were taken with either LDSS3c, Binospec, IMACS, or the Blue Channel spectrographs \citep{schmidt89} on the MMT 6.5-m telescope. We reduced these spectra using standard IRAF routines with the {\tt twodspec} package. The spectra were bias-subtracted and flat-fielded, the sky background was modeled and subtracted from each image, and the one-dimensional spectra were optimally extracted, weighed by the inverse variance of the data. Wavelength calibration was applied using an arc lamp spectrum taken near the time of each science image. Relative flux calibration was applied to each spectrum using a standard star taken close to the time of observation. Lastly, the spectra were corrected for Galactic extinction in the same way as the public spectra described above.

Spectroscopically, LSNe are not a uniform sample but span a wide range of features. Some LSNe are blue like SLSNe, while others are red and closely resemble SNe Ic. Representative spectra of LSNe are shown in Figure~\ref{fig:spectra}, where we include the closest spectrum to peak for each LSN that has a spectrum taken within $\pm 10$ days of peak. We find that LSNe appear to create a smooth continuum, from blue and SLSN-like to red and SN Ic-like, without a clear threshold or distinction that allows us to separate them neatly into either class. An in-depth study of the spectral properties of LSNe will be presented in a future paper.

\section{Light Curve Modeling}\label{sec:modeling}

\begin{deluxetable*}{llll}
	\tablecaption{\mosfit Parameter Definitions \label{tab:parameters}}
	\tablehead{\colhead{Parameter} & \colhead{Prior} & \colhead{Units} & \colhead{Definition}}
	\startdata
	$M_{\text{ej}}$        & $[0.1, 100]$              &  M$_\odot$     & Ejecta mass  \\
	%$M_{\text{Ni}}$        & \nodata                   &  \nodata       & Radioactive nickel mass  \\
	$f_{\text{Ni}}$        & $\log((0, 0.5])$          &                & Nickel mass as a fraction of the  ejecta mass  \\
	$v_{\text{ej}}$        & $\log([10^3, 10^5])$      &  km s$^{-1}$   & Ejecta velocity  \\
	%$E_{k}$                & \nodata                   &  \nodata       & Ejecta kinetic energy   \\
	$M_{\text{NS}}$        & $1.7 \pm 0.2$             &  M$_\odot$     & Neutron star mass   \\
	$P_{\text{spin}}$      & $[0.7, 30]$               &  ms            & Magnetar spin   \\
	$B_{\perp}$            & $\log((0, 15])$           &  $10^{14}$ G   & Magnetar magnetic field strength \\
	$\theta_{\text{BP}}$   & $[0, \pi/2]$              &  rad           & Angle of the dipole moment \\
	$t_{\text{exp}}$       & $[0, 200]$                &  days          & Explosion time relative to first data point  \\
	$T_{\text{min}}$       & $[3000, 10000]$           &  K             & Photosphere temperature floor  \\
	$\lambda$              & $[2000, 6000]$            &  \AA           & Flux below this wavelength is suppressed \\
	$\alpha$               & $[0, 5]$                  &                & Slope of the wavelength suppression \\
	$n_{H,\text{host}}$    & $\log([10^{16},10^{23}])$ &  cm$^{-2}$     & Column density in the host galaxy \\
	%$A_{V, \text{host}}$   & \nodata                   &  \nodata       & Extinction in the host galaxy   \\
	$\kappa$               & $[0.01, 0.34]$            &                & Optical opacity \\
	$\kappa_{\gamma}$      & $\log([0.01, 0.5])$       &  cm$^2$g$^{-1}$& Gamma-ray opacity  \\
	$\sigma$               & $[10^{-3}, 10^2]$         &                & Uncertainty required for $\chi^2_r=1$ \\
	\enddata
    \tablecomments{Parameters used in the \mosfit model, their priors, units, and definitions. Priors noted in $\log$ have a log-flat prior, priors without it are flat in linear space, and priors with a center and error bars have a Gaussian distribution.}
\end{deluxetable*}

To explore the properties of LSNe, and to enable a robust comparison to SLSNe and SNe Ic/Ic-BL, we model the light curves of all the LSNe, SLSNe, and SNe Ic/Ic-BL in our sample in a uniform way. To achieve this we use the Modular Open-Source Fitter for Transients ({\tt MOSFiT}) package, a flexible Python code that uses the {\tt emcee} \citep{Foreman13} Markov chain Monte Carlo (MCMC) implementation to fit the light curves of transients using a variety of different power sources \citep{guillochon18}. Since SNe Ic/Ic-BL are known to be powered by radioactive decay \citep{Filippenko97, Taddia19_broadlined}, and SLSNe are likely powered by a magnetar central engine \citep{Kasen10,Woosley10,Nicholl17_mosfit}, we model all light curves using a combined magnetar central engine plus radioactive decay model (designated {\tt slsnni}). By fitting all SNe with the same model we can evaluate which power source best reproduces their light curves, without imposing our own assumption based on properties such as peak magnitude or spectral classification.

The \mosfit setup for the magnetar model \citep{Kasen10,Woosley10} is described in detail in \citet{Nicholl17_mosfit}, while the radioactive component implementation is taken from \cite{Nadyozhin94}. The magnetar model imposes a constraint that penalizes models in which the total kinetic energy is higher than the magnetar energy plus neutrino energy, minus radiative losses. In the {\tt slsnni} implementation we use here, we relax this constraint to allow for additional energy from the radioactive decay component, effectively allowing the total kinetic energy to be higher. The extra luminosity comes from the maximum allowed energy from burning pure helium into nickel, or $\sim 10^{52} \times (M_{\rm Ni} / {\rm M}_\odot)$ erg, where $M_{\rm Ni}$ is the total nickel mass synthesized during the explosion.

The magnetar model in \mosfit has a modified blackbody SED, where flux bluewards of a cutoff wavelength $\lambda_0$ is suppressed by a factor proportional to $(\lambda / \lambda_0)^{\alpha}$ with a fixed $\alpha = 1$ and a variable $\lambda_0$, in order to account for the UV absorption seen in SLSNe (e.g., \citealt{Yan18}). In our modified {\tt slsnni} model we allow the power law index $\alpha$ of the suppression to vary in addition to $\lambda_0$, so that we can fit all SNe with the same uniform model, regardless of the location or steepness of the suppression. We are thus also able to fit the light curves of SNe that are reddened due to line-blanketing from radioactive decay.

\begin{figure*}
	\begin{center}
		\includegraphics[width=0.9\textwidth]{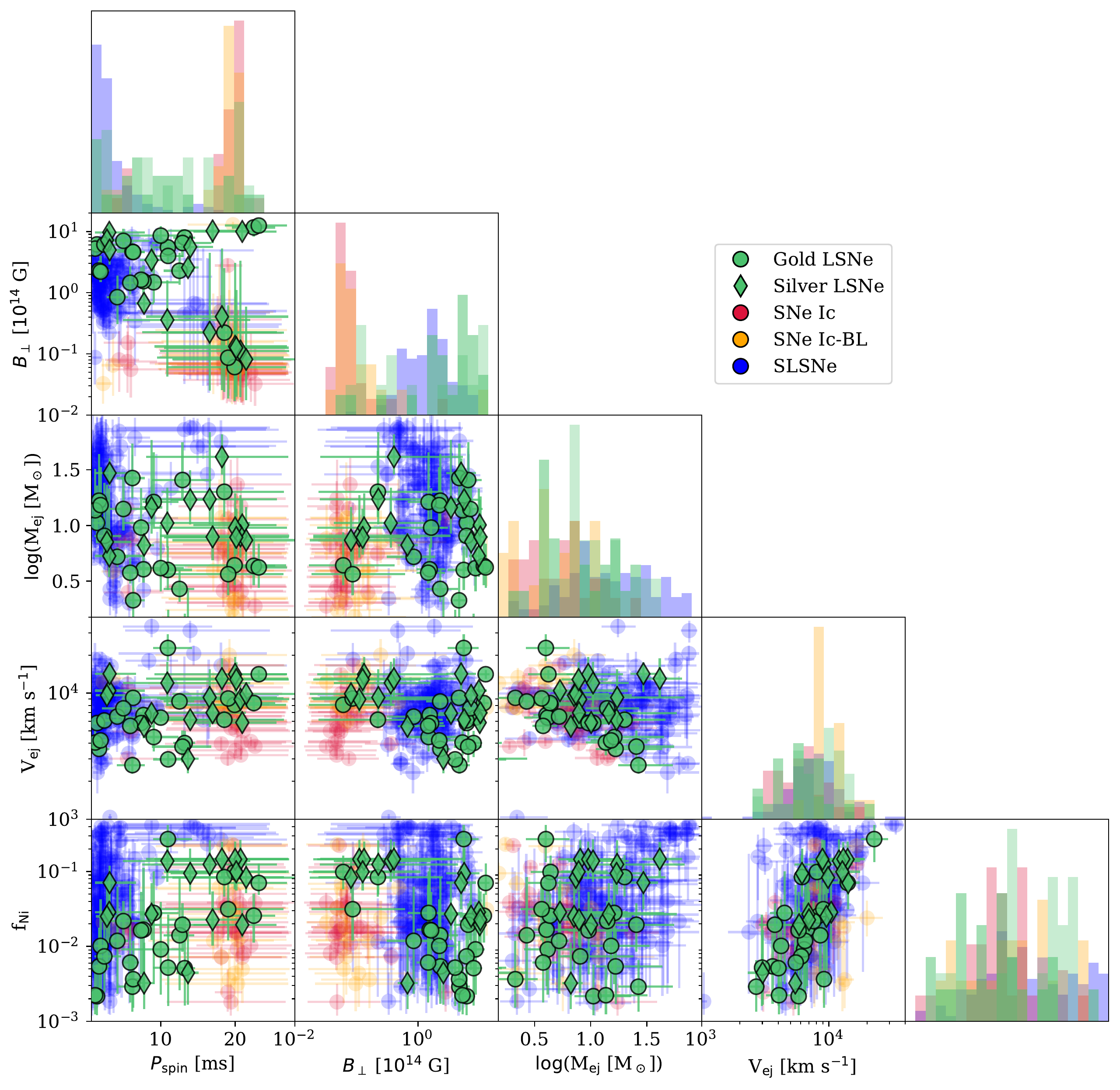}
		\caption{Best-fit \mosfit parameters for the SLSNe, SNe Ic, SNe Ic-BL, and LSNe populations. The green circles represent the Golden LSNe sample, while the diamonds represent the Silver sample. Other SNe types are shown as faded circles for comparison. We see that SLSNe and SNe Ic/Ic-BL separate very well in terms of most parameters, while LSNe span almost the entire range of allowable models, further emphasizing their intermediate nature. \label{fig:triangle}}
	\end{center}
\end{figure*}

\begin{figure}
	\begin{center}
		\includegraphics[width=\columnwidth]{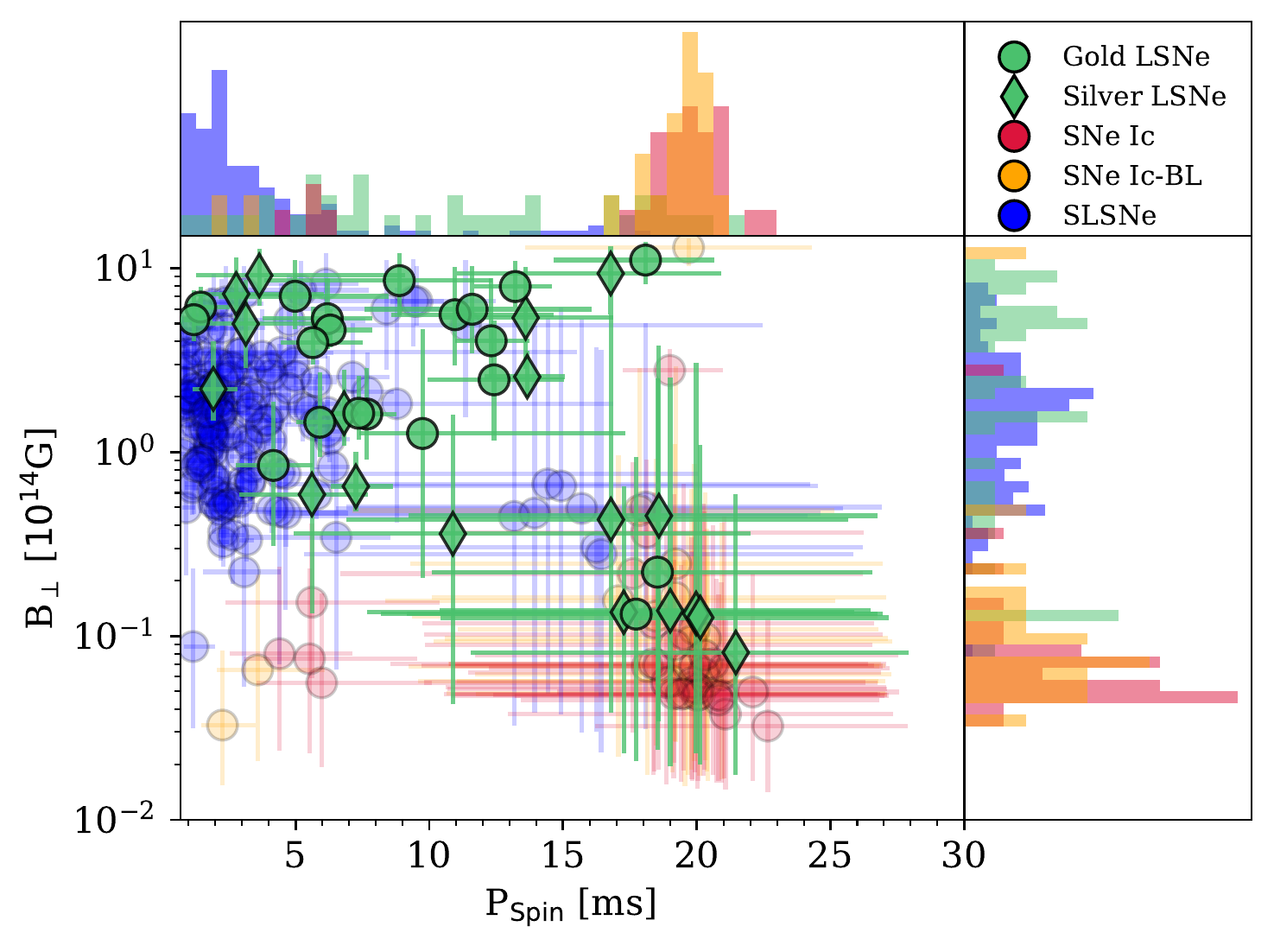}
		\caption{Best-fit Magnetar spin period and magnetic field values for the LSNe, SLSNe, SNe Ic, and SNe Ic-BL populations. The upper left corner, where most SLSNe lie, is dominated by powerful fast-spinning magnetars, whereas in the bottom right corner, populated by SNe Ic/Ic-BL, the magnetar power is low and the observed optical emission is dominated by radioactive decay. Some LSNe appear powered by magnetars similar to those in SLSNe, some have intermediate power magnetars, and some have weak or no magnetar contribution. \label{fig:magnetar}}
	\end{center}
\end{figure}

We retain similar model priors to those \cite{Nicholl17_mosfit} used to model SLSNe, with two modifications to accommodate the wider range of SNe modeled here. First, we impose a conservative upper limit on the total nickel mass fraction of $f_{\rm Ni} < 0.5$, higher than the typical ranges that SNe Ic/Ic-BL reach \citep{Barbarino20, Taddia19_broadlined}. And second, since we are unable to constrain the value for the neutron star mass $M_{\rm NS}$ in any model, we impose a Gaussian prior of $M_{\rm NS} = 1.7 \pm 0.2 $ M$_\odot$, similar to previous studies \citep{Blanchard20}, and motivated by the typical masses of neutron stars \citep{Ozel16}. The actual choice of prior for $M_{\rm NS}$ has no effect on the output parameters since the mass of the neutron star has a negligible effect on the output light curves. In Table~\ref{tab:parameters} we list all the model parameters, their priors, units, and definitions.

We fit the multi-band light curves of all LSNe and list the best-fit values and uncertainties in Table~\ref{tab:mosfit}. The uncertainties presented here represent only the statistical errors on the fits. In Table~\ref{tab:derived} we list additional parameters calculated from the posteriors of the fitted parameters. We measure the peak $r$-band magnitude of each SN from its light curve model to quantify the peak, even for the SNe that do not have $r$-band observations available. Table~\ref{tab:derived} also lists an estimated explosion date in MJD and a rise time, defined as the time from explosion date to maximum $r$-band brightness. In the same table, we list estimates for the total kinetic energy, $E_k = (3/10) M_{\rm ej} V_{\rm ej}^2$ and total nickel mass synthesized in the explosion, $M_{\rm Ni} = f_{\rm Ni} \times M_{\rm ej}$.

We explore which areas of parameter space LSNe could exist in by generating 10,000 LSNe input light curves based on the priors listed in Table~\ref{tab:parameters} and then selecting the output objects with a peak magnitude $M_r$ between $-19$ and $-20$ mag. We find the input and output distributions to be similar at the $\sim 90$\% level for almost every parameter, meaning if LSNe exist within our defined prior, we would be able to recover them. The one exception are SNe that have both a $P_{\text{spin}}> 9$ ms and a $f_{\text{Ni}} < 0.03$, as they all fall below our lower luminosity threshold of $M_r = -19$ and would likely appear to be normal SNe Ic.

\section{Modeling Results}
\label{sec:results}

\begin{figure*}[]
	\begin{center}
		\includegraphics[width=\columnwidth]{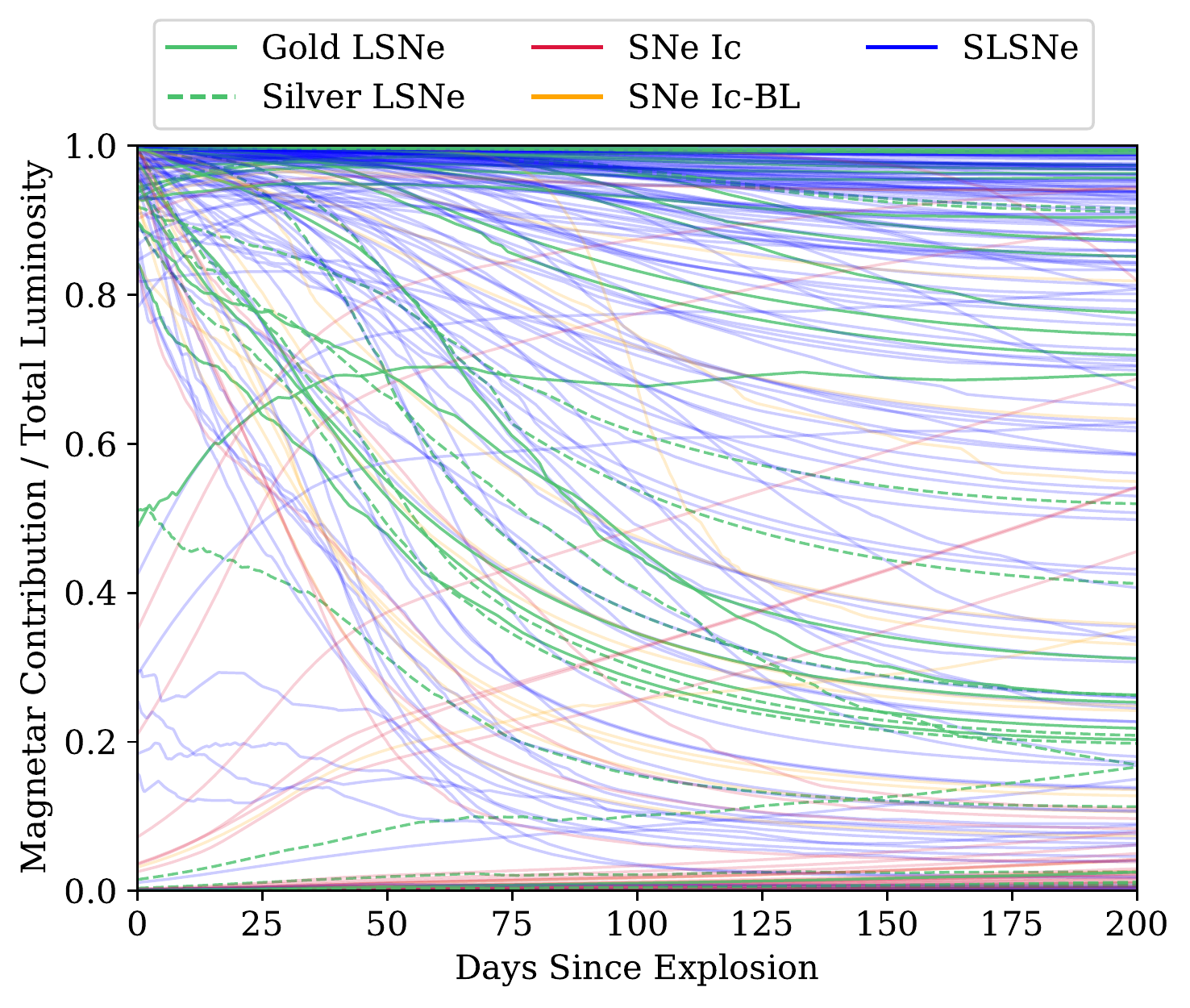}
		\includegraphics[width=\columnwidth]{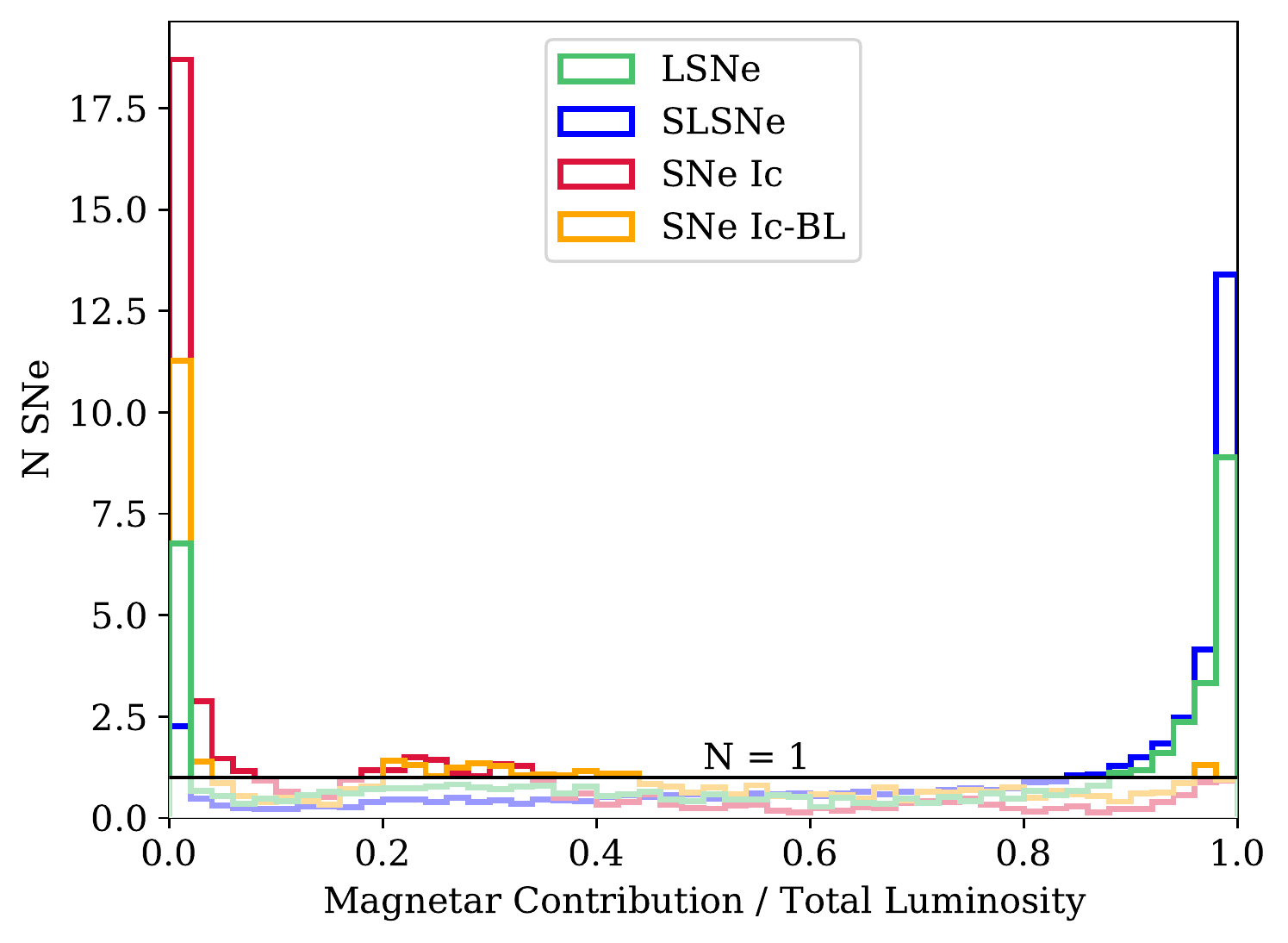}
		\caption{\textit{Left}: Fractional contribution of the magnetar component to the total output luminosity of the light curve (magnetar plus radioactive decay) for all the SNe in our sample as a function of days since explosion. \textit{Right}: Posterior model distribution for the same fractional magnetar contribution, but integrated over the first 200 days of the SN. We include 150 samples for each SN and normalize the populations by their respective sample sizes. SLSNe appear magnetar dominated and mostly stay as such throughout their evolution, whereas SNe Ic are mostly radioactively dominated. While some LSNe appear magnetar dominated, some seem to be powered entirely by radioactive decay. \label{fig:power}}
	\end{center}
\end{figure*}

In Figure~\ref{fig:triangle} we show the distribution of the most relevant physical parameters $P_{\text{spin}}$, $B_{\perp}$, $M_{\text{ej}}$, $v_{\text{ej}}$, and $f_{\text{Ni}}$ for the LSNe, SLSNe, SNe Ic, and SNe Ic-BL populations. In general, we find that SLSNe and SNe Ic/Ic-BL separate well in terms of most parameters; while LSNe span the whole range of allowed parameter space. Some LSNe have magnetar parameters ($P_{\text{spin}}$, $B_{\perp}$) that overlap the SLSNe population, consistent with powerful central engines, while some LSNe overlap the SNe Ic/Ic-BL population. The latter have weak or no evidence for magnetars, but instead appear powered by radioactive decay, as evidenced by their high $f_{\text{Ni}}$ values. The ejecta masses of LSNe span a wide range, from $\sim 1.5$ M$_\odot$ and up to $\sim 30$ M$_\odot$. We find the ejecta velocity estimates among all types of SNe to be very similar.

\subsection{Magnetar Parameters}

In order to quantify how different the parameter distributions of LSNe are from those of SLSNe and SNe Ic/Ic-BL, we implement a two-sample Kolmogorov-Smirnov (KS). A KS metric of $D = 0.0$ indicates the two populations are drawn from the same distribution, and $D = 1.0$ means there is no overlap between the distributions. We find a KS metric (and $p$ value) for the LSN distribution of $P_{\text{spin}}$ of $D = 0.63\ (< 10^{-3})$ and $D = 0.62\ (< 10^{-3})$ when compared to the SLSNe and SNe Ic/Ic-BL distributions, respectively. This indicates that the spin period distribution of LSNe is not similar to that of either SLSNe or SNe Ic/Ic-BL, as LSNe span a wider range of $P_{\text{spin}}$ values than either SLSNe or SNe Ic/Ic-BL. A similar result is found for the magnetic field strength, where we find values of $D = 0.34\ (< 10^{-3})$ and $D = 0.75\ (< 10^{-3})$ when LSNe are compared to SLSNe and SNe Ic/Ic-BL, respectively. This suggests that the LSNe $B_{\perp}$ distribution is very different from that of SNe Ic/Ic-BL, and still distinct (but less so) from the SLSN population.

In Figure~\ref{fig:magnetar} we focus on the distribution of magnetar parameters ($P_{\text{spin}}$, and $B_{\perp}$). While 90\% of SLSNe have $P_{\text{spin}}$ values $\lesssim 6.9$ ms, and 90\% of SNe Ic/Ic-BL have values of $P_{\text{spin}} \gtrsim 17$ ms (i.e., there is no evidence that they have a rapidly spinning magnetar engine), LSNe span the whole range of $P_{\text{spin}} \approx 2 -23$ ms. Similarly, while only $\sim 10$\% of SLSNe have spin periods $P_{\text{spin}} \approx 7 -17$ ms (and 3\% of SNe Ic/Ic-BL), 38\% of LSNe lie in this intermediate range. We note that some LSNe are best fit by magnetars with very strong magnetic fields and slow spin periods. The slow spin period makes it such that the contribution from the magnetar is negligible and the magnetic field strength becomes irrelevant in terms of its effect on the output light curve for SNe with slow spin periods.

In Figure~\ref{fig:power} we explore the relative contribution of the magnetar engine component to the total luminosity of the SNe. The left panel of Figure~\ref{fig:power} shows how this magnetar contribution evolves as a function of time. We find that 88\% of SLSNe have a significant magnetar contribution $\gtrsim 80$\% soon after explosion, and all SLSNe have at least some magnetar contribution above $10$\%. On the other hand, 60\% of SNe Ic/Ic-BL have a magnetar contribution $\lesssim 10$\%, and only 31\% have a magnetar contribution $\gtrsim 50$\% in the few days after explosion. LSNe span a wide range of magnetar contributions that overlap with both SLSNe and SNe Ic/Ic-BL. While 80\% of LSNe have a magnetar contribution $\gtrsim10$\%, 75\% of them have at least a 50\% magnetar contribution.

Along the same lines, in the right panel of Figure~\ref{fig:power} we show the same fractional magnetar contribution but integrated over the first 200 days after explosion for every SN. This histogram includes 150 samples for each SN, one for each realization (or walker) of the model light curves. We mark the threshold that corresponds to $N = 1$ SN, above which we can consider the measurements to be significant. Above this threshold, 92\% of SLSNe samples have a magnetar contribution $> 80$\%, and only 8\% have a contribution $< 10$\%. Conversely, only 5\% of SNe Ic-BL (and no SNe Ic) have a magnetar contribution $> 80$\%; and 45\% and 71\% of SNe Ic-BL and SNe Ic have a magnetar contribution $< 10$\%, respectively. LSNe show a more noticeable bifurcation, where 74\% of samples have a magnetar contribution $> 80$\% and 27\% have a magnetar contribution $< 10$\%.

\subsection{Ejecta Parameters}\label{sec:results_mass}

The parameter distributions for nickel mass fraction $f_{\text{Ni}}$ and ejecta mass $M_{\text{ej}}$ show less stark differences between LSNe and the other populations than the magnetar parameters. We find a KS metric (and $p$ value) for the distribution of $f_{\text{Ni}}$ values of SLNe of $D = 0.23\ (0.06)$ and $D = 0.25\ (0.09)$ compared to the SLSNe and SNe Ic/Ic-BL distributions, respectively. In terms of $M_{\text{ej}}$, we find a metric of $D = 0.39\ (< 10^{-3})$ and $D = 0.19\ (0.34)$ for the same populations. With the exception of the distribution of $M_{\text{ej}}$ between LSNe and SLSNe ($p < 10^{-3}$), we can not rule out the null hypothesis that these populations are drawn for the same distribution based on the KS metric. Finally, we measure values of $D = 0.14\ (0.5)$ and $D = 0.14\ (0.74)$ for the distribution of $v_{\text{ej}}$ values for LSNe compared to the SLSNe and SNe Ic/Ic-BL populations; suggesting that all populations of SNe are consistent with being drawn for the same distribution of $v_{\text{ej}}$ values.

\begin{figure}
	\begin{center}
		\includegraphics[width=\columnwidth]{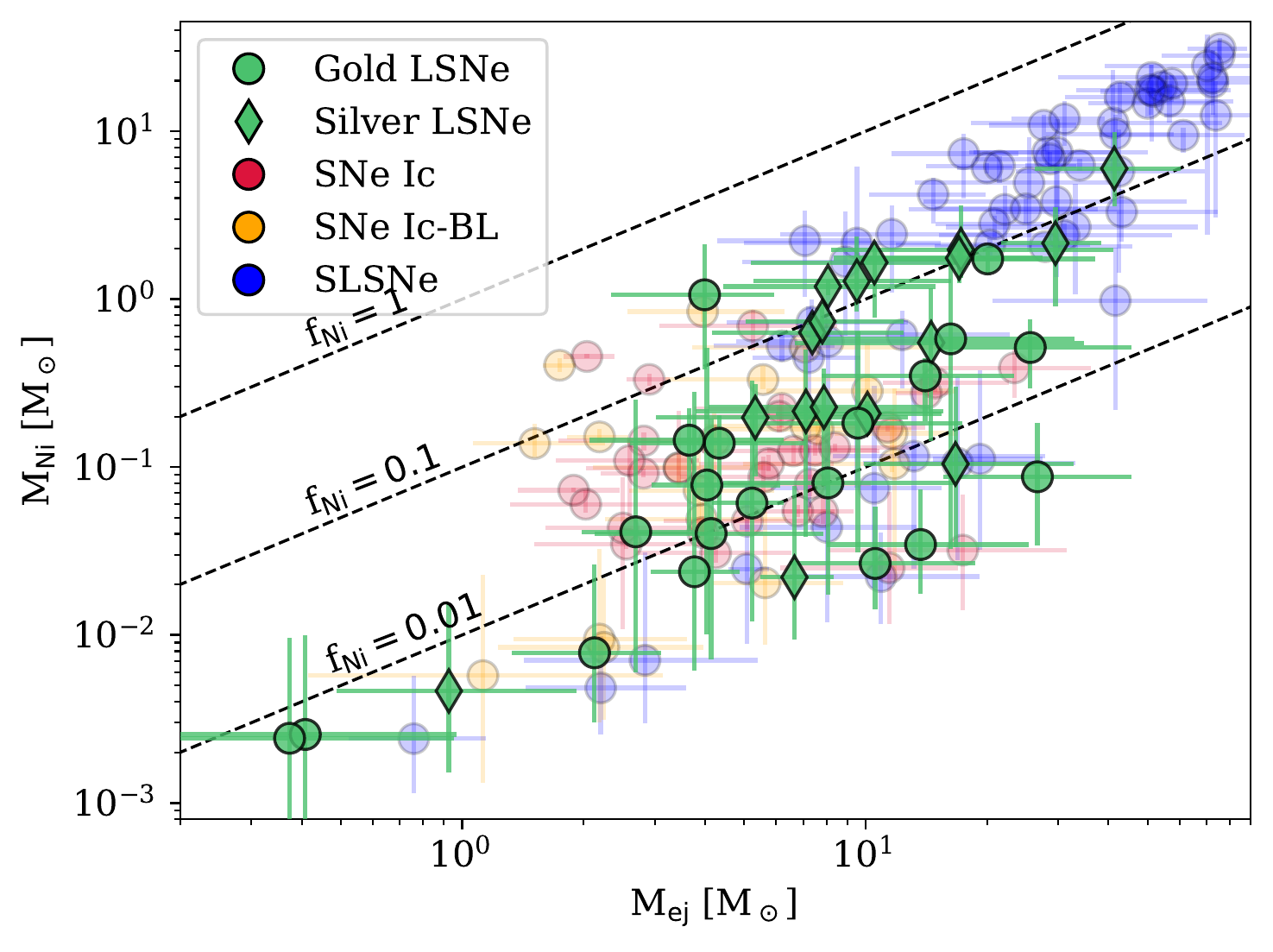}
		\caption{Best fit nickel mass as a function of ejecta mass for the LSNe, SLSNe, SNe Ic, and SNe Ic-BL populations. For clarity, we exclude the SLSNe with unconstrained nickel mass fractions (i.e., those for which even $f_{\rm Ni}=0.5$ still provides a sub-dominant contribution to the light curve). The dashed lines indicate $f_{\rm Ni}=0.01, 0.1, 1$ . LSNe mostly occupy a similar range to SNe Ic/Ic-BL, but also overlap the parameter space occupied by SLSNe. The higher luminosities of LSNe compared to SNe Ic/Ic-BL indicates an additional contribution from magnetar engines. \label{fig:mejecta}}
	\end{center}
\end{figure}

In Figure~\ref{fig:mejecta} we show the values of nickel mass as a function of ejecta mass. The main difference between SLSNe and SNe Ic is not necessarily the nickel mass or even the nickel mass fraction, but rather how well constrained this parameter is. Effectively all SNe Ic/Ic-BL have well-constrained nickel mass fractions, usually $f_{\rm Ni}\lesssim 0.1$, which translates to a total nickel mass of $M_{\rm Ni} \lesssim 0.5$ M$_\odot$. On the other hand, $\sim 50$\% of SLSNe and LSNe have an unconstrained nickel mass fractions with an uncertainty on $\log(f_{\rm Ni}) \gtrsim 0.5$. For some SLSNe, the posterior reaches the $f_{\rm Ni} = 0.5$ limit imposed by the prior. In these situations, the light curves are dominated by the magnetar component, and the value for $f_{\rm Ni}$ can therefore not be constrained. As expected, LSNe have nickel masses that span a wide range of possibilities, while some appear magnetar-dominated, others are best fit by a radioactively powered light curve. We find that the values of $f_{\rm Ni}$ for the radioactively dominated LSNe span the same range as SNe Ic/Ic-BL, $f_{\rm Ni} \approx 0.01 - 0.1$.

\begin{figure}
	\begin{center}
		\includegraphics[width=\columnwidth]{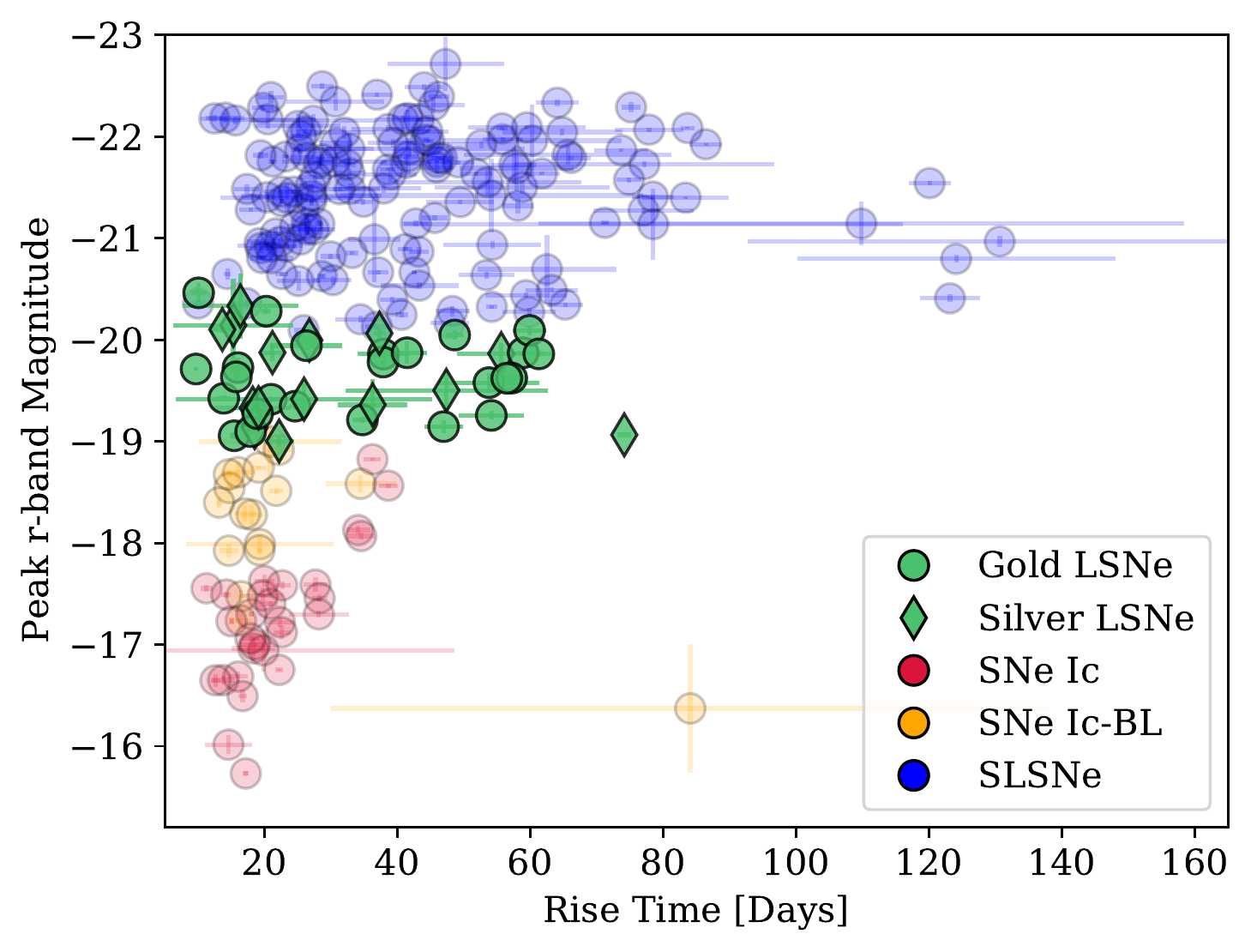}
		\caption{Rise time as a function of peak $r$-band absolute magnitude for the LSNe, SLSNe, SNe Ic, and SNe Ic-BL populations. We find that, by definition, LSNe occupy a very clear space in terms of peak magnitude, but they also have intermediate rise times between those of SLSNe and SNe Ic/Ic-BL. \label{fig:peak_magnitudes}}
	\end{center}
\end{figure}

Figure~\ref{fig:peak_magnitudes} shows how the rise time of LSNe compares to those of SLSNe and SNe Ic/Ic-BL. LSNe occupy a very distinct space in terms of peak magnitude (by definition), but also tend to have intermediate rise times between SLSNe and SNe Ic. The rise times of LSNe span the range from $\sim 20$ to $\sim 65$ days, similar but slightly shorter than the $\sim 20$ to 90 days for SLSNe, but significantly wider than the $\lesssim 30$ days of SNe Ic/Ic-BL. We explore the possibility that this ``broadening'' of allowed rise times is caused by the underlying power source but conclude this does not appear to be the case, since we find no strong correlation between the dominant power source and the rise time of the transient. Instead, the parameter controlling the rise-time difference appears to be the ejecta mass.

\begin{figure}
	\begin{center}
		\includegraphics[width=\columnwidth]{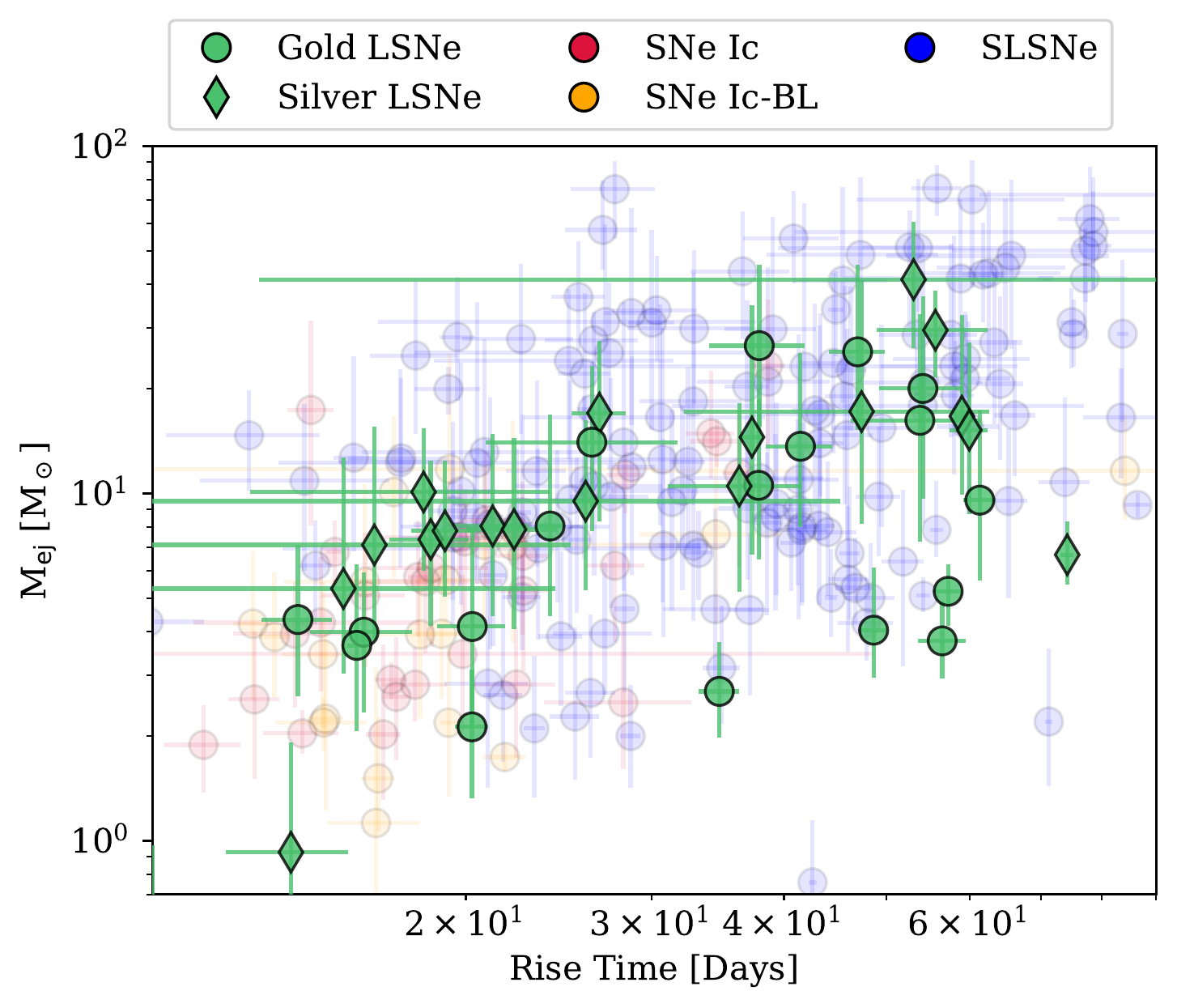}
		\caption{Ejecta mass as a function of rise time for the LSNe, SLSNe, SNe Ic, and SNe Ic-BL population. We see that, like previously established for SLSNe (e.g., \citealt{Nicholl15, Konyves21}) and SNe Ic/Ic-BL (e.g., \citealt{Dessart16}), LSNe with longer rise times tend to have larger ejecta masses. \label{fig:rise_mejecta}}
	\end{center}
\end{figure}

The rise times of CCSNe has been shown to correlate with ejecta mass, as higher ejecta masses lead to longer rise times (e.g., \citealt{Dessart16}). This correlation is to be expected since the diffusion timescale is proportional to $\propto (M_{\rm ej} / v_{\rm ej})^{1/2}$, and shown to be evident in SLSNe (e.g., \citealt{Nicholl15, Konyves21}). In Figure~\ref{fig:rise_mejecta} we show this trend for all SNe in our sample and see that it appears to also apply to LSNe. None of the fast-evolving LSNe with rise times $< 25$ days have high ejecta masses above $20$ M$_\odot$, and most of the slowly evolving LSNe with rise times $> 30$ days have ejecta masses between $10-40$ M$_\odot$. Out of the 40 LSNe, four (SN\,2019dwa, SN\,2019stc, SN\,2021uvy, and DES16C3cv) appear to deviate from this trend, since they have rise times between 30 and 70 days but ejecta masses $< 6$ M$_\odot$. 

\begin{figure}[h!]
	\begin{center}
		\includegraphics[width=\columnwidth]{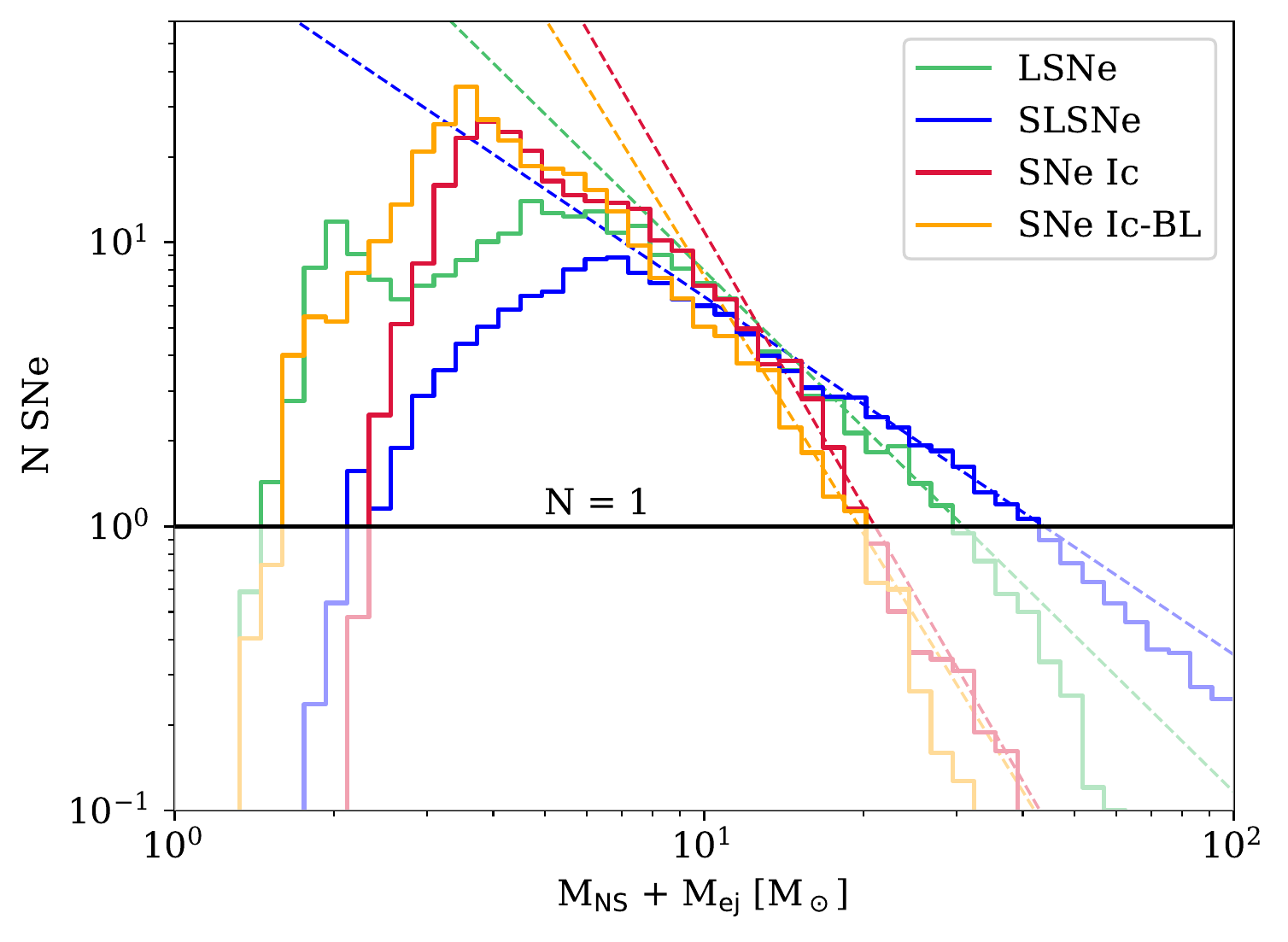}
		\caption{Pre-explosion mass distribution for the LSNe, SLSNe, SNe Ic, and SNe Ic-BL populations, normalized by their respective sample sizes. We estimate the progenitor mass by summing the best fit value of the ejecta mass with the mass of the remnant neutron star. SLSNe have a sharp drop-off at the low-mass end, where SNe Ic peak, and LSNe are intermediate between these two populations. The same trend is seen in the high-mass end: SLSNe extend to very high masses of $\sim 40$ M$_\odot$, LSNe appear to max out at $\sim 30$ M$_\odot$, and SNe Ic/Ic-BL have almost no objects above $\sim 20$ M$_\odot$. Dashed lines are linear fits to the distributions above $10$ M$_\odot$ and below the $N = 1$ SN threshold. The peak in the LSNe distribution at $\sim 2$ M$_\odot$ is driven entirely by two objects with tightly constrained $M_{\text{ej}}$ values (iPTF16asu and SN\,2021lwz). \label{fig:mass_distribution}}
	\end{center}
\end{figure}

In Figure~\ref{fig:mass_distribution} we show the pre-explosion mass distribution for the LSNe, SLSNe, SNe Ic, and SNe Ic-BL populations, calculated by summing the posteriors for the ejecta mass and neutron star mass of each SN. The histogram includes 150 samples for each SN, one for each model realization (or walker). We indicate with a shaded region the threshold where the number of samples is equivalent to one SN, below which the measurements are not significant. The distribution of LSNe pre-explosion masses is intermediate to those of SLSNe and SNe Ic/Ic-BL. This is particularly evident at the high mass end, where the populations appear clearly distinct. LSNe extend to masses of $\sim 30$ M$_\odot$, higher than the $\sim 20$ M$_\odot$ limit for SNe Ic/Ic-BL, but not as high as SLSNe, which extend up to $\sim 40$ M$_\odot$. While SLSNe have a sharp drop off at the low-mass end below $\sim 2$ M$_\odot$, the mass distribution for LSNe extends as low as those of SNe Ic/Ic-BL, down to $\sim 1.5$ M$_\odot$. To quantify this distinction we fit a linear slope of the form $d N / d \log M \propto M ^ \alpha$ to the mass distribution above $10$ M$_\odot$ and up to where the distributions reach the $1$ SN threshold and find best fit values of $\alpha$ for the different populations of: $\alpha = -1.26 \pm 0.04$ (SLSNe), $\alpha = -1.83 \pm 0.07$ (LSNe), $\alpha = -3.21 \pm 0.23$ (SN Ic), $\alpha = -3.02 \pm 0.26$ (SN Ic-BL). LSNe have a steeper slope than SLSNe, but shallower than SNe Ic/Ic-BL, suggesting their progenitors might be a mix of SLSN-like and SNe Ic-like. The value we obtain for SLSNe perfectly matches the one found by \cite{Blanchard20} of $\alpha = -1.26 \pm 0.06$. We find the peaks of the distributions to be $\sim 3.5$ M$_\odot$ (SNe Ic-BL),  $\sim 4.0$ M$_\odot$ (SNe Ic),  $\sim 5.6$ M$_\odot$ (LSNe), and $\sim 6.6$ M$_\odot$ (SLSNe). Our estimate for the peak of the SLSNe distribution is higher than the $\sim 4$ M$_\odot$ found by \cite{Blanchard20} as a result of our models allowing for an additional energy component from radioactive decay that is not present in their models.

\subsection{Comparison To Other Studies}

We compare our best fit $M_{\rm ej}$ values obtained from \mosfit to the results from several independent studies. \cite{Taddia19_broadlined} and \cite{Barbarino20} fit the bolometric light curves of SNe Ic-BL and SNe Ic, respectively, with an Arnett model \citep{Arnett82} to measure the ejecta masses of these SNe. \cite{Konyves21} used a combination of bolometric light curve models and ejecta velocity measurements to estimate the ejecta mass of SLSNe. \cite{Jerkstrand17_slsn} modeled the spectra of SLSNe with the SUMO spectral synthesis code \citep{Jerkstrand12} to select the best ejecta mass estimates from a grid of models. And finally, \cite{Mazzali16} modeled the spectra of SLSNe using a Monte Carlo spectral synthesis code \citep{Mazzali93} to infer the ejecta masses of SLSNe.

We find our estimates of $M_{\rm ej}$ for SLSNe and SNe Ic/Ic-BL to be in good agreement with values estimated by these studies. Five of the LSNe in this work are included in these studies. We find an ejecta mass for SN\,2011kl of $M_{\rm ej} = 3.9 \pm 1.8$ M$_\odot$, consistent with the estimate of $M_{\rm ej} = 2 - 3$ M$_\odot$ from \cite{Mazzali16}. We estimate $M_{\rm ej} = 9.6^{+7.8}_{-4.0}$ M$_\odot$ for PTF12gty, within $2\sigma$ of the average ejecta mass estimate of $M_{\rm ej} = 20.7 \pm 6.0$ M$_\odot$ found by \cite{Konyves21}, but consistent with their ``Equation 8'' estimate of $M_{\rm ej} = 14.68$ M$_\odot$. The ejecta masses we find for iPTF13dnt ($9.5^{+6.8}_{-4.2}$ M$_\odot$), iPTF16asu ($0.4^{+0.6}_{-0.2}$ M$_\odot$), and iPTF17cw ($4.3^{+2.8}_{-1.7}$ M$_\odot$) are all within $1\sigma$ of the estimates from \cite{Taddia19_broadlined} of $7.0 \pm 2.6$ M$_\odot$, $0.9 \pm 0.1$ M$_\odot$, and $4.5 \pm 1.8$ M$_\odot$, respectively.

\begin{longrotatetable}
\begin{deluxetable}{cccccccccccc}
    \tablecaption{Fitted Parameters \label{tab:mosfit}}
    \tablehead{\colhead{Name} & \colhead{$B_{\perp}$} & \colhead{$\lambda_P$} & \colhead{$\lambda$}  & \colhead{$f_{\text{Ni}}$} & \colhead{$\kappa$} & \colhead{$\log{(\kappa_{\gamma})}$} & \colhead{$M_{\text{ej}}$} & \colhead{$P_{\text{spin}}$} & \colhead{$T_{\text{min}}$} & \colhead{$\theta_{\text{PB}}$} & \colhead{$V_{\text{ej}}$} \\
                    &  ($10^{14}$ G) &   &   (1000 \AA) &      &      &   (cm$^2$g$^{-1}$)   &   (M$_\odot$)  &  (ms)   &  (1000K) &  (rad) &  (1000 km s$^{-1}$)}
    \startdata
    DES14C1rhg      & $8.5 \pm 3.2$         & $2.0 \pm 1.7$         & $2.73 \pm 0.55$           & $0.01^{+0.03}_{-0.01}$    & $0.05^{+0.05}_{-0.03}$    & $0.07^{+0.19}_{-0.05}$    & $4.1^{+3.7}_{-2.0}$   & $10.0^{+3.1}_{-5.3}$  & $5.1 \pm 1.54$            & $1.06 \pm 0.34$           & $6.39 \pm 0.55$       \\
    DES15C3hav      & $6.1^{+3.4}_{-1.6}$   & $0.2 \pm 0.2$         & $4.01 \pm 1.31$           & $0.01^{+0.02}_{-0.0}$     & $0.04^{+0.03}_{-0.02}$    & $0.09^{+0.19}_{-0.06}$    & $8.1^{+8.8}_{-3.6}$   & $2.4^{+2.3}_{-1.3}$   & $7.56 \pm 0.76$           & $0.96 \pm 0.42$           & $6.07 \pm 0.64$       \\
    DES16C3cv       & $0.8^{+1.0}_{-0.5}$   & $3.2^{+1.0}_{-0.6}$   & $3.43 \pm 0.19$           & $0.01^{+0.05}_{-0.01}$    & $0.28 \pm 0.05$           & $0.01^{+0.0}_{-0.0}$  & $5.2 \pm 1.1$         & $4.2 \pm 1.4$         & $5.04 \pm 1.44$           & $0.29 \pm 0.15$           & $6.52 \pm 0.52$       \\
    iPTF13dnt       & $0.1^{+2.9}_{-0.1}$   & $2.8 \pm 1.7$         & $4.05 \pm 1.35$           & $0.15^{+0.11}_{-0.07}$    & $0.09^{+0.08}_{-0.05}$    & $0.05^{+0.21}_{-0.03}$    & $9.5^{+6.8}_{-4.2}$   & $20.0^{+7.0}_{-10.8}$     & $3.79^{+0.83}_{-0.51}$    & $0.77 \pm 0.56$           & $14.08 \pm 4.52$      \\
    iPTF16asu       & $7.9^{+3.0}_{-1.8}$   & $2.5 \pm 0.3$         & $5.49 \pm 0.13$           & $0.01^{+0.02}_{-0.0}$     & $0.05^{+0.06}_{-0.02}$    & $0.03^{+0.04}_{-0.01}$    & $0.4^{+0.6}_{-0.2}$   & $13.2 \pm 1.5$            & $7.78^{+1.45}_{-3.49}$    & $1.02 \pm 0.37$           & $4.02 \pm 0.96$       \\
    iPTF17cw        & $11.6 \pm 2.9$            & $2.7^{+1.2}_{-0.5}$   & $5.17 \pm 0.31$           & $0.03^{+0.03}_{-0.01}$    & $0.03^{+0.02}_{-0.01}$    & $0.09^{+0.19}_{-0.06}$    & $4.3^{+2.8}_{-1.7}$   & $22.5 \pm 2.9$            & $6.11 \pm 0.53$           & $1.21 \pm 0.3$    & $8.34 \pm 1.24$       \\
    OGLE15xl        & $0.4^{+1.2}_{-0.3}$   & $2.2 \pm 1.7$         & $3.93 \pm 1.32$           & $0.14^{+0.13}_{-0.07}$    & $0.1^{+0.14}_{-0.06}$     & $0.23 \pm 0.15$           & $10.5 \pm 6.5$            & $10.9^{+11.1}_{-6.0}$     & $5.35^{+2.53}_{-1.37}$    & $0.8 \pm 0.55$    & $12.04 \pm 4.89$      \\
    PS15cvn         & $0.1^{+0.5}_{-0.1}$   & $3.9 \pm 0.7$         & $4.79 \pm 0.11$           & $0.03^{+0.04}_{-0.02}$    & $0.07 \pm 0.01$           & $0.01^{+0.0}_{-0.0}$  & $3.7^{+2.6}_{-1.6}$   & $19.1 \pm 8.5$            & $6.34 \pm 0.22$           & $0.64 \pm 0.56$       & $9.07 \pm 0.66$       \\
    PTF10gvb        & $12.5^{+1.8}_{-2.9}$  & $1.5^{+2.3}_{-1.1}$   & $3.33^{+1.43}_{-0.89}$    & $0.07^{+0.06}_{-0.02}$    & $0.08^{+0.05}_{-0.03}$    & $0.21^{+0.18}_{-0.11}$    & $4.2 \pm 1.9$         & $23.2 \pm 4.2$            & $4.85 \pm 0.2$            & $1.22 \pm 0.25$           & $14.11 \pm 1.09$      \\
    PTF10iam        & $2.6^{+1.0}_{-0.6}$   & $3.4 \pm 0.9$         & $5.83 \pm 0.16$           & $<0.01$   & $0.02 \pm 0.01$           & $0.1^{+0.2}_{-0.07}$  & $0.9^{+1.0}_{-0.4}$   & $13.7 \pm 1.5$            & $6.61 \pm 2.0$            & $1.06 \pm 0.37$           & $3.02 \pm 0.86$       \\
	PTF11img     	& $0.1^{+0.8}_{-0.1}$ 	& $1.9^{+2.0}_{-0.9}$ 	& $4.65 \pm 0.67$		 	& $0.22^{+0.11}_{-0.07}$ 	& $0.12^{+0.07}_{-0.04}$ 	& $0.02^{+0.01}_{-0.0}$ 	& $3.9^{+2.3}_{-1.4}$ 	& $18.5 \pm 8.7$		 	& $5.52 \pm 0.44$		 	& $0.71 \pm 0.55$		 	& $16.54 \pm 3.12$		\\
    PTF12gty        & $1.6^{+1.0}_{-0.5}$   & $1.3^{+2.1}_{-1.0}$   & $2.95^{+1.2}_{-0.67}$     & $0.02^{+0.04}_{-0.01}$    & $0.16^{+0.11}_{-0.07}$    & $0.15^{+0.21}_{-0.09}$    & $9.6^{+7.8}_{-4.0}$   & $7.4 \pm 0.8$         & $5.26 \pm 0.33$           & $0.96 \pm 0.41$           & $5.79 \pm 0.49$       \\
    PTF12hni        & $7.0^{+4.1}_{-2.4}$   & $1.6^{+1.7}_{-0.9}$   & $4.36^{+0.99}_{-1.76}$    & $0.02^{+0.02}_{-0.01}$    & $0.05^{+0.03}_{-0.02}$    & $0.1^{+0.2}_{-0.08}$  & $14.1 \pm 7.8$            & $5.0 \pm 3.2$         & $5.48^{+0.55}_{-0.3}$     & $1.08 \pm 0.41$           & $7.58 \pm 1.63$       \\
    SN\,1991D           & $9.7 \pm 3.2$         & $2.1 \pm 1.7$         & $3.54 \pm 1.13$           & $0.03^{+0.05}_{-0.02}$    & $0.05 \pm 0.04$           & $0.04^{+0.05}_{-0.03}$    & $5.3^{+7.4}_{-2.3}$   & $3.1^{+4.8}_{-1.9}$   & $4.57^{+0.38}_{-0.25}$    & $1.07 \pm 0.34$       & $10.45 \pm 1.8$       \\
    SN\,2003L           & $0.2^{+4.2}_{-0.2}$   & $2.2 \pm 1.6$         & $4.15 \pm 1.35$           & $0.12^{+0.14}_{-0.08}$    & $0.09^{+0.13}_{-0.06}$    & $0.08^{+0.22}_{-0.06}$    & $17.2^{+23.9}_{-9.0}$     & $16.6 \pm 9.0$            & $6.58 \pm 2.41$           & $0.8 \pm 0.53$            & $9.24^{+6.22}_{-3.85}$\\
    SN\,2007ce          & $10.0 \pm 2.7$            & $2.6 \pm 0.2$         & $5.43 \pm 0.07$           & $0.02 \pm 0.01$           & $0.02^{+0.01}_{-0.0}$     & $0.21 \pm 0.15$           & $10.1 \pm 4.7$            & $20.9 \pm 5.6$            & $7.26 \pm 0.17$           & $1.16 \pm 0.31$           & $5.84 \pm 1.0$        \\
    SN\,2009cb          & $7.2^{+4.2}_{-2.5}$   & $2.2 \pm 1.8$         & $2.99 \pm 0.64$           & $0.03^{+0.04}_{-0.02}$    & $0.04 \pm 0.02$           & $0.06^{+0.2}_{-0.05}$     & $7.1^{+8.4}_{-3.3}$   & $2.8^{+3.3}_{-1.5}$   & $4.74 \pm 1.5$            & $1.07 \pm 0.42$       & $9.59 \pm 1.88$       \\
    SN\,2010ay          & $0.1^{+3.2}_{-0.1}$   & $3.3 \pm 1.1$         & $4.7 \pm 0.25$            & $0.15^{+0.12}_{-0.07}$    & $0.02^{+0.02}_{-0.01}$    & $0.03^{+0.07}_{-0.02}$    & $8.1^{+6.8}_{-3.6}$   & $20.7^{+6.6}_{-10.8}$     & $5.14 \pm 1.46$           & $0.73 \pm 0.57$       & $8.93 \pm 2.31$       \\
    SN\,2011kl          & $5.5^{+4.0}_{-2.5}$   & $2.1 \pm 1.9$         & $2.41 \pm 0.34$           & $0.27 \pm 0.14$           & $0.26^{+0.05}_{-0.09}$    & $0.01^{+0.0}_{-0.0}$  & $3.9 \pm 1.8$         & $11.0^{+3.2}_{-2.0}$  & $9.35 \pm 0.49$           & $0.93 \pm 0.44$       & $22.8 \pm 5.69$       \\
    SN\,2012aa          & $4.7^{+2.2}_{-1.2}$   & $4.1 \pm 0.7$         & $4.84 \pm 0.16$           & $<0.0$    & $0.03^{+0.03}_{-0.01}$    & $0.07^{+0.22}_{-0.05}$    & $26.7^{+19.0}_{-11.1}$    & $6.2 \pm 1.9$         & $6.45 \pm 0.49$           & $1.05 \pm 0.39$           & $2.69 \pm 0.41$       \\
    SN\,2013hy          & $5.6^{+4.3}_{-2.7}$   & $2.6^{+1.5}_{-0.9}$   & $3.25^{+0.27}_{-0.17}$    & $0.09^{+0.11}_{-0.04}$    & $0.02^{+0.02}_{-0.01}$    & $0.07^{+0.19}_{-0.05}$    & $17.0 \pm 9.6$            & $14.0 \pm 2.9$            & $7.88 \pm 0.42$           & $0.97 \pm 0.44$   & $6.16 \pm 0.71$       \\
    SN\,2018beh         & $6.1^{+1.8}_{-1.2}$   & $0.1^{+0.2}_{-0.1}$   & $3.07^{+1.7}_{-0.85}$     & $<0.0$    & $0.07^{+0.05}_{-0.03}$    & $0.09^{+0.2}_{-0.06}$     & $10.6^{+8.1}_{-4.1}$  & $1.5^{+1.1}_{-0.5}$   & $4.4 \pm 1.06$            & $1.19 \pm 0.32$           & $5.79 \pm 0.18$       \\
    SN\,2018don         & $0.7 \pm 0.3$         & $4.6^{+0.3}_{-0.5}$   & $4.59 \pm 0.06$           & $<0.01$   & $0.26 \pm 0.05$           & $0.01^{+0.0}_{-0.0}$  & $6.7 \pm 1.4$         & $7.8 \pm 1.2$         & $3.91^{+0.37}_{-0.15}$    & $0.83 \pm 0.27$           & $5.45 \pm 0.33$       \\
    SN\,2018fcg         & $4.6^{+1.4}_{-0.9}$   & $0.5^{+1.5}_{-0.3}$   & $3.78 \pm 1.52$           & $<0.01$   & $0.09^{+0.05}_{-0.03}$    & $0.22 \pm 0.13$           & $2.1 \pm 0.9$         & $6.3^{+1.6}_{-1.0}$   & $4.72 \pm 0.13$           & $1.13 \pm 0.33$           & $9.17^{+0.53}_{-0.82}$\\
    SN\,2019cri         & $6.5^{+4.3}_{-2.7}$   & $2.4 \pm 1.3$         & $5.17 \pm 0.44$           & $0.02^{+0.02}_{-0.01}$    & $0.03^{+0.03}_{-0.02}$    & $0.07^{+0.21}_{-0.05}$    & $25.6^{+19.9}_{-10.3}$    & $12.9 \pm 4.0$            & $6.05 \pm 0.88$           & $0.99 \pm 0.44$           & $3.77 \pm 1.11$       \\
    SN\,2019dwa         & $2.3^{+2.2}_{-0.9}$   & $1.7^{+2.3}_{-1.4}$   & $3.04^{+1.11}_{-0.74}$    & $0.01^{+0.07}_{-0.01}$    & $0.26 \pm 0.07$           & $0.01^{+0.01}_{-0.0}$     & $2.7 \pm 0.9$         & $12.5 \pm 2.0$            & $4.67 \pm 0.4$            & $0.94 \pm 0.42$   & $8.59 \pm 0.79$       \\
    SN\,2019gam         & $5.0^{+3.7}_{-2.1}$   & $2.4 \pm 1.7$         & $4.46^{+0.68}_{-0.3}$     & $0.07 \pm 0.04$           & $0.28^{+0.05}_{-0.08}$    & $0.02^{+0.02}_{-0.01}$    & $29.5 \pm 8.4$            & $3.1^{+3.5}_{-2.0}$   & $9.28^{+0.53}_{-1.13}$    & $1.09 \pm 0.36$       & $14.23 \pm 3.68$      \\
    SN\,2019hge         & $2.3^{+1.0}_{-0.5}$   & $0.2^{+0.2}_{-0.1}$   & $3.69 \pm 1.28$           & $0.01^{+0.01}_{-0.0}$     & $0.05^{+0.04}_{-0.02}$    & $0.06^{+0.21}_{-0.04}$    & $16.7^{+16.0}_{-6.8}$     & $1.7^{+1.4}_{-0.7}$   & $5.42 \pm 1.71$           & $1.04 \pm 0.39$           & $3.61 \pm 0.32$       \\
    SN\,2019J           & $5.2^{+2.3}_{-1.2}$   & $1.1^{+1.9}_{-0.8}$   & $2.63^{+1.31}_{-0.48}$    & $<0.0$    & $0.06^{+0.06}_{-0.03}$    & $0.06^{+0.19}_{-0.05}$    & $13.7^{+11.7}_{-5.7}$     & $1.2^{+0.7}_{-0.4}$   & $9.29^{+0.5}_{-0.79}$     & $1.01 \pm 0.4$            & $4.06 \pm 0.95$       \\
    SN\,2019moc         & $0.1^{+0.4}_{-0.0}$   & $4.2 \pm 0.7$         & $4.61 \pm 0.13$           & $0.1^{+0.06}_{-0.04}$     & $0.02 \pm 0.01$           & $0.05^{+0.18}_{-0.04}$    & $4.4^{+3.0}_{-1.8}$   & $19.9 \pm 8.0$        & $4.26^{+0.58}_{-0.28}$    & $0.63 \pm 0.54$           & $8.08 \pm 1.11$       \\
    SN\,2019obk         & $3.4^{+2.7}_{-1.3}$   & $0.8^{+2.1}_{-0.7}$   & $3.28^{+1.46}_{-0.92}$    & $0.03^{+0.04}_{-0.01}$    & $0.03^{+0.04}_{-0.02}$    & $0.09^{+0.19}_{-0.06}$    & $14.5^{+20.3}_{-7.9}$     & $8.8 \pm 1.5$         & $6.49^{+1.32}_{-0.71}$    & $0.92 \pm 0.46$           & $6.92 \pm 1.23$       \\
    SN\,2019pvs         & $1.5^{+3.7}_{-0.8}$   & $1.8^{+2.1}_{-1.4}$   & $3.2 \pm 0.94$            & $0.03^{+0.07}_{-0.02}$    & $0.03^{+0.05}_{-0.02}$    & $0.16^{+0.2}_{-0.11}$     & $16.2^{+16.7}_{-9.0}$     & $9.1^{+3.3}_{-1.5}$   & $8.71 \pm 1.03$           & $0.8 \pm 0.53$            & $4.48^{+2.17}_{-1.04}$\\
    SN\,2019stc         & $1.5^{+1.2}_{-0.6}$   & $1.1^{+1.9}_{-0.7}$   & $4.34^{+0.89}_{-1.64}$    & $0.02^{+0.1}_{-0.01}$     & $0.24 \pm 0.07$           & $0.01^{+0.01}_{-0.0}$     & $4.0^{+2.1}_{-1.1}$   & $7.7 \pm 1.0$         & $5.51^{+0.52}_{-0.96}$    & $0.92 \pm 0.44$           & $6.71 \pm 1.09$       \\
    SN\,2019unb         & $2.2^{+1.8}_{-0.7}$   & $0.2^{+0.4}_{-0.1}$   & $3.58 \pm 1.51$           & $0.01^{+0.03}_{-0.01}$    & $0.06^{+0.06}_{-0.03}$    & $0.1^{+0.19}_{-0.07}$     & $15.2^{+12.0}_{-6.5}$     & $1.9 \pm 0.8$         & $8.41 \pm 1.33$           & $0.98 \pm 0.47$       & $4.22 \pm 0.5$        \\
    SN\,2019uq          & $10.2 \pm 3.5$            & $2.0 \pm 1.8$         & $3.07 \pm 0.85$           & $0.02^{+0.03}_{-0.01}$    & $0.03^{+0.03}_{-0.01}$    & $0.05^{+0.21}_{-0.04}$    & $7.9^{+6.6}_{-3.8}$   & $17.0 \pm 5.3$            & $5.86^{+1.13}_{-1.83}$    & $1.11 \pm 0.36$           & $6.99 \pm 1.26$       \\
    SN\,2019wpb         & $0.1^{+0.5}_{-0.1}$   & $3.0 \pm 1.5$         & $4.3^{+0.52}_{-0.93}$     & $0.08^{+0.06}_{-0.03}$    & $0.02^{+0.02}_{-0.01}$    & $0.07^{+0.22}_{-0.05}$    & $7.4^{+5.0}_{-3.2}$   & $21.5^{+6.5}_{-9.9}$  & $4.08 \pm 0.81$           & $0.74 \pm 0.57$           & $9.88 \pm 2.0$        \\
    SN\,2021lei         & $0.1^{+1.0}_{-0.1}$   & $2.7 \pm 1.4$         & $4.79 \pm 0.47$           & $0.1 \pm 0.05$            & $0.04^{+0.04}_{-0.02}$    & $0.08^{+0.19}_{-0.06}$    & $7.8^{+4.6}_{-2.8}$   & $20.1 \pm 8.4$            & $4.31 \pm 1.0$            & $0.73 \pm 0.58$           & $12.8 \pm 4.19$       \\
    SN\,2021lwz         & $4.0^{+1.7}_{-0.8}$   & $3.8 \pm 1.0$         & $3.59 \pm 0.36$           & $0.01^{+0.02}_{-0.0}$     & $0.03^{+0.02}_{-0.01}$    & $0.1^{+0.2}_{-0.08}$  & $0.4^{+0.6}_{-0.2}$   & $11.0 \pm 2.0$            & $6.55 \pm 2.45$           & $1.04 \pm 0.37$       & $2.98 \pm 0.31$       \\
    SN\,2021uvy         & $1.5^{+1.3}_{-0.5}$   & $0.2^{+0.6}_{-0.2}$   & $2.89^{+1.67}_{-0.71}$    & $0.01^{+0.06}_{-0.0}$     & $0.25 \pm 0.07$           & $0.01^{+0.0}_{-0.0}$  & $3.8 \pm 1.0$         & $5.9 \pm 1.0$         & $9.64^{+0.27}_{-0.78}$    & $0.91 \pm 0.48$           & $5.5 \pm 0.46$        \\
    SN\,2021ybf         & $0.2^{+2.3}_{-0.2}$   & $1.8^{+2.1}_{-1.4}$   & $3.44 \pm 1.01$           & $0.08^{+0.08}_{-0.04}$    & $0.05^{+0.05}_{-0.03}$    & $0.07^{+0.2}_{-0.05}$     & $20.1^{+16.9}_{-10.4}$    & $18.5 \pm 8.2$            & $4.39 \pm 1.02$           & $0.74 \pm 0.54$   & $6.15 \pm 1.01$       \\
    \enddata
    \tablecomments{Full list of the best-fit parameters from the \mosfit model to all the LSNe in our sample. The definitions and priors of all parameters are given in Table~\ref{tab:parameters}. Additional parameters derived from these posteriors are listed in Table~\ref{tab:derived}. The only parameter excluded from this table is the mass of the neutron star $M_{\text{NS}}$, since it is effectively equal to the prior for all objects.}
\end{deluxetable}
\end{longrotatetable}

\subsection{Summary}

We determined that LSNe have ejecta masses in between those of SLSNe and SNe Ic/Ic-BL, and magnetar parameters ($P_{\text{spin}}$, $B_{\perp}$) that span the entire range of allowed parameter space, emphasizing their intermediate nature and the contribution to their luminosity from both magnetar engines and radioactive decay. While SLSNe appear to have fast spins and strong magnetic fields, SNe Ic/Ic-BL have weak or no magnetars. This agrees with the idea that SLSNe are powered by a magnetar central engine, whereas there is no evidence for a significant magnetar contribution in SNe Ic/Ic-BL. In terms of their pre-explosion masses, LSNe extend to higher masses than SNe Ic/Ic-BL, but not as massive as SLSNe, and while SLSNe have a sharp drop off at the low-mass end, the ejecta masses of LSNe extend as low as those of SNe Ic/Ic-BL. We find that LSNe tend to be powered either by an over-abundant production of $^{56}$Ni or by weak magnetar engines.

\begin{deluxetable*}{cccccccc}
    \tablecaption{Additional Parameters \label{tab:derived}}
    \tablehead{\colhead{Name} & \colhead{Absolute $r$-Mag} & \colhead{Explosion Date} & \colhead{Rise Time}& $A_{V, \text{host}}$ & $E_{k}$         & $M_{\text{Ni}}$ & WAIC \\
                              &                            & (MJD)                    & (Days)             & (mag)                & ($10^{51}$ erg) & (M$_\odot$)     &      }
    \startdata
    DES14C1rhg   & $-19.40^{+0.03}_{-0.02}$ & $56983.2 \pm 2.2$         & $21.0 \pm  1.6$ & $<0.05$                 & $1.3^{+1.4}_{-0.6}$   & $0.04^{+0.15}_{-0.03}$    & 51 \\
    DES15C3hav   & $-19.35 \pm 0.04$        & $57303.1 \pm 1.0$         & $24.6 \pm  1.0$ & $0.71 \pm 0.23$         & $1.7^{+1.9}_{-0.8}$   & $0.08^{+0.13}_{-0.06}$    & 64 \\
    DES16C3cv    & $-19.62 \pm 0.03$        & $57578.1 \pm 2.9$         & $57.2 \pm  1.9$ & $<0.11$                 & $1.3 \pm 0.4$       & $0.06^{+0.27}_{-0.05}$    & 105 \\
    iPTF13dnt    & $-19.40^{+0.07}_{-0.28}$ & $56529.3^{+9.9}_{-27.2}$  & $26.0 \pm 19.3$ & $<0.06$                 & $11.8^{+16.7}_{-7.4}$   & $1.28^{+1.09}_{-0.43}$    & 16 \\
    iPTF16asu    & $-20.46 \pm 0.10$        & $57513.4 \pm 1.1$         & $10.1 \pm  1.2$ & $<0.02$                 & $0.04^{+0.09}_{-0.02}$    & $0.003^{+0.007}_{-0.002}$   & 135 \\
    iPTF17cw     & $-19.40^{+0.03}_{-0.02}$ & $57754.1 \pm 0.7$         & $13.9 \pm  1.1$ & $<0.02$                 & $1.8^{+1.5}_{-0.8}$   & $0.14^{+0.06}_{-0.09}$    & 32 \\
    OGLE15xl     & $-19.36 \pm 0.25$        & $57320.8 \pm 2.9$         & $36.3 \pm  5.2$ & $<0.21$                 & $8.7^{+17.4}_{-6.1}$    & $1.65^{+0.55}_{-0.88}$    & 34 \\
    PS15cvn      & $-19.64 \pm 0.04$        & $57326.6 \pm 0.4$         & $15.8 \pm  0.3$ & $<0.01$                 & $0.8^{+0.2}_{-0.1}$   & $0.14 \pm 0.09$       & 116 \\
    PTF10gvb     & $-19.10 \pm 0.04$        & $55314.7 \pm 0.9$         & $17.9 \pm  0.9$ & $<0.04$                 & $5.0 \pm 2.4$       & $0.31 \pm 0.06$       & 70 \\
    PTF10iam     & $-20.10 \pm 0.03$        & $55338.3^{+1.0}_{-1.7}$   & $13.7 \pm  1.8$ & $<0.04$                 & $0.05^{+0.12}_{-0.03}$    & $0.005^{+0.011}_{-0.003}$   & 56 \\
	PTF11img     & $-19.27 \pm 0.03$		& $55743.9 \pm 1.2$		 	& $19.0 \pm 1.1$ & $<0.04$			   	    & $6.5^{+6.9}_{-3.4}$ 	 & $0.84^{+0.2}_{-0.11}$ 	 & 52 \\
    PTF12gty     & $-19.86 \pm 0.03$        & $56071.0 \pm 2.1$         & $61.3 \pm  2.2$ & $<0.09$                 & $1.9^{+1.9}_{-0.8}$   & $0.18^{+0.46}_{-0.15}$    & 92 \\
    PTF12hni     & $-19.94 \pm 0.04$        & $56126.1 \pm 5.3$         & $26.3 \pm  5.4$ & $<0.08$                 & $4.8^{+4.1}_{-2.7}$   & $0.35^{+0.09}_{-0.16}$    & 71 \\
    SN\,1991D    & $-20.10^{+0.34}_{-0.58}$ & $48263.7 \pm 6.8$         & $15.3 \pm  9.0$ & $<0.06$                 & $3.6^{+4.5}_{-1.8}$   & $0.2 \pm 0.11$      & 21 \\
    SN\,2003L    & $-19.50 \pm 0.20$        & $52635.9^{+3.2}_{-9.9}$   & $47.4 \pm 15.2$ & $<0.33$                 & $10.4^{+21.5}_{-8.3}$   & $1.97^{+1.64}_{-0.63}$    & -3 \\
    SN\,2007ce   & $-19.33 \pm 0.10$        & $54200.1 \pm 5.2$         & $18.2 \pm  5.8$ & $<0.02$                 & $2.0 \pm 1.0$       & $0.21 \pm 0.05$       & 140 \\
    SN\,2009cb   & $-20.33 \pm 0.32$        & $54881.7^{+6.3}_{-10.0}$  & $16.4 \pm  8.8$ & $0.04^{+0.62}_{-0.04}$  & $4.3^{+4.5}_{-2.4}$   & $0.21^{+0.28}_{-0.18}$    & 33 \\
    SN\,2010ay   & $-19.87 \pm 0.08$        & $55249.8 \pm 0.5$         & $21.2 \pm  1.7$ & $<0.05$                 & $3.7^{+6.5}_{-2.2}$   & $1.19 \pm 0.13$       & 14 \\
    SN\,2011kl   & $-19.73 \pm 0.10$        & $55903.8 \pm 2.9$         & $16.0 \pm  1.8$ & $<0.06$                 & $12.1^{+17.3}_{-7.2}$   & $1.06^{+1.05}_{-0.68}$    & 11 \\
    SN\,2012aa   & $-19.86 \pm 0.06$        & $55921.0 \pm 3.1$         & $37.9 \pm  3.9$ & $<0.04$                 & $1.1^{+1.0}_{-0.4}$   & $0.09^{+0.1}_{-0.05}$   & 65 \\
    SN\,2013hy   & $-20.00 \pm 0.04$        & $56515.9 \pm 2.2$         & $26.8 \pm  1.6$ & $<0.03$                 & $3.8 \pm 2.5$       & $1.76^{+0.25}_{-0.51}$    & 109 \\
    SN\,2018beh  & $-19.78 \pm 0.01$        & $58212.9 \pm 0.4$         & $37.8 \pm  0.4$ & $<0.03$                 & $2.1^{+1.7}_{-0.9}$   & $0.03^{+0.03}_{-0.01}$    & 211 \\
    SN\,2018don  & $-19.06 \pm 0.01$        & $58198.6 \pm 1.0$         & $74.1 \pm  1.0$ & $0.01^{+0.33}_{-0.01}$  & $1.2 \pm 0.3$       & $0.02^{+0.06}_{-0.01}$    & 967 \\
    SN\,2018fcg  & $-20.28 \pm 0.03$        & $58336.8 \pm 0.8$         & $20.3 \pm  0.7$ & $<0.02$                 & $1.0^{+0.6}_{-0.3}$   & $0.01^{+0.02}_{-0.0}$   & 235 \\
    SN\,2019cri  & $-19.10^{+0.08}_{-0.05}$ & $58560.4 \pm 2.5$         & $47.0 \pm  2.9$ & $<0.08$                 & $2.1^{+2.8}_{-1.2}$   & $0.52 \pm 0.23$       & 54 \\
    SN\,2019dwa  & $-19.21 \pm 0.03$        & $58576.2 \pm 1.1$         & $34.7 \pm  1.5$ & $<0.03$                 & $1.2 \pm 0.5$       & $0.04^{+0.21}_{-0.04}$    & 79 \\
    SN\,2019gam  & $-19.90^{+0.13}_{-0.08}$ & $58607.1^{+4.1}_{-7.7}$   & $55.6 \pm  6.7$ & $<0.09$                 & $36.3 \pm 21.3$       & $2.15 \pm 1.3$      & 52 \\
    SN\,2019hge  & $-19.87 \pm 0.02$        & $58625.6 \pm 1.1$         & $58.9 \pm  1.1$ & $0.51 \pm 0.19$         & $1.3^{+1.2}_{-0.5}$   & $0.1^{+0.2}_{-0.07}$    & 145 \\
    SN\,2019J    & $-19.87 \pm 0.11$        & $58487.1 \pm 3.0$         & $41.5 \pm  3.0$ & $<0.08$                 & $1.3^{+1.4}_{-0.5}$   & $0.03^{+0.04}_{-0.02}$    & 28 \\
    SN\,2019moc  & $-19.06 \pm 0.04$        & $58691.9^{+0.1}_{-0.2}$   & $15.4 \pm  0.6$ & $<0.02$                 & $1.7^{+1.6}_{-0.8}$   & $0.43 \pm 0.03$       & 81 \\
    SN\,2019obk  & $-20.06 \pm 0.05$        & $58689.1 \pm 1.0$         & $37.3 \pm  0.9$ & $<0.04$                 & $4.2^{+6.6}_{-2.5}$   & $0.55 \pm 0.51$       & 45 \\
    SN\,2019pvs  & $-19.58 \pm 0.10$        & $58678.6^{+4.7}_{-7.7}$   & $53.8 \pm  7.6$ & $<0.05$                 & $1.9^{+5.5}_{-1.2}$   & $0.58^{+1.33}_{-0.55}$    & 10 \\
    SN\,2019stc  & $-20.05 \pm 0.04$        & $58735.6 \pm 1.5$         & $48.6 \pm  1.4$ & $<0.04$                 & $1.0^{+0.9}_{-0.4}$   & $0.08^{+0.44}_{-0.07}$    & 64 \\
    SN\,2019unb  & $-20.09 \pm 0.05$        & $58763.7 \pm 1.7$         & $59.9 \pm  2.5$ & $0.53 \pm 0.26$         & $1.5^{+1.5}_{-0.6}$   & $0.17^{+0.47}_{-0.14}$    & 63 \\
    SN\,2019uq   & $-19.00 \pm 0.06$        & $58482.8 \pm 2.0$         & $22.2 \pm  2.2$ & $<0.05$                 & $2.3^{+2.3}_{-1.3}$   & $0.23 \pm 0.15$       & 31 \\
    SN\,2019wpb  & $-19.14 \pm 0.06$        & $58808.9 \pm 0.9$         & $18.5 \pm  1.6$ & $<0.02$                 & $4.1^{+4.5}_{-2.3}$   & $0.63 \pm 0.08$       & 43 \\
    SN\,2021lei  & $-19.33 \pm 0.04$        & $59326.3 \pm 1.2$         & $19.1 \pm  1.4$ & $<0.05$                 & $8.6 \pm 7.0$       & $0.74 \pm 0.16$       & 50 \\
    SN\,2021lwz  & $-19.70^{+0.02}_{-0.01}$ & $59341.4 \pm 0.1$         & $ 9.7 \pm  0.1$ & $0.50 \pm 0.07$         & $0.02^{+0.04}_{-0.01}$    & $0.002^{+0.007}_{-0.002}$   & 336 \\
    SN\,2021uvy  & $-19.62 \pm 0.03$        & $59391.6 \pm 1.7$         & $56.5 \pm  2.9$ & $0.62 \pm 0.17$         & $0.7 \pm 0.3$       & $0.02^{+0.26}_{-0.02}$    & 67 \\
    SN\,2021ybf  & $-19.26 \pm 0.04$        & $59419.1 \pm 4.7$         & $54.1 \pm  4.9$ & $<0.05$                 & $4.4^{+4.7}_{-2.5}$   & $1.74^{+0.22}_{-0.33}$    & 81 \\    \enddata
    \tablecomments{Parameters derived from the \mosfit light curve models: the peak absolute magnitude in observer $r$-band (after correcting for extinction and cosmological K-correction of $+ 2.5\times \log(1 + z)$), the average rest-frame days from explosion to peak in $r$-band, the intrinsic host extinction in $V$ band, the kinetic energy, the total nickel mass, and the Watanabe-Akaike Information Criterion (WAIC) for fit quality \citep{Watanabe10, Gelman14}.}
\end{deluxetable*}

\section{LSN sub-groups}
\label{sec:grouping}

LSNe span a wide range of both observational properties and model parameters. Significant differences within the LSNe population exist and they are unlikely to all be a product of the same physical process. In this section, we attempt to group LSNe into distinct sub-groups with uniform spectral and photometric properties based on the labels designated in Table~\ref{tab:classes}. We label each LSN with either a ``Superluminous" label if their spectra are SLSN-like or ``Normal" if their spectra are Ic-like. Similarly, we use a ``Fast" label if their rise times are $\lesssim 25$ days like most SNe Ic or ``Slow" if they are $\gtrsim 35$ days like most SLSNe. Given their intermediate nature, we resort to labeling some LSNe as having ``Ambiguous" spectra consistent with either a SLSN or a SN Ib/c, or a light curve with ``Medium" rise time between 25 and 35 days. This breakdown leads to four main groups: \textit{Slow SLSN-like}, \textit{Fast SLSN-like}, \textit{Slow Ic-like}, and \textit{Fast Ic-like}. We include an \textit{Other} group for the LSNe that do not clearly fit into any of the previous four groups. Of the 40 Gold and Silver LSNe presented here, Slow SLSN-like make up 23\% (N = 9), Fast SLSN-like 18\% (N = 7), Slow Ic-like 15\% (N = 6), Fast Ic-like 30\% (N = 12), and 15\% (N = 6) are in the Other group.

\begin{figure*}[]
	\begin{center}
		\includegraphics[width=0.9\textwidth]{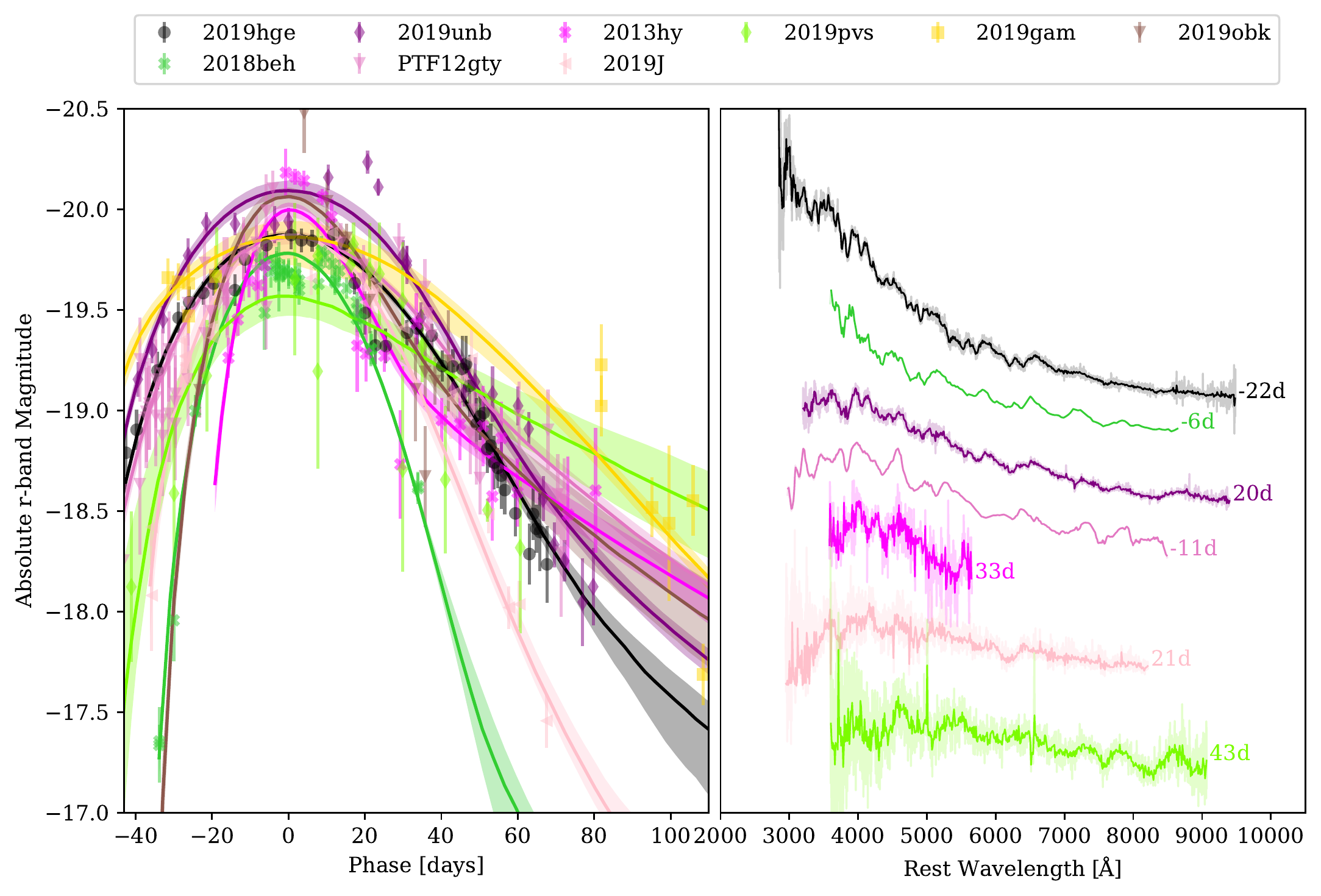}
		\caption{Light curves and spectra of the Slow SLSN-like LSNe group. Only $r$-band light curves and their respective \mosfit models are shown. Spectra are arbitrarily scaled. Individual references are listed in the Appendix. \label{fig:slow_slsn}}
	\end{center}
\end{figure*}

\subsection{Slow SLSN-like}

This is the group most similar to normal SLSNe (Figure~\ref{fig:slow_slsn}). The objects we group here are: SN\,2013hy, SN\,2018beh, SN\,2019gam, SN\,2019hge, and SN\,2019J, SN\,2019obk, SN\,2019pvs, SN\,2019unb, and PTF12gty. These LSNe have spectra that closely resemble those of SLSNe, and broad light curves reminiscent of SLSNe. Nevertheless, they are dimmer than typical SLSNe. This group of LSNe has the most energetic magnetars of the LSNe population, with typical spin periods $\lesssim 10$ ms and magnetar magnetic fields $\gtrsim 10^{14}$ G. Their magnetic fields are similar to those of normal SLSNe, which span the $\sim 0.4-60\times10^{14}$ G range, but their spin periods extend to higher values than the $P_{\rm spin} \lesssim 6$ ms found in most SLSNe; likely the reason for their lower luminosity. The fact that some of these LSNe have magnetars with stronger magnetic fields than some normal SLSNe suggests magnetic field strength might not the dominant parameter powering SLSNe, particularly for slow spin periods. Compared to the rest of the LSN population, this group has the largest ejecta masses, all $\gtrsim 10 $ M$_\odot$. The power behind their light curves is also greatly dominated by the magnetar component, where all but two SNe have a magnetar contribution $> 90$\%; SN\,2013hy and SN\,2019pvs are the exception with a magnetar contribution of $\sim 50$\%.

Almost all LSNe in this group have relatively low ejecta velocities $V_{\rm ej}\lesssim 6000$ km s$^{-1}$, which is likely contributing to their low peak luminosity, as low ejecta velocities increase the diffusion time, distributing the output luminosity over a larger stretch of time. Two exceptions are SN\,2019gam and SN\,2019obk, which have ejecta velocities of $\approx 14000$ km s$^{-1}$ and $\approx 8200$ km s$^{-1}$, respectively. It is possible these two objects are just normal SLSNe but are marginally under-luminous due to either the slow spin period of $\approx 9$ ms for SN\,2019obk, or low magnetic field of $\approx 0.7\times10^{14}$ G for SN\,2019gam.

This group of LSNe appear powered by magnetar engines, have light curve durations similar to those of SLSNe, and have spectra consistent with SLSNe. Therefore, these LSNe can be considered to be the faintest SLSNe known, extending down to $M_r \sim -19.5$ mag, and likely under-luminous due to their slower spin periods than typical SLSNe.

\subsection{Fast SLSN-like}

\begin{figure*}
	\begin{center}
		\includegraphics[width=0.9\textwidth]{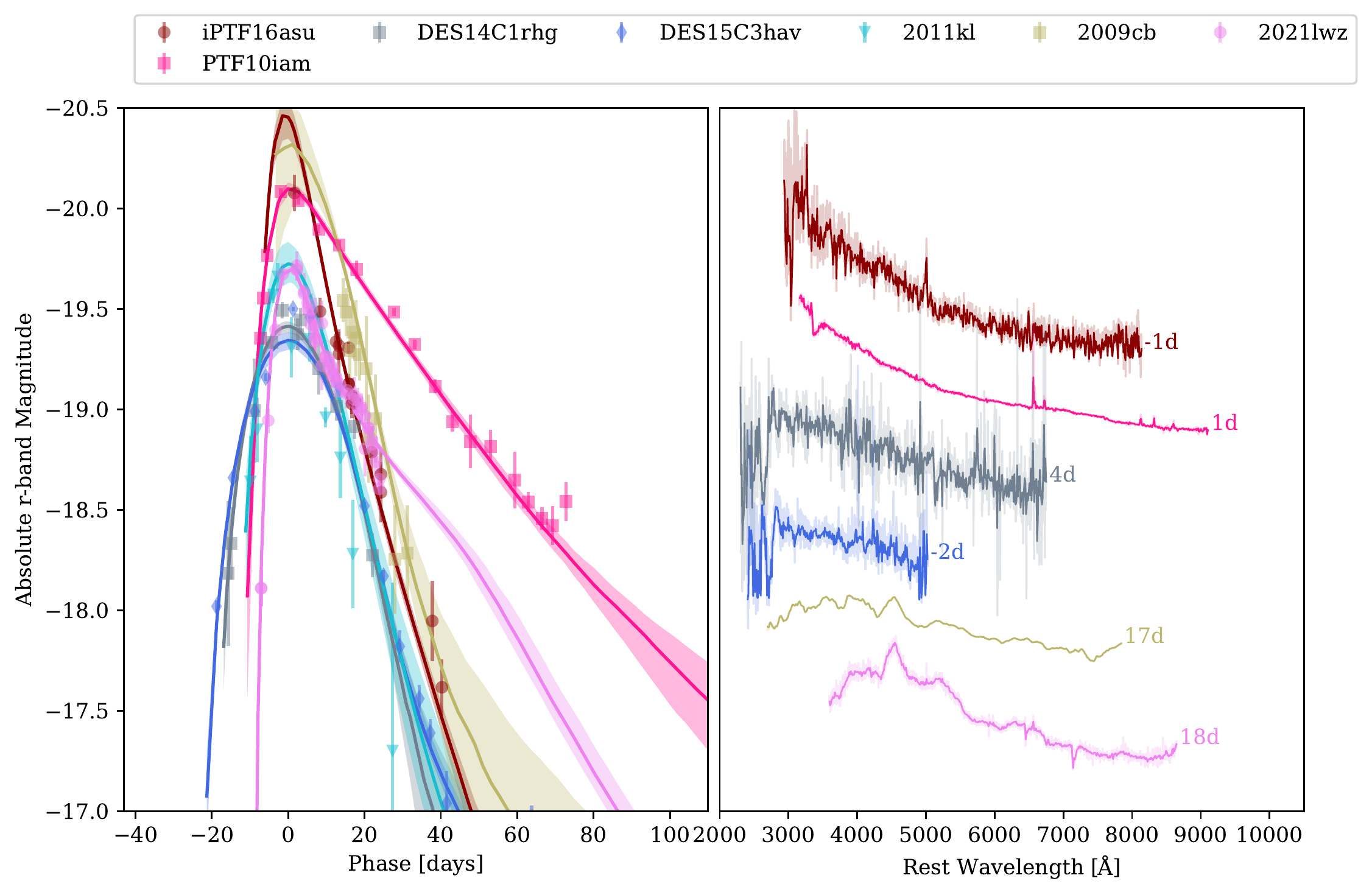}
		\caption{Light curves and spectra for the Fast SLSN-like LSNe group. Only $r$-band light curves and their respective \mosfit models are shown. Spectra are arbitrarily scaled. Individual references are listed in the Appendix. \label{fig:fast_slsn}}
	\end{center}
\end{figure*}

We find seven LSNe that appear spectroscopically consistent with being SLSNe, but have light curves that are much faster than normal SLSNe (Figure~\ref{fig:fast_slsn}). These objects are: iPTF16asu, PTF10iam, DES14C1rhg, DES15C3hav, SN\,2011kl, SN\,2021lwz, SN\,2009cb. Even though the decline time of PTF10iam is relatively long, we include it in this group given that it has a rise time of $\sim 14$ days, among the fastest of all LSNe. All objects in this group have very strong magnetic fields $\gtrsim 4\times10^{14}$ G (higher than normal SLSNe), but slow spin periods $\gtrsim 5$ ms, leading to their low luminosities and fast time-scales.

The light curves of Fast SLSN-like LSNe also appear largely dominated by a magnetar engine, where all have magnetar contributions $> 90$\%, except for SN\,2011kl, which has a $\sim 75$\% magnetar contribution. Almost all the SNe in this group have low ejecta masses $M_{\rm ej} \lesssim 4$ M$_\odot$ (except for SN\,2009cb with $M_{\rm ej} \approx 7$ M$_\odot$), which lies in stark contrast to the LSNe in the Slow SLSN-like group, which have ejecta masses $\gtrsim 10 $ M$_\odot$. This is to be expected, since higher ejecta masses correlate to longer diffusion times, as discussed in \S\ref{sec:results_mass}.

LSNe in this group appear to be an extension of the SLSN population, given their spectra and physical parameters, but they have relatively fast-evolving light curves due to their low ejecta masses.

\begin{figure*}
	\begin{center}
		\includegraphics[width=0.9\textwidth]{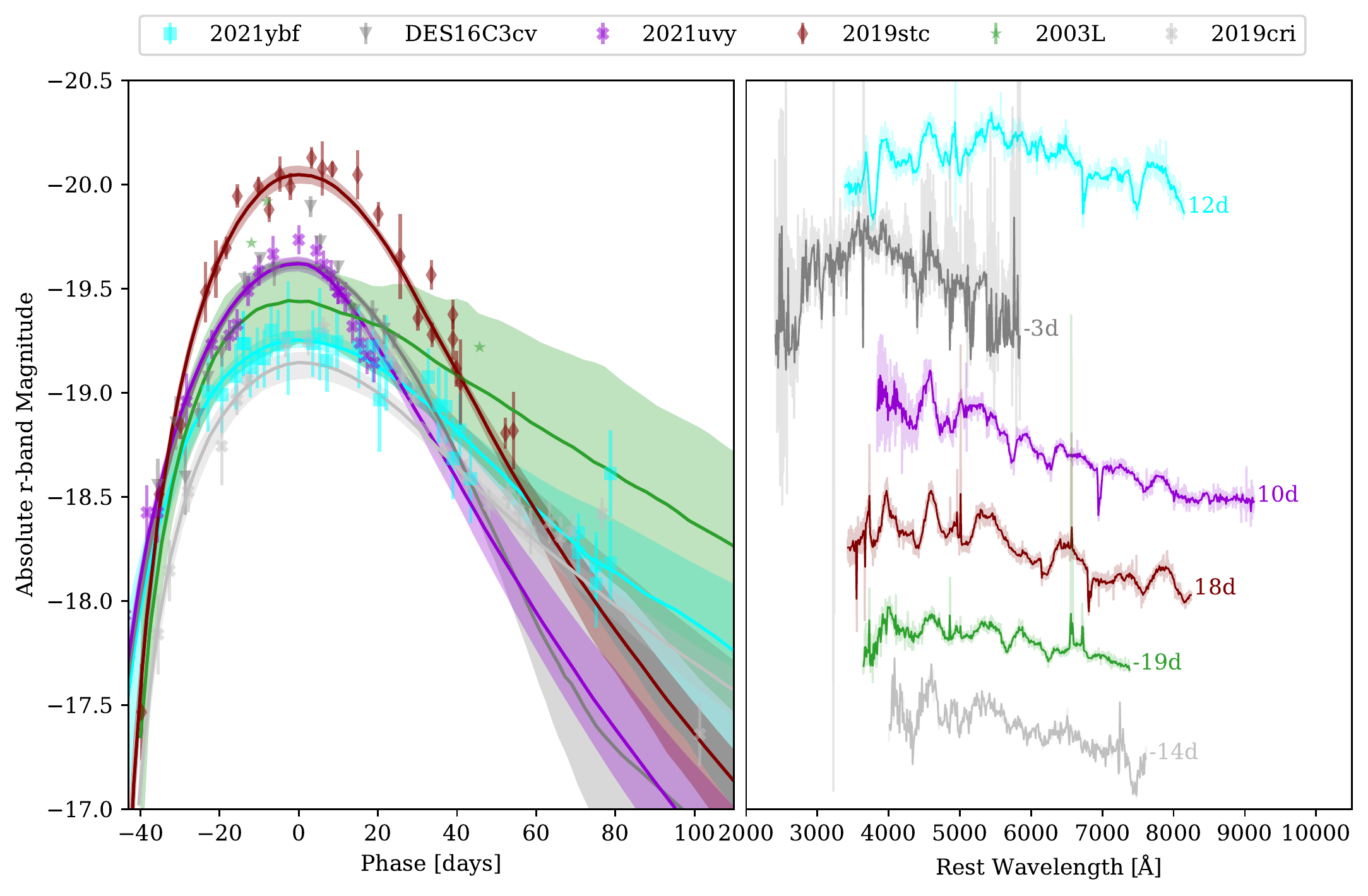}
		\caption{Light curves and spectra of the Slow Ic-like LSNe group. Only $r$-band light curves and their respective \mosfit models are shown. Spectra are arbitrarily scaled. Individual references are listed in the Appendix. \label{fig:slow_Ic}}
	\end{center}
\end{figure*}

\subsection{Slow Ic-like}

The LSne in this group have red spectra like normal SNe Ic but light curves that are as broad as normal SLSNe (Figure~\ref{fig:slow_Ic}). Six objects fall in this category: SN\,2021ybf, DES16C3cv, SN\,2021uvy, SN\,2003L, SN\,2019stc and SN\,2019cri. Unlike the previous two groups of Slow and Fast SLSN-like LSNe, the parameters of Slow Ic-like LSNe appear to bifurcate rather than cluster in a particular region of parameter space. 

While the light curves of DES16C3cv, SN\,2019cri, SN\,2019stc, and SN\,2021uvy are best fit by an almost pure magnetar model, SN\,2003L and SN\,2021ybf appear entirely radioactively powered. The only distinguishing feature is that the four magnetar powered LSNe all show either a prominent secondary light curve peak, or in the case of SN\,2019cri, a late time flattening that could be indicative of the start of a secondary peak. Neither of the radioactively powered SNe show evidence for a secondary peak.

SN\,2003L and SN\,2021ybf have respective nickel mass fractions of $f_{\rm Ni} \approx 0.12$ and $f_{\rm Ni} \approx 0.08$, which explains why they are brighter than normal SNe Ic, which tend to have values of $f_{\rm Ni} \lesssim 0.04$. Additionally, their respective high ejecta masses of $M_{\rm ej} \approx 17$ M$_\odot$ and $M_{\rm ej} \approx 20$ M$_\odot$ explain their slow evolution.

Conversely, DES16C3cv, SN\,2019stc, SN\,2021uvy, and SN\,2019cri all appear magnetar dominated, but powered by weaker magnetar engines than normal SLSNe. The first three have spin periods between $4-8$ ms and magnetic fields between $4-8\times10^{14}$ G, and SN\,2019cri has a relatively high magnetic field of $\approx 6.5\times10^{14}$ G but a spin period of 13 ms, leading to its relatively low luminosity. We presented an in-depth analysis of SN\,2019stc in \cite{Gomez21_2019stc}, where we found the source to have a SLSN-like light curve, but a spectrum that is identical to those of normal SNe Ic. We concluded that a combination of radioactive decay and a magnetar central engine was required to power the luminous first peak while preserving a red Ic-like spectra. A similar interplay of power sources could be responsible for the luminous nature of these objects, while still preserving Ic-like spectra.

The LSNe in this group divide into two groups. The SNe in one set (SN\,2003L and SN\,2021ybf) appear to be radioactively powered and are more luminous than normal SNe Ic due to their high nickel fractions, while their high ejecta masses lead to their slow evolution. The second set (DES16C3cv, SN\,2019stc, SN\,2021uvy, and SN\,2019cri) are dominated by magnetars, but retain SNe Ic-like spectra. This could be a consequence of an interplay of power sources; while the magnetar component makes the SNe more luminous, the radioactive component makes the spectra appear Ic-like.

\begin{figure*}[]
	\begin{center}
		\includegraphics[width=0.9\textwidth]{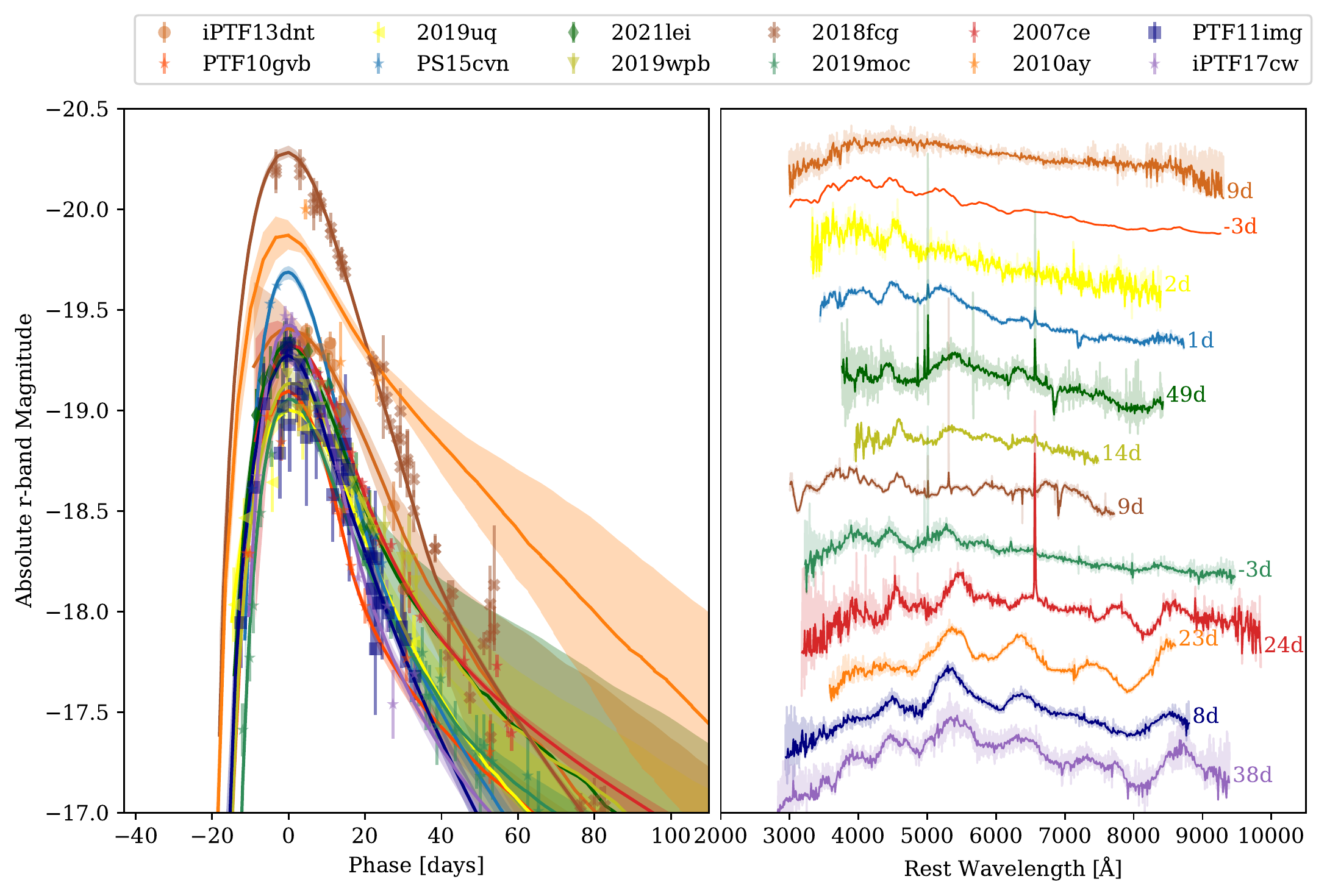}
		\caption{Light curves and spectra of the Fast Ic-like LSNe group. Only $r$-band light curves and their respective \mosfit models are shown. Spectra are arbitrarily scaled. Individual references are listed in the Appendix. \label{fig:fast_Ic}}
	\end{center}
\end{figure*}

\subsection{Fast Ic-like}

This group is defined as LSNe with SN Ic-like spectra that also evolve rapidly, like SNe Ic, yet are significantly brighter than normal SNe Ic (Figure~\ref{fig:fast_Ic}). This is the most populous group, with twelve SNe: iPTF13dnt, PTF11img, SN\,2019uq, PS15cvn, SN\,2019moc, SN\,2021lei, SN\,2019wpb, SN\,2007ce, and SN\,2010ay, SN\,2018fcg, PTF10gvb, and iPTF17cw. We note that SN\,2018fcg peaked at $M_r \sim -20.3$ mag and is technically outside the LSN definition, but we include it in the sample due to its strong spectral resemblance to normal SNe Ic and intermediate nature. These SNe have the slowest spin periods of all LSNe, with $P_{\rm spin} \gtrsim 15$ ms. For spin periods this slow, if the SNe have magnetars, these would have little to no contribution to the light curves. The main difference in terms of parameters between these SNe and normal SNe Ic is that these have significantly higher nickel mass fractions ($f_{\rm Ni} \gtrsim 0.07$) compared to the normal SNe Ic population ($f_{\rm Ni} \lesssim 0.04$). Two exceptions are PS15cvn and SN\,2019uq, which both have nickel mass fractions $f_{\rm Ni} \lesssim 0.04$. SN\,2019uq is a Silver LSN with poor photometric coverage, the dimmest object in this group, and possibly just a normal SN Ic. On the other hand, PS15cvn is the third brightest object in this group, likely due to its relatively high nickel mass of $M_{\rm Ni} \approx 0.15$ M$_\odot$.

These LSNe are the most similar to normal SNe Ic in terms of power sources, light curve durations, and spectral properties. These SNe can therefore be considered to be the brightest SNe Ic, where their high luminosity is due to an over-abundance of nickel.

\subsection{Other Objects}

\begin{figure*}
	\begin{center}
		\includegraphics[width=0.9\textwidth]{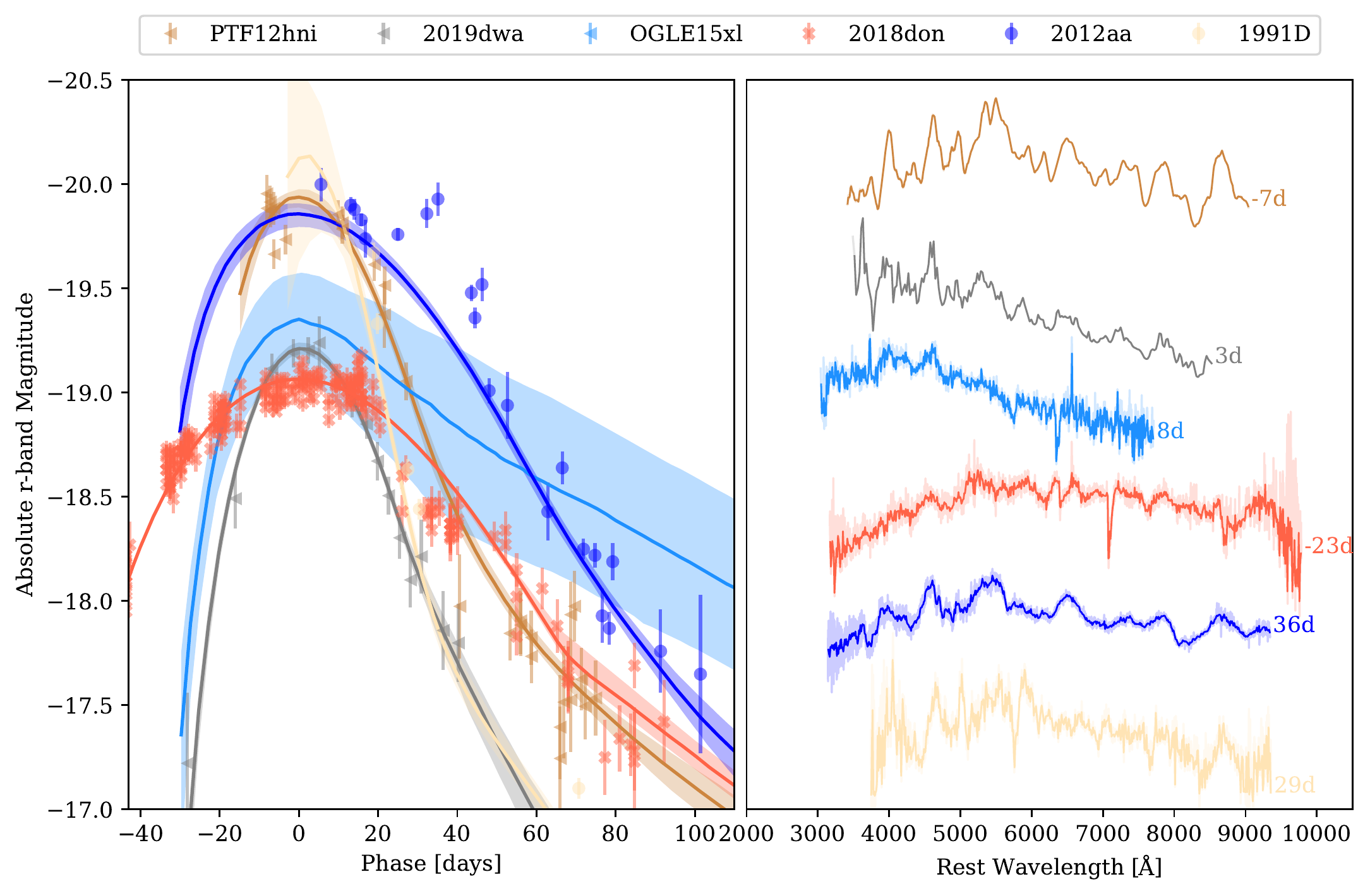}
		\caption{Light curves and spectra of the Other group. Only $r$-band light curves and their respective \mosfit models are shown. Spectra are arbitrarily scaled. Individual references are listed in the Appendix. \label{fig:other_sne}}
	\end{center}
\end{figure*}

Some LSNe do not neatly fit into any of the previous four groups, and we include them here. There are six SNe in this group: SN\,2019dwa, SN\,2018don, SN\,2019stc, PTF11hrq, OGLE15xl, and PTF12hni (Figure~\ref{fig:other_sne}). The classification of these SNe is uncertain, either due to a lack of data, or because their nature is intermediate to two of the four groups presented here.

First is OGLE15xl, its light curve fits are of poor quality due to the fact only one band is available, and its spectrum lies in-between that of SLSNe and SNe Ic.

SN\,2019dwa and PTF12hni are intermediate in every sense. Both objects have spectra intermediate to SLSNe and SNe Ic; bluer than normal SNe Ic but not quite as much as normal SLSNe. Both SNe also have an intermediate light curve evolution. Therefore, these do not easily fit into any of the defined groups. \cite{Quimby18} reached the same conclusion regarding the classification of PTF12hni.

SN\,1991D is not spectroscopically similar to any other LSN, since it has clear evidence for helium in its spectra and was previously classified as a Type-Ib SN by \cite{Benetti02}. SN\,2012aa is another peculiar object in this category, which although lacking in helium, was previously identified by \cite{Yan17} to have late time signatures of hydrogen.

Finally, SN\,2018don was presented in \cite{Lunnan20_four} as a SLSN with possibly substantial host galaxy extinction of $A_V \approx 0.4$ mag. Correcting for this amount of extinction would place the SN well into the normal SLSN regime. Due to the uncertainty in the extinction value, we avoid grouping this object into one of the four groups.

\subsection{Summary}

\begin{figure}
	\begin{center}
		\includegraphics[width=\columnwidth]{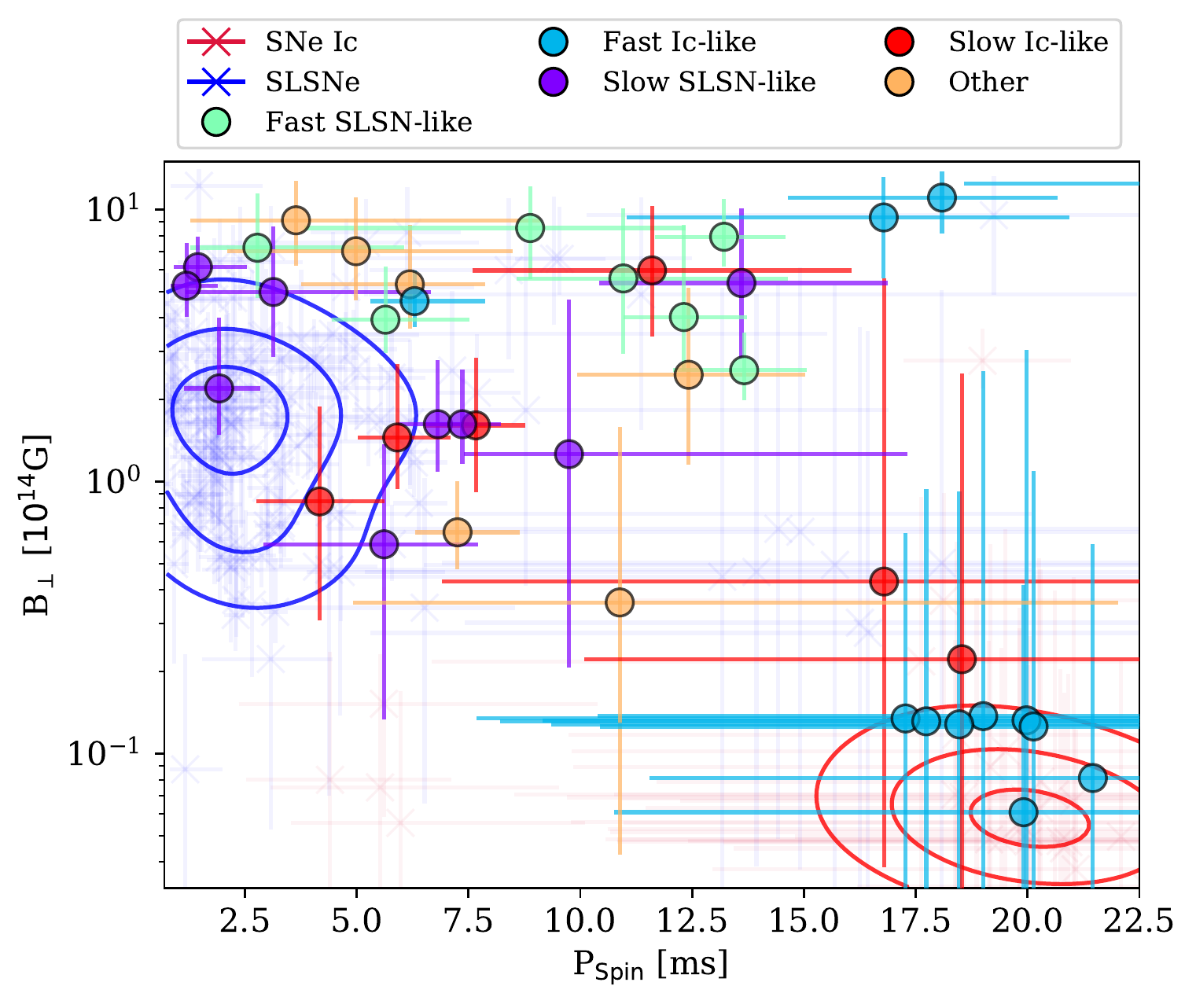}
		\caption{Same as Figure~\ref{fig:magnetar}, but with LSNe labeled based on their distinct groups. We see that LSNe do not perfectly separate into distinct classes, even after sub-diving them. But some trends so start to appear: Slow SLSN-like LSNe mostly reside in the SLSNe dominated parameter space; Fast Ic-like LSNe mostly overlap with the existing SNe Ic population; Fast SLSN-like and Slow Ic-like LSNe lie outside the typical parameter space occupied by SLSNe or SNe Ic. \label{fig:magnetar_class}}
	\end{center}
\end{figure}

In Figure~\ref{fig:magnetar_class} we show how the LSNe, now labeled in terms of their groupings, lie in the $B_{\perp}$ vs $P_{\text{spin}}$ parameter space and how they compare to the SLSNe and SNe Ic/Ic-BL populations. We find that after grouping LSNe into distinct classes, some separation does begin to appear. Mainly, Slow SLSN-like LSNe seem to occupy mostly the same parameter space as SLSNe, having powerful magnetars; whereas Fast Ic-like LSNe overlap mostly with the existing SNe Ic population with weak or no evidence for magnetars. Slow Ic-like LSNe on the other hand, still span a wide range of parameter space. And Fast Ic-like LSNe have strong magnetic fields, but spin periods that are slow enough to make the magnetar contribution negligible. In conclusion, brighter objects tend to have SLSN-like spectra, while dimmer objects more closely resemble SNe Ic.

\begin{figure}
	\begin{center}
		\includegraphics[width=\columnwidth]{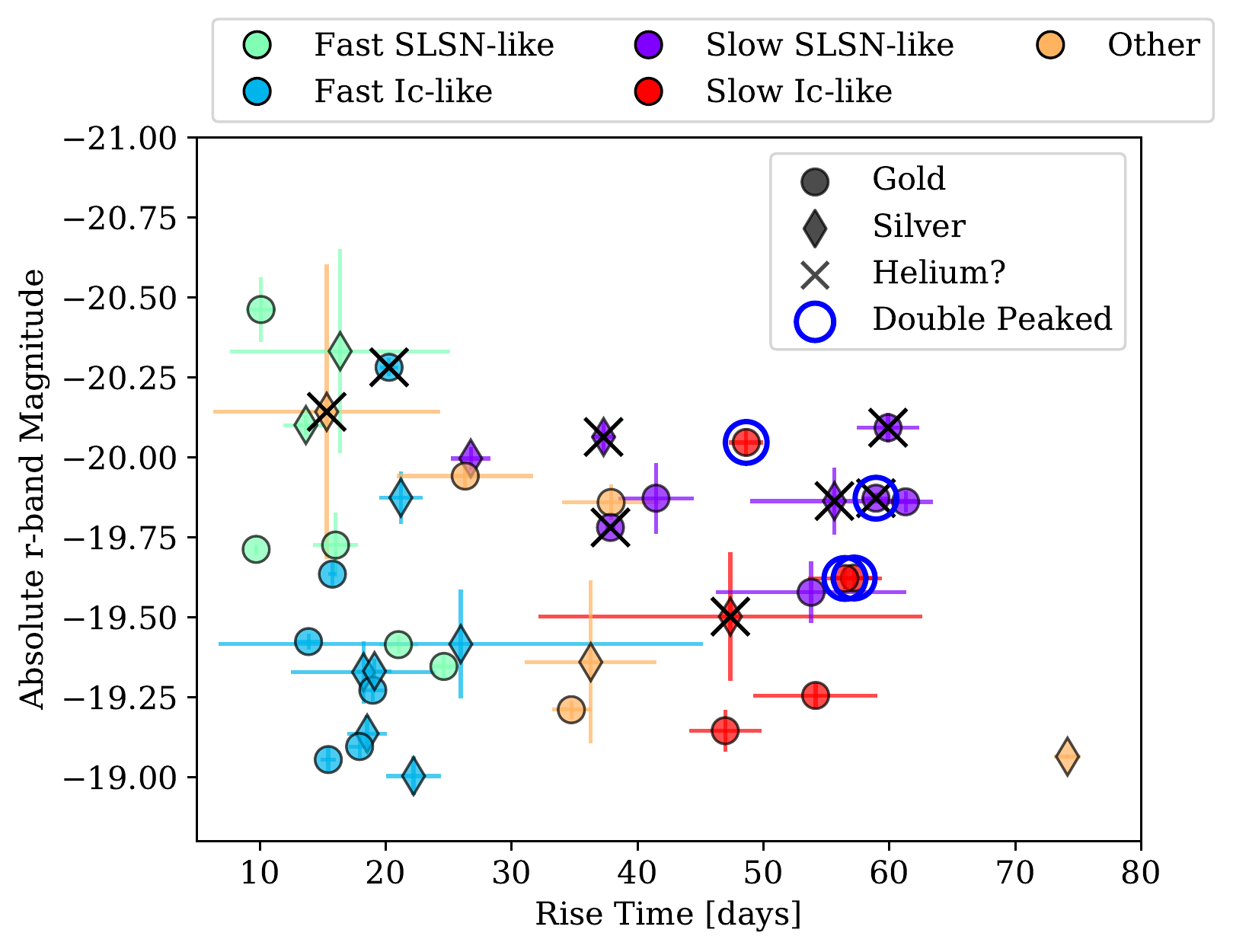}
		\caption{Peak $r$-band magnitude as a function of rise time for LSNe, labeled by their distinct groupings. We see that LSNe do appear to separate well in terms of rise time and absolute peak magnitude, forming four distinct quadrants. \label{fig:luminosity}}
	\end{center}
\end{figure}

The different groups of LSNe separate well in terms of their rise-time and peak luminosity. In Figure~\ref{fig:luminosity} we show that Fast SLSN-like LSNe have short rise times and high peak luminosities, Fast Ic-like LSNe also have short rise times but with low peak luminosities, Slow SLSN-like LSNe have long rise times and high peak luminosities, and finally, Slow Ic-like LSNe have long rise times but low peak luminosities. In the same plot, we can see that all objects with a double-peaked light curve lie in the quadrant with slow rise times and high peak luminosities. And all the objects that show possible helium are among the brighter LSNe, most of them also having long rise times.

\section{Observational Properties}\label{sec:properties}

\subsection{Spectral Features}

We explore the presence of helium in LSNe and find that SN\,1991D, SN\,2003L, SN\,2019gam, SN\,2019hge, SN\,2019obk, SN\,2019unb, SN\,2018beh, and SN\,2018fcg show tentative evidence for helium in their spectra. These SNe are all brighter than $M_r \sim -19.5$ mag and have rise times spanning $\sim 15$ to $\sim 60$ days. \cite{Yan20} presented a sample of seven SLSNe with possible detections of helium, three of which had peak magnitudes $M_r \lesssim -20.5$ mag and are therefore outside the LSNe definition, the remaining four (SN\,2019gam, SN\,2019hge, SN\,2019unb, SN\,2019obk) all lie in the Slow SLSN-like LSNe group. The excitation of Helium in SLSNe requires either non-thermal radiation, potentially from a magnetar central engine \citep{Dessart12}, or interaction with helium-rich circumstellar material \citep{Yan15}.

SLSNe tend to show a broad W-shaped absorption feature found around 4200\AA, and 4450\AA\ not seen in SNe Ic \citep{Quimby18}. We find that no LSN has this distinctive \ion{O}{2} W-shaped line, which could simply be a result of these SNe not reaching sufficiently high temperatures at early time to excite these ions or very rapid cooling of the ejecta.

\subsection{Double-peaked Light Curves}

Four LSNe (SN\,2019stc, DES16C3cv, SN\,2021uvy, and SN\,2019hge) show a double-peaked light curve structure. All four are relatively bright with magnitudes $M_r \lesssim -19.5$ mag, and have rise times greater than $\sim 45$. SN\,2019hge is the only LSN that shows both a double-peaked structure and the presence of helium. The interplay between the distinct power sources of radioactive decay and a magnetar engine may be responsible for the double-peaked structure observed in some LSNe.

\subsection{Relative Rates}

Given the fact that we selected the population of LSNe presented here from non-uniform surveys, we can not draw definitive conclusions regarding their absolute rate. Nevertheless, we can estimate their relative rates. We apply a volumetric correction to the population of LSNe, SLSNe, and SNe Ic/Ic-BL to account for Malmquist bias following the method of \cite{Cia18} and find that LSNe are more common than SLSNe but less common than either SNe Ic or SNe Ic-BL.

We compare our sample of LSNe to the ZTF Bright Transient Survey (BTS; \citealt{Perley20_BTS}) to estimate their observational rates. The BTS aims to spectroscopically classify every bright transient found by ZTF, with an completeness of $\sim 93$\% down to 18.5 mag. At the time of comparison the BTS\footnote{\url{https://sites.astro.caltech.edu/ztf/bts/bts.php}} list of SN-like events has 3636 objects brighter than $m_r = 19$ mag, which have already been pruned to only include SN-like objects with well-sampled light curves and low extinction $A_V < 1$ mag. We find that LSNe are rare, making up only $0.3 \pm 0.1$\% of SN-like transients from the BTS survey, or $0.4 \pm 0.1$\% of all CCSNe observed by this magnitude-limited survey. For comparison, SLSNe make up $\sim 1$\% of all the SN-like transients in the BTS survey.

\section{Conclusions}\label{sec:conclusions}

We have presented the first comprehensive study of all the stripped-envelope CCSNe that lie in the intermediate regime between SLSNe and SNe Ic, allowing us to better place SLSNe in the context of CCSNe, how they relate to other SNe, and the nature of their progenitors. We analyzed a sample of 40 luminous supernovae (LSNe), defined as stripped-envelope core-collapse SNe with a peak $r$-band magnitude between $M_r = -19$ and $-20$ mag, bounded by SLSNe on the bright end, and by SNe Ic/Ic-BL on the dim end. Observationally, we find that:

\begin{itemize}
  \item LSNe have intermediate rise times between $\approx 20 - 65$ days.
  \item The spectra of LSNe span a continuum, from blue and SLSN-like to red and SNe Ic-like.
  \item Brighter LSNe tend to have SLSN-like spectra, while dimmer LSNe resemble SNe Ic.
  \item No LSN shows the distinctive W-shaped \ion{O}{2} absorption feature found in some SLSNe.
  \item LSNe are rare and make up $\sim 0.3$\% of all SNe-like transients from a magnitude-limited survey; or $\sim 0.4$\% of all observed CCSNe.
  \item In absolute terms, LSNe are likely more common than SLSNe, but less common than SNe Ic/Ic-BL.
  \item LSNe with possible helium are brighter than $\sim -19.5$ mag.
  \item LSNe with a double-peaked light curve are brighter that $\sim -19.5$ mag and have long rise times $\gtrsim 45$ days.
\end{itemize}

We modeled the light curves of all 40 LSNe, as well as a sample of 149 SLSNe and 61 SNe Ic/Ic-BL in a uniform way with a combined magnetar plus radioactive decay model to compare their physical parameters. From our models we find that:

\begin{itemize}
  \item Around 25\% of LSNe appear to be radioactively powered, while the rest have at least a 50\% contribution from a magnetar engine.
  \item The nickel fractions for the radioactively dominated LSNe span a range of $f_{\rm Ni} \approx 0.01 - 0.1$, similar to SNe Ic/Ic-BL.
  \item The pre-explosion masses of LSNe extend to $\sim 30$ M$_\odot$, higher than SNe Ic/Ic-BL, but not as high as SLSNe. The slope of the high-end mass distribution of LSNe is also intermediate to SLSNe and SNe Ic/Ic-BL, as is the peak of their distribution.
  \item The pre-explosion masses of LSNe can be as low as those of SNe Ic/Ic-BL, $\sim 1.5$ M$_\odot$.
  \item Like SLSNe and SNe Ic/Ic-BL, LSNe with larger ejecta masses have longer rise times.
\end{itemize}

We attempt to separate LSNe into distinct groups and find a natural breakdown in terms of their spectral similarity to either SLSNe or SNe Ic, and whether their light curves evolve fast like SNe Ic or slowly like SLSNe. We present four main groups of LSNe: \textit{Slow SLSN-like}, \textit{Fast SLSN-like}, \textit{Slow Ic-like}, and \textit{Fast Ic-like}. From these sub-groups, we find that:

\begin{itemize}
  \item Slow SLSN-like LSNe are the most similar to normal SLSNe. They are less luminous due to their slow spin periods, and long-lasting due to their large ejecta masses. These are effectively the lowest luminosity SLSNe known.
  \item Fast SLSN-like LSNe evolve rapidly due to their low ejecta masses, but their strong magnetars make them more luminous than normal SNe Ic.
  \item Slow Ic-like LSNe bifurcate into two groups: a population of radioactively powered SNe with higher nickel fractions than typical SNe Ic/Ic-BL; and a magnetar-powered population with magnetars weaker than normal SLSNe, and spectra that still resemble SNe Ic likely due to the presence of radioactive decay.
  \item Fast Ic-like LSNe are the most similar to normal SNe Ic but are more luminous due to their relatively high nickel fractions and masses. These can be considered the most luminous SNe Ic known.
\end{itemize}

We have shown that some LSNe are an extension towards either the dimmest SLSNe (SLow SLSN-like LSNe) or the brightest SNe Ic known (Fast Ic-like LSNe), while other LSNe appear to have a more complex nature, borrowing a combination of properties from SLSNe and SNe Ic. We have analyzed in a systematic way all the SNe that occupy the up to now mostly unexplored link between SLSNe and SNe Ic. This work opens the door for subsequent studies that will focus on the details of the spectroscopic features of LSNe and how they relate to SLSNe and SNe Ic, as well as an in-depth study of their host galaxies and environments. Looking further ahead, the Legacy Survey of Space and Time \citep{Ivezic19}, scheduled to commence in 2023, will increase the transient discovery rate by about 2 orders of magnitude, and will allow us to explore the cutting edges of parameter space, providing a more comprehensive view of the relation between various types of CCSNe.

\acknowledgments
We thank Y.~Beletsky for carrying out some of the Magellan observations. S.G. is partly supported by an STScI Postdoctoral Fellowship. The Berger Time-Domain Group at Harvard is supported by NSF and NASA grants. MN is supported by the European Research Council (ERC) under the European Union’s Horizon 2020 research and innovation programme (grant agreement No.~948381) and by a Fellowship from the Alan Turing Institute. This paper includes data gathered with the 6.5 meter Magellan Telescopes located at Las Campanas Observatory, Chile. Observations reported here were obtained at the MMT Observatory, a joint facility of the University of Arizona and the Smithsonian Institution. This research has made use of NASA’s Astrophysics Data System. This research has made use of the SIMBAD database, operated at CDS, Strasbourg, France. IRAF is written and supported by the National Optical Astronomy Observatories, operated by the Association of Universities for Research in Astronomy, Inc. under cooperative agreement with the National Science Foundation. Operation of the Pan-STARRS1 telescope is supported by the National Aeronautics and Space Administration under grant No. NNX12AR65G and grant No. NNX14AM74G issued through the NEO Observation Program. This work has made use of data from the European Space Agency (ESA) mission {\it Gaia} (\url{https://www.cosmos.esa.int/gaia}), processed by the {\it Gaia} Data Processing and Analysis Consortium (DPAC, \url{https://www.cosmos.esa.int/web/gaia/dpac/consortium}). Funding for the DPAC has been provided by national institutions, in particular the institutions participating in the {\it Gaia} Multilateral Agreement. This work makes use of observations from Las Cumbres Observatory global telescope network.

\facilities{ADS, TNS}
\software{Astropy \citep{Astropy18}, extinction \citep{Barbary16}, Matplotlib \citep{matplotlib}, emcee \citep{Foreman13}, NumPy \citep{Numpy}, FLEET \citep{Gomez20_FLEET}, MOSFiT \citep{guillochon18}, PyRAF \citep{science12}, SAOImage DS9 \citep{Smithsonian00}, corner \citep{foreman16}, HOTPANTS \citep{Becker15}, SciPy \citep{Walt11}, PYPHOT (\url{https://github.com/mfouesneau/pyphot}).}

\clearpage
\newpage

\bibliography{references}

\clearpage
\newpage
\appendix

In Table~\ref{tab:spectroscopy} we list all the new previously unpublished spectra used in this work. In Table~\ref{tab:photometry} we list all the photometry used for this work, both previously published and new.

\begin{deluxetable}{cccccc}[h!]
    \tablecaption{Log of Optical Spectroscopic Observations \label{tab:spectroscopy}}
    \tablewidth{0pt}
    \tablehead{
        \colhead{Supernova} & \colhead{UT Date} & \colhead{Phase} & \colhead{Exposure Time} & \colhead{Instrument + Telescope} \\
                            &                   & \colhead{(d)}   & \colhead{(s)}           &                                       
    }
    \startdata
    SN\,2018fcg             & 2018 Sep 8        & +9              & 1200                    & Blue Channel + MMT              \\
    SN\,2019pvs             & 2019 Nov 4        & +43             & 1500                    & LDSS3c + Magellan               \\
    SN\,2021lei             & 2021 Jul 7        & +49             & 2700                    & IMACS + Magellan                \\
    SN\,2021lwz             & 2021 Jun 6        & +18             & 1500                    & Binospec + MMT                  \\
    SN\,2021uvy             & 2021 Sep 7        & +10             & 1620                    & LDSS3c + Magellan               \\
    SN\,2021ybf             & 2021 Oct 7        & +12             & 2700                    & Binospec + MMT                  \\
    \enddata
\end{deluxetable}

\begin{deluxetable}{cccccccccccc}[h!]
    \tablecaption{Log of Photometric Observations \label{tab:photometry}}
    \tablewidth{0pt}
    \tablehead{
\colhead{Supernova}&\colhead{MJD}&\colhead{Mag}&\colhead{Raw}&\colhead{MagErr}&\colhead{Telescope}&\colhead{Instrument}&\colhead{Filter}&\colhead{UL}&\colhead{System}&\colhead{Ignore}&\colhead{Source}
    }
    \startdata
2021uvy & 59606.082031 & 20.0442 & 20.2629 & 0.06620 & FLWO & KeplerCam & g & False & AB & False & This Work \\ 
2021uvy & 59606.085937 & 19.9689 & 20.1264 & 0.06510 & FLWO & KeplerCam & r & False & AB & False & This Work \\ 
2021uvy & 59606.093750 & 20.1336 & 20.2547 & 0.09170 & FLWO & KeplerCam & i & False & AB & False & This Work \\ 
2021uvy & 59609.085937 & 20.0729 & 20.2916 & 0.06770 & FLWO & KeplerCam & g & False & AB & False & This Work \\ 
2021uvy & 59609.093750 & 20.0595 & 20.2170 & 0.05790 & FLWO & KeplerCam & r & False & AB & False & This Work \\ 
2021uvy & 59609.101562 & 20.4580 & 20.5790 & 0.12510 & FLWO & KeplerCam & i & False & AB & False & This Work \\ 
    \enddata
    \tablecomments{Sample of table format, the full table available in machine readable format. For each supernova we provide the MJD, the Mag (corrected for galactic extinction and in some cases host flux contribution), the Raw magnitude without such corrections, The magnitude error MagErr, the telescope, instrument, and band used, as well as whether or not the measurement is an Upper Limit. We also include an ``Ignore" column on whether or not we excluded the data for the \mosfit model. Source indicates the source of the photometry calculations, not necessarily the source of the original images.}
\end{deluxetable}

We provide a list of all the LSNe used for this work, separated in groups. First, a confident sample of ``Golden" LSNe (\S\ref{sec:golden_lsne}) that have good photometric and spectroscopic coverage. Then, a list of ``Silver" LSNe (\S\ref{sec:silver_lsne}), objects that either have some missing data. Both the Golden and Silver groups were used for all our analyses. Then, we list a sample of ``Bronze" LSNe (\S\ref{sec:bronze_lsne}); objects that are either severely lacking in data. For completeness, we provide a list of peculiar objects that even though they meet the luminosity threshold for LSNe, are very distinct in nature and likely unrelated to LSNe (\S\ref{sec:not_lsne}). Finally, we list the Type Ic and Ic-BL SNe used for comparison in this work (\S\ref{sec:compare}). \\

%%%%%%%%%%%%%%%%%%%%%%%%%%%%%%%%%%%%%%%%%%%%%%%%%%%%%%%%%%%%%%%%%%%%%
%%%%%%%%%%%%%%%%%%%%%%%%%%%% Golden LSNe %%%%%%%%%%%%%%%%%%%%%%%%%%%%
%%%%%%%%%%%%%%%%%%%%%%%%%%%%%%%%%%%%%%%%%%%%%%%%%%%%%%%%%%%%%%%%%%%%%

\section{``Golden" Luminous Supernovae}\label{sec:golden_lsne}

\subsection{2011kl}
SN\,2011kl (=GRB111209A) was classified as a SLSN-I with an associated GRB by \cite{Greiner15}, the authors determine a redshift of $z = 0.677$ based on afterglow spectra. Their photometry is corrected for the GRB afterglow contribution, host extinction, galactic extinction, and host galaxy flux. The resulting light curve is brighter than normal SNe Ic, but not quite in the regime of SLSNe. \cite{Mazzali16} also classify the spectra as a SLSN-I.

\subsection{2012aa}
SN\,2012aa (=PSN J14523348-0331540 =Howerton-A20) was presented by \cite{Roy16} as a SN in between a Type Ibc and a SLSN-I. We include two early photometry points from the CRTS that \cite{Roy16} calibrate to V-band. We subtract a nominal magnitude of $m_V = 19.11$ mag from those data to account for the host flux. Including spectra from \cite{Shivvers19}, we agree with the peculiar spectral classification and find that SN\,2012aa is overall a good spectral match to SNe Ic-BL, but also shares some similarities with the SLSN-I 2016wi, both showing late time signs of hydrogen \cite{Yan17}.

\subsection{2018beh}
SN\,2018beh (=ZTF18aahpbwz =ASASSN-18ji =PS18ats =ATLAS18nvb) was classified as a SN Ib/c by \cite{Mcbrien18} and as a SN Ic by \cite{Dahiwale20_beh}. We include one upper limit from ATLAS and one from ASAS-SN. We include our own PSF photometry of FLWO, LCO, and ZTF images after doing difference imaging to subtract the host flux. We exclude detections before MJD = 58212 from the MOSFiT fit, since these seem to be from a pre-explosion feature. The spectra match that of a SLSN-I at early times. The light curve is dim for a SLSN-I. Comparison spectra are from the TNS \citep{Dahiwale20_beh, Mcbrien18}.

\subsection{2018fcg}
SN\,2018fcg (=ZTF18abmasep =Gaia18cms =ATLAS18ucc) was classified as a SLSN-I by \citep{Fremling18_fcg} and \cite{Lunnan18_fcg}. We include photometry from \textit{Gaia}, ATLAS, and ZTF. In addition, we include our own PSF photometry of FLWO images after doing difference imaging to subtract the host flux. We do not include photometry from \cite{Hernandez18}, since these are not corrected for the host flux. We exclude the $r$-band points from ZTF after MJD=58420 since they are very discrepant from all other sources of photometry. Our own spectra of SN\,2018fcg from the Blue Channel and Binospec spectrographs are more consistent with a SN Ic spectrum, yet the light curve is most similar to fast SLSN-like LSNe.

\subsection{2019cri}
SN\,2019cri was classified as a SN Ic by \cite{Fremling19_19cri} and \cite{Prentice19_19cri}. The spectra look like a SN Ic, but the light curve is much broader than typical SNe Ic, closer to the broadness of SLSN-I. We include photometry from ATLAS, Gaia, PS1, and ZTF. Comparison spectra were obtained from the TNS \citep{Fremling19_19cri, Prentice19_19cri}. This object was presented as the equivalent of a LSN in \cite{Prentice21}.

\subsection{2019dwa}
SN\,2019dwa (=ZTF19aarfyvc =Gaia19bxj) was classified as a SN Ic by \cite{Fremling_19dwa}. We include photometry from the GSA and ZTF. \cite{Fremling_19dwa} report a redshift of $z = 0.09$, but we find a redshift of $z = 0.082$ to be a better match to the SN features. The spectra are consistent with either particularly blue SNe Ic or particularly red SLSN-I. The light curve is as bright as the brightest normal SNe Ic, but broader. The comparison spectrum was obtained from the TNS \citep{Fremling_19dwa}. This object was presented as the equivalent of a LSN in \cite{Prentice21}.

\subsection{2019hge}
SN\,2019hge (=ZTF19aawfbtg =Gaia19est =ATLAS19och =PS19elv) was classified as a SLSN-Ib by \cite{Yan20}. We include photometry from ZTF, PS1, and the GSA. But we exclude the photometry after MJD = 58650 from the MOSFiT fit, since the SN shows a late-time re-brightening, which effectively removes all the GSA data from the fit. The spectra match those of SLSN-I. This object is only marginally dimmer than most SLSN-I, and could be just a dim SLSN-I with possible helium. Additional comparison spectra are from \cite{Dahiwale19_hge} and \cite{Prentice19_hge}. This object was presented as the equivalent of a LSN in \cite{Prentice21}.

\subsection{2019J}
SN\,2019J (=ZTF19aacxrab =PS18crs =ATLAS19cay) was classified as a SLSN-I by \cite{Fremling19_19J}. We include photometry from PS1, ATLAS, and our own PSF photometry of ZTF images after doing difference imaging to subtract the host flux. The spectra match a SLSN-I, but the light curve is dim for a SLSN-I. The comparison spectrum was obtained from the TNS \citep{Fremling19_19J}.

\subsection{2019moc}
SN\,2019moc (=ZTF19ablesob =Gaia19dks =ATLAS19rgu) was classified as a SN Ic-BL by \cite{Dahiwale19_moc} and as a SN Ic by \cite{Nicholl19_moc}. We include photometry from ZTF, \textit{Gaia}, and ATLAS. This source is only marginally in the LSNe range with a peak absolute magnitude of $M_r = 19.1$ mag and has overall close resemblance to a normal SN Ic.

\subsection{2019pvs}
We classified SN\,2019pvs (=ZTF19abuogff =PS19fbe) as a SLSN-I as part of FLEET \citep{Gomez21_TNS}. We determine a redshift of $z = 0.167$ based on host emission lines. We include photometry from PS1, and our own PSF photometry of FLWO and ZTF images after doing difference imaging to subtract the host flux. The light curve is as broad as normal SLSN-I, but dimmer. The late time spectra match either a SLSN-I or a SN Ic. We include two spectra of SN\,2019pvs we obtained with the LDSS spectrograph.

\subsection{2019stc}
We classified SN\,2019stc (=ZTF19acbonaa) as a transitional object between a SLSN-I and a SN Ic; having a SN Ic spectrum but light curve features and environment resembling a SLSN-I \citep{Gomez21_2019stc}. The light curve shows a pronounced second peak, we include photometry up to MJD = 58853 (excluding the secondary peak) and spectra from \citep{Gomez21_2019stc}. \cite{Yan20_2019stc} classified SN\,2019stc as a SLSN-I.

\subsection{2019unb}
SN\,2019unb (=ZTF19acgjpgh =Gaia19fbu =PS19isr =ATLAS19bari) was classified as a SLSN-Ib by \cite{Yan20}. We include photometry from ZTF, the GSA, PS1, and ATLAS. This object is only marginally dimmer than most SLSN-I, and could be just a dim SLSN-I with possible helium. Comparison spectra are from \cite{Prentice19} and \cite{Dahiwale20_unb}. This object was presented as the equivalent of a LSN in \cite{Prentice21}.

\subsection{2021lwz}
SN\,2021lwz was classified as a SLSN-I by \cite{Perley21}. The spectra are blue and SLSN-I like, but also share features with normal SNe Ic. The peak magnitude is dimmer than typical SLSNe. We include photometry from ZTF, ATLAS, and our own PSF photometry of FLWO images. Comparison spectra are from \cite{Perley21}.

\subsection{2021uvy}
SN\,2021uvy was classified as a SLSN-I by \cite{Poidevin21_uvy}, as a SN Ib/c by \cite{Ridley21}, and as a peculiar SN Ib by \cite{Chu21_uvy}. We include photometry from ZTF. We find the spectra are a good match to SNe Ib/c. The light curve of this object has a pronounced second peak begging at MJD = 59476, which we exclude from the \mosfit fit. We include two spectra we obtained with the Binospec spectrograph on MMT and the LDSS3c spectrographs.

\subsection{2021ybf}
SN\,2021ybf was classified as a SLSN-I by \cite{Bruch21}. We include photometry from ZTF. We obtained spectra of the source with the Binospec spectrograph and find it to be most consistent with a SN Ic, the peak luminosity is also similar to SNe Ic. Nevertheless, the light curve is much broader than for typical SNe Ic.

\subsection{DES14C1rhg}
DES14C1rhg was classified as a ``gold" SLSN-I by \cite{Angus19}. We include photometry and spectra from \cite{Angus19}, but exclude detections before MJD = 56990 from the \mosfit fit since the SN shows a slow pre-brightening. The spectra are noisy but broadly consistent with a SLSN-I. The light curve is slightly brighter and broader than a normal SN Ic.

\subsection{DES15C3hav}
DES15C3hav was classified as a ``gold" SLSN-I by \cite{Angus19}. We include photometry and spectra from \cite{Angus19}, but exclude detections before MJD = 57305 from the \mosfit fit since these are due to a particularly red pre-SN feature. The spectra are noisy but broadly consistent with a SLSN-I. The light curve is slightly brighter and broader than a normal SN Ic.

\subsection{DES16C3cv}
DES16C3cv was classified as a ``silver" SLSN-I by \cite{Angus19}. We include photometry from \cite{Angus19}, which has a very pronounced second peak. We exclude the first $i$-band point from the \mosfit model since this is inconsistent with the rest of the light curve rise. We also impose a conservative cut and remove data after MJD = 57720 from the fit to exclude the second peak from the fit. The spectrum of DES16C3cv was taken near the first peak when SLSNe tend to be very blue. Despite the spectrum being quite noisy, it does not appear particularly blue but is rather consistent with a SN Ic spectrum.

\subsection{iPTF16asu}
iPTF16asu was classified as a SLSN-I by \cite{Whitesides17}. The authors correct the photometry for extinction using $E(B-V) = 0.029$ and report all magnitudes in the AB system. The early spectra match a SLSN-I, yet the late time spectra match those of SNe Ic-BL. The light curve reaches a peak magnitude close to those of SLSN-I, but the decline is very fast, similar to SNe Ic. This object was included in the SNe Ic-BL sample from \cite{Taddia19_broadlined}.

\subsection{iPTF17cw}
iPTF17cw was classified as a SN Ic-BL by \cite{Taddia19_broadlined}. The spectra at late times resemble SNe Ic-BL, yet at early times the spectra are bluer than normal SNe Ic. The light curve is on the bright end for SNe Ic. We include photometry and spectra from \cite{Taddia19_broadlined}.

\subsection{PS15cvn}
PS15cvn (=MLS151203:000922-000407 =iPTF15dqg) was classified as a SN Ic by \cite{Kangas15} and then reclassified as a SN Ic-BL by \cite{Taddia19_broadlined}. We include photometry from PS1, the CRTS, and \cite{Taddia19_broadlined}. The spectra match a SN Ic-BL and are a good match to SN\,2007ce \cite{Quimby07_07ce}. The peak magnitude of the light curve is in between those of SLSN-I and SNe Ic. Spectra are from \cite{Taddia19_broadlined}.

\subsection{PTF10gvb}
PTF10gvb was classified as a possible SLSN-I by \cite{Quimby18}, who noted the spectra are consistent with either a SLSN-I or a SN Ic-BL. The host was studied in \cite{Taggart19}. The source was presented as a SN Ic-BL in the \cite{Taddia19_broadlined} sample. We include photometry from \cite{Taddia19_broadlined}. The light curve of PTF10gvb resembles a normal SN Ic. This object is only marginally in the LSNe range, with a a peak absolute magnitude of $M_r = -19.0$. Most evidence points towards PTF10gvb being a slightly luminous but otherwise normal SN Ic-BL.

\subsection{PTF11img}
PTF11img was classified as a SN Ic-BL by \cite{Taddia19_broadlined}, but its peak magnitude is within the LSNe range. We include photometry \cite{Taddia19_broadlined}.

\subsection{PTF12gty}
PTF12gty was classified as a SLSN-I by \cite{Quimby18}. \cite{Barbarino20} classified PTF12gty as a SN Ic due to the similarity in spectral features. We argue the spectrum is closer to a SLSN-I, despite sharing some features with early SNe Ic. The light curve of PTF12gty is as bright and broad as dim SLSN-I. We include photometry from \cite{Cia18}, which the authors correct for extinction using $E(B-V) = 0.058$. Comparison spectra are from \cite{Quimby18}.

\subsection{PTF12hni}
PTF12hni was classified as a SLSN-I by \cite{Quimby18}. We include photometry from \cite{Cia18}, which the authors correct for extinction using $E(B-V) = 0.052$. The SN shows a re-brightening after $\sim 75$ days, which is why we exclude detections after MJD = 56250 from our MOSFiT fit. The features of the spectrum are consistent with a SLSN-I, but redder than normal SLSN-I and closer to SNe Ic. The shape of the light curve is in between those of SLSN-I and SNe Ic. Comparison spectra are from \cite{Quimby18}.\\

%%%%%%%%%%%%%%%%%%%%%%%%%%%%%%%%%%%%%%%%%%%%%%%%%%%%%%%%%%%%%%%%%%%%%
%%%%%%%%%%%%%%%%%%%%%%%%%%%% Silver LSNe %%%%%%%%%%%%%%%%%%%%%%%%%%%%
%%%%%%%%%%%%%%%%%%%%%%%%%%%%%%%%%%%%%%%%%%%%%%%%%%%%%%%%%%%%%%%%%%%%%

\section{``Silver" Luminous Supernovae}\label{sec:silver_lsne}

\subsection{1991D}
SN\,1991D was classified as a luminous SN Ib by \cite{Benetti02}. We adopt a redshift of $z = 0.04179$, established from a spectrum of the host galaxy \citep{Maza89}. At this redshift, the peak magnitude of the SN is $M_R \sim -19.7$, much brighter than normal SNe Ib. We include photometry from \cite{Benetti02} and spectroscopy from \cite{Matheson01}. The spectra at late times resemble those of SLSNe, with strong spectral similarities to the LSN 2019pvs. Unlike SN\,2019pvs, SN\,199D has a much narrower light curve, fading by 2mag in under $\sim 20$ days. Moreover, SN\,199D has strong helium lines, much like the SLSN-Ib sample from \cite{Yan20}, despite fading much faster than these. We assign a Silver label due to the lack of photometry during the rise or peak of the light curve, which prevents us from accurately estimating its peak magnitude and physical parameters. This also means that the peak date might have been much earlier than estimated.

\subsection{2003L}
SN\,2003L was classified as a SN Ic by \cite{Valenti03} and reported in \cite{Soderberg05} as a SN Ibc. At the time, SN\,2003L was the 2nd most luminous SN Ib/c discovered, just after SN\,1998bw. We include photometry from the CRTS and the VSNET \citep{Nogami97}. The optical spectra obtained from \cite{Shivvers19} resemble that of a SN Ib, with clear Helium lines, but the light curve is much broader and brighter than normal SNe Ib. We assign a Silver label due to the very sparse light curve.

\subsection{2007ce}
SN\,2007ce was classified as a SN Ic by \cite{Quimby07_07ce} and reclassified in \cite{Modjaz14} as a SN Ic-BL. We include photometry from \cite{Bianco14}. The spectra from \cite{Modjaz14} and \cite{Shivvers19} mostly match SNe Ic-BL spectra, but also resemble the spectra of the SLSN-I 2018avk \cite{Lunnan20_four}. We assign a Silver label due to the very sparse light curve and lack of coverage during the light curve peak.

\subsection{2009cb}
SN\,2009cb (=CSS090319:125916+271641 =PTF09as) was classified as a SLSN-I by \cite{Quimby18}. We adopt a redshift of $z = 0.1867$, determined from a study of the host galaxy by \cite{Perley16}, who found a redshift of $z = 0.1867$. We use photometry from \cite{Cia18}, which the authors correct for extinction using $E(B-V) = 0.0065$. We assign a Silver label since the SN was caught after peak, so there is no photometry available during the rise of peak, but our \mosfit model predicts a peak magnitude of $M_r \sim -20$ mag, within the range of SLSNe. The spectra match either a SN Ic or a SLSN-I, but given the uncertainty in the peak date, we might be underestimating the phase, which would make a SLSN-I classification more plausible.

\subsection{2010ay}
SN\,2010ay (=CSS100317:123527+270403) was classified as a SNe Ic-BL by \cite{Sanders12}, who noted this was one of the most luminous SNe Ib/c ever discovered. We only include photometry from \cite{Sanders12}. Additional spectra from \cite{Shivvers19} also resemble SNe Ic-BL. We assign a Silver label due to the very sparse light curve and lack of photometry during the rise.

\subsection{2013hy}
SN\,2013hy (=DES13S2cmm) was classified as a SLSN-I by \cite{Papadopoulos15} and as a ``Gold" SLSN-I by \cite{Angus19}. \cite{Ouyed15} argue SN\,2013hy was actually a quark nova in a high mass X-ray binary. We only include photometry from \cite{Angus19}. The light curve is almost as bright as normal SLSNe, but evolves as fast as SNe Ic. The spectra have very low S/N, but overall match the shape and general features match those of SLSN-I. We assign a Silver label due to the poor spectral coverage from a single low S/N spectrum.

\subsection{2018don}
SN\,2018don (=ZTF18aajqcue =PS18aqo) was classified as a SLSN-I by \cite{Lunnan20_four}, but noting this could also be a SN\,2007bi-like event with considerable reddening, or a long-lived and luminous SN Ic. The authors correct the photometry for foreground extinction using $E(B-V) = 0.009$, but find a total extinction of $E(B-V) = 0.4$ when accounting for the intrinsic host extinction. We include photometry from \cite{Lunnan20_four} and additional photometry from PS1, which we correct for extinction by the same value of $E(B-V) = 0.009$. Additional spectra from \citep{Fremling19} are consistent with SNe Ic. The light curve is as broad as normal SLSN-I, but much dimmer, maybe a consequence of the strong reddening. We assign a Silver label due to the uncertainty in the extinction value, which could be very high, making this a normal SLSN-I.

\subsection{2019gam}
SN\,2019gam (=ZTF19aauvzyh =ATLAS19lsz) was classified as a SLSN-Ib/IIb by \cite{Yan20}. We include photometry from ATLAS and our own PSF photometry of ZTF images after doing difference imaging to subtract the host flux. This object is only marginally dimmer than most SLSN-I, and could be just a dim SLSN-I with possible helium. We assign a Silver label since there is no photometry coverage during peak and the spectral data are not available for download, preventing further spectral comparisons.

\subsection{2019obk}
SN\,2019obk (=ZTF19abrbsvm =PS19eqz =ATLAS19tvm) was classified as a SLSN-Ib by \cite{Yan20}. We include photometry from ATLAS and our own PSF photometry of ZTF images after doing difference imaging to subtract the host flux. We assign a Silver label since the spectral data are not available for download, preventing further spectral comparisons and our light curve model is unable to reproduce the fast decline of the $r$-band light curve.

\subsection{2019uq}
SN\,2019uq (=ZTF19aadgqvd =ATLAS19bos) was classified as a SN Ic by \cite{Piranomonte19}. We include photometry from ATLAS and our own PSF photometry of ZTF images after doing difference imaging to subtract the host flux. The spectra are not a perfect match to normal SLSN-I, but instead match the peculiar SLSN-Ib class such as SN\,2020qef \cite{Terreran20} and the SLSN-I 2016wi \cite{Yan17}. We assign a Silver label due to the poor photometric coverage. Comparison spectra are from \cite{Piranomonte19}.

\subsection{2019wpb}
SN\,2019wpb (=ZTF19acxxxec =PS19heh) was classified as a SN Ic by \cite{Prentice19_wpb}. We adopt a redshift of $z = 0.06779$ from a spectrum of the host galaxy from SDSS \citep{Stoughton02}. The spectra match that of a SN Ic. We assign a Silver label since there is only one datapoint in ATLAS-c band during the rise.

\subsection{2021lei}
SN\,2021lei was classified as a SN Ic by \cite{Leloudas21}. We obtained spectra using the IMACS spectrograph, and include photometry from ZTF and ATLAS. We assign a Silver label due to the peculiar colors of the SN, which our model is not able to reproduce accurately.

\subsection{iPTF13dnt}
iPTF13dnt was classified as a SN Ic-BL by \cite{Taddia19_broadlined}. We include photometry and spectra from \cite{Taddia19_broadlined}. The spectrum is mostly flat but consistent with a SN Ic. The light curve is brighter and with a slower decline than normal SNe Ic. We assign a Silver label since this object lacks photometry before peak and only has one spectrum with very weak features. 

\subsection{OGLE15xl}
OGLE15xl was classified as an unknown kind of SN by \cite{Breton15}, and as a SLSN-I by Inserra et al., in prep. Only OGLE I-band is available. We determine a redshift of $z = 0.198$ is a better fit to the host emission lines than the redshift reported by OGLE of $z = 0.2$. The light curve is broad but dim for a SLSN-I. The spectra are a better match to SLSN-I than to SNe Ic. We obtain the comparison spectrum from WISeREP. We assign a Silver label since only a single datum of photometry is available.

\subsection{PTF10iam}
PTF10iam was classified by \cite{Arcavi16} as an intermediate supernova between SLSN-I and normal Type I SNe. The source has possible signatures of hydrogen, which is why we assign a Silver label. The spectra are blue and consistent with a SLSN-I, but dimmer than normal SLSN-I. We include photometry and spectra from \cite{Arcavi16}.

%%%%%%%%%%%%%%%%%%%%%%%%%%%%%%%%%%%%%%%%%%%%%%%%%%%%%%%%%%%%%%%%%%%%%
%%%%%%%%%%%%%%%%%%%%%%%%%%%% Bronze LSNe %%%%%%%%%%%%%%%%%%%%%%%%%%%%
%%%%%%%%%%%%%%%%%%%%%%%%%%%%%%%%%%%%%%%%%%%%%%%%%%%%%%%%%%%%%%%%%%%%%

\section{``Bronze" Luminous Supernovae}\label{sec:bronze_lsne}

\subsection{1992ar}
SN\,1992ar was classified as a SN Ic by \cite{Hamuy92} and presented in \cite{Clocchiatti00} as a luminous SN Ic. At the reported redshift of $z = 0.145$ the peak magnitude of the SN is $M_V \sim -19$ mag, the most luminous for a SN Ic at the time, but not quite in the SLSN-I regime. We include photometry from \cite{Clocchiatti00}. We assign a Bronze label due to the lack of photometry during the rise or peak of the light curve, and overall very sparse light curve. Moreover, the spectral data are not available for download, preventing further comparisons.

\subsection{2006jx}
SN\,2006jx (=SDSS-II SN 15170) was classified as a possible SN Ib at a redshift of $z = 0.25$ by \cite{Bassett06}, but later reclassified as part of the Sloan Digital Sky Survey-II Supernova Survey as a SN Ia at a redshift of $z = 0.395$ by \cite{Sako18}. We include photometry from \cite{Sako18}. At the redshift of $z = 0.395$ from \cite{Sako18} the peak magnitude of the SN is $M_R \sim -20.4$, within the range of SLSN-I. We assign a Bronze label since there are no public spectra of the source and we are unable to confirm either classification.

\subsection{2007tx}
SN\,2007tx (=ESSENCEy122) was classified as a SN Ia by \cite{Silverman07} and later reclassified as a SN Ic at a redshift of $z = 0.6764$ by \cite{Narayan16}. At this redshift the peak absolute magnitude of the SN is uncertain but in the range of $M_R = -12.2 \pm 0.4$. We assign a Bronze label since there is only one public photometry datum of the source and the public spectrum has very low S/N.

\subsection{2009bh}
SN\,2009bh (=PTF09q) was classified as a possible SLSN-I by \cite{Quimby18} and as a SN Ic by \cite{Kasliwal09}. The peak absolute magnitude of $M_R \sim -18.6$ is well below the SLSN-I range but within the SNe Ic range. We assign a Bronze label since there is only one datum of photometry available.

\subsection{2009ca}
SN\,2009ca was originally classified as a SN Ic by \cite{Stritzinger09} and then reclassified as a SN Ic-BL by \cite{Taddia18_Carnegie}, who point out the SN is significantly brighter than most SNe Ic-BL. We include photometry from \cite{Stritzinger18} and some early photometry from \cite{Pignata09}. We assign a Bronze label since there are no photometry data points before or during peak and no public spectra of the source.

\subsection{2011cs}
SN\,2011cs (=CSS110427:120801+491333) was classified as a SN Ic by \cite{Drake11}. At the quoted redshift of $z = 0.1$, the peak magnitude of the SN is $\sim -20$ in unfiltered C band. We include photometry from the CRTS. Nevertheless, we assign a Bronze label since there are no public spectra of the source and only unfiltered photometry available.

\subsection{2012gh}
SN\,2012gh (=CSS121026:164652+565105) was classified as a SN Ic by \cite{Tomasella12}. We include photometry from \cite{Mahabal12} and one epoch from the CRTS. Most of the CRTS data is too uncertain to be reliable with high scatter. We do include upper limits from before and after peak from the CRTS. We assign a Bronze label since there are only two epochs of photometry available.

\subsection{2016gkm}
SN\,2016gkm (=Gaia16bjf =PS18jq =iPTF16gkm) was classified as a SN Ib/c by \cite{Galbany19}. We include photometry from PS1 and the GSA. We do not include existing photometry from the Cambridge Photometry Calibration Server (CPCS; \citealt{Zielinski19}) since this appears highly discrepant. We assign a Bronze label due to the lack of good photometric coverage and no detections during or before the peak.

\subsection{2017dgk}
SN\,2017dgk was classified as a SN Ic at $z = 0.058$ by \cite{Gutierrez17}, but we find a better match to a redshift of $z = 0.065$ using narrow emission lines from the host galaxy. We include photometry from \cite{Cacella17}. We assign a Bronze label since there is a single datum of photometry of the source, and this is in unfiltered band with no quoted error bars.

\subsection{2017iwh}
SN\,2017iwh (=Gaia17dfz) was classified as a SN Ic by \cite{Angus17_iwh}. We include photometry from the GSA. We assign a Bronze label since there is no photometry coverage during the rise or peak and only single datum of photometry exists.

\subsection{2018hom}
SN\,2018hom (=ZTF18acbwxcc =Gaia18dfj =ATLAS18xyr) was classified as a SN Ic-BL by \cite{Fremling18_hom}. We include photometry from ZTF, the GSA, and ATLAS. We subtract a nominal $m_G = 18.07$ magnitude from the GSA photometry to account for the host flux, estimated from the average pre-explosion photometry. We adopt a redshift of $z = 0.0297$, from a spectrum of the host in the LEDA catalog \citep{Makarov14} as opposed to the redshift of $z = 0.0362$ reported by \cite{Fremling18_hom}. This new redshift makes it such that the source is no longer particularly bright or broad, but barely on the high margin for SNe Ic, hence the Bronze label.

\subsection{2019dgw}
SN\,2019dgw (=ZTF19aapwnmb) was classified as a SN Ib by \cite{Fremling19_dgw}. We include our own PSF photometry from ZTF images after doing difference imaging to subtract the host flux. We assign a Bronze label since the light curve only has one datum of photometry during the rise, and the single available spectrum has low S/N and shows large broad features which could be artifacts.

\subsection{PS110ahq}
PS1-10ahq (=PSc080762) was classified as a SN Ic by \cite{McCrum15}. We include MDS photometry from \cite{Hosseinzadeh20, Villar20}, excluding data before MJD = 55460, since those are from well before explosion. We assign a Bronze label given the lack of photometry before peak and lack of publicly available spectra. The SN also has close resemblance to a normal SN Ic with a peak magnitude of $M_r = 19.2$ mag.

\subsection{PTF10cs}
PTF10cs was classified as a SN Ic-BL by \cite{Taddia19_broadlined}. We include photometry from \cite{Taddia19_broadlined}. We assign a Bronze label since there are only three public photometry data points of the source and no spectra available for download.

\subsection{PTF11hrq}
PTF11hrq was classified as a SLSN-I by \cite{Quimby18}. The host galaxy of PTF11hrq was studied by \cite{Perley16}, who found a redshift of $z = 0.057$. We include photometry from \cite{Cia18}, which the authors correct for extinction using $E(B-V) = 0.013$. There are no early spectra available, which makes distinguishing between SLSN-I and SNe Ic spectra challenging. We assign a Bronze label since there is no photometry during the rise or peak of the SN. Our MOSFiT model predicts a peak magnitude between that of SLSN-I and SNe Ic. \\

\subsection{PTF12grr}
PTF12grr was classified as a SN Ic-BL by \cite{Taddia19_broadlined}. We include photometry from \cite{Taddia19_broadlined}. We assign a Bronze label due to the very poor light curve coverage and the fact there are no spectra of the source available for download.

\subsection{SNLS04D4ec}
SNLS04D4ec was classified by \cite{Arcavi16} as an intermediate supernova between SLSN-I and normal Type I SNe. Nevertheless, we assign a Bronze label since there are no spectra of the source during peak, but only of the host galaxy. We only include photometry from \cite{Arcavi16}.

\subsection{SNLS05D2bk}
SNLS05D2bk was classified by \cite{Arcavi16} as an intermediate supernova between SLSN-I and normal Type I SNe. Nevertheless, we assign a Bronze label since there are no spectra of the source during peak, but only of the host galaxy. We only include photometry from \cite{Arcavi16}.

\subsection{SNLS06D1hc}
SNLS06D1hc was classified by \cite{Arcavi16} as an intermediate supernova between SLSN-I and normal Type I SNe. Nevertheless, we assign a Bronze label since there are no spectra of the source during peak, but only of the host galaxy. We only include photometry from \cite{Arcavi16}. \\

%%%%%%%%%%%%%%%%%%%%%%%%%%%%%%%%%%%%%%%%%%%%%%%%%%%%%%%%%%%%%%%%%%%%%%
%%%%%%%%%%%%%%%%%%%%%%%%%%%% Not LSNe %%%%%%%%%%%%%%%%%%%%%%%%%%%%%%%%
%%%%%%%%%%%%%%%%%%%%%%%%%%%%%%%%%%%%%%%%%%%%%%%%%%%%%%%%%%%%%%%%%%%%%%
\section{Excluded Luminous Supernovae}\label{sec:not_lsne}

These objects were excluded from our sample since they have been well studied before and appear to be very different in nature to the LSNe population.

\subsection{1998bw}
SN\,1998bw (=GRB980425A =AAVSO 1927-53) was classified as a luminous SNe Ic-BL with an associated GRB by \cite{Galama98}, \cite{Wang98}, and \cite{Kulkarni98}. We exclude this object from our sample because even though SN\,1998bw passes the threshold for having a luminosity in between that of SNe Ic and SLSNe, this is a well known luminous SNe Ic-BL with very broad spectral features and significantly different to any SN in the LSNe sample.

\subsection{2018gep}
SN\,2018gep (=ZTF18abukavn =ATLAS18vah) was classified as a SN Ic-BL by \cite{Burke18} and presented in \cite{Ho19}. We exclude this object from our sample since it was thoroughly studied in \cite{Ho19} and appears to be a peculiar transient with different photometric and spectroscopic features to the sample presented here, related to fast transients such as AT\,2018cow more than LSNe.

\subsection{ASASSN15no}
ASASSN-15no was presented in \cite{Benetti18} as a SN with features of SLSN-I, normal SNe Ib/c, but also signs of interaction, such as hydrogen emission lines. The spectral data are not available for download, but visual inspection suggests the spectrum is roughly consistent with a SLSN-I. The peak luminosity and decline rate are between those of SLSN-I and SNe Ic. We exclude this object since \cite{Benetti18} explored it in detail revealing this SN to be very different in nature to the sample presented here, with signs on interaction.

\subsection{2021lji}
SN\,2021lji was classified as a SLSN-I by \cite{Charalampopoulos21}, who determine a redshift of $z = 0.12$ from the supernova features. We obtained a late time spectrum with the Binospec spectrograph that shows clear host emission lines, from which we derive a redshift of $z = 0.092$. We exclude this object from our sample since the new redshift means it matches a normal core-collapse SN. \\

%%%%%%%%%%%%%%%%%%%%%%%%%%%%%%%%%%%%%%%%%%%%%%%%%%%%%%%%%%%%%%%%%%%%%%
%%%%%%%%%%%%%%%%%%%%%%%%%%% Comparison LSNe %%%%%%%%%%%%%%%%%%%%%%%%%%
%%%%%%%%%%%%%%%%%%%%%%%%%%%%%%%%%%%%%%%%%%%%%%%%%%%%%%%%%%%%%%%%%%%%%%

\section{Comparison Supernovae}\label{sec:compare}

We include a sample of SNe Ic and SNe Ic-BL for comparison with our sample of LSNe.

We include the SNe Ic-BL iPTF13ebw, iPTF13u, iPTF14dby, iPTF14gaq, iPTF16gox, iPTF17axg, iPTF16ilj, PTF10aavz, PTF10ciw, PTF10qts, PTF10tqv, PTF10vgv, PTF10xem, PTF11cmh, PTF11lbm, PTF12as, PTF12eci, and iPTF16coi from \cite{Taddia19_broadlined}. For iPTF13u we exclude the single $B$-band data point due to its high discrepancy from the rest of the data.

We include the SNe Ib/c PTF09dh, PTF09ut, PTF10bip, PTF10lbo, PTF10osn, PTF10tqi, PTF10yow, PTF10zcn, PTF11bli, PTF11bov, PTF11hyg, PTF11jgj, PTF11klg, PTF11lmn, PTF11mwk, PTF11rka, PTF12dtf, PTF12dcp, PTF12cjy, PTF12fgw, PTF12hvv, PTF12jxd, PTF12ktu, PTF13ab, PTF13aot, PTF13cuv, PTF13dht, PTF13djf, PTF14gao, PTF14gqr, PTF14fuz, PTF14jhf, PTF14ym, PTF15acp, PTF15cpq, PTF15dtg, PTF16flq, and PTF16hgp from \cite{Barbarino20}.

We also include the entire samples of SLSNe from \cite{Nicholl17_mosfit, Villar18, Blanchard20, Gomez20} and newer SLSNe that will be presented in Gomez et al., in prep: 2002gh, 2005ap, 2006oz, 2007bi, 2009jh, 2010gx, 2010hy, 2010kd, 2010md, 2011ke, 2011kg, 2012il, 2013dg, 2015bn, 2016aj, 2016ard, 2016eay, 2016inl, 2016wi, 2017beq, 2017dwh, 2017egm, 2017ens, 2017gci, 2017jan, 2018avk, 2018bgv, 2018bsz, 2018bym, 2018cxa, 2018fd, 2018ffj, 2018ffs, 2018gkz, 2018hti, 2018ibb, 2018kyt, 2018lfe, 2019cca, 2019cdt, 2019cwu, 2019dlr, 2019eot, 2019gam, 2019gfm, 2019gqi, 2019hge, 2019hno, 2019itq, 2019kcy, 2019kwq, 2019kws, 2019kwt, 2019kwu, 2019lsq, 2019neq, 2019nhs, 2019obk, 2019pud, 2019sgh, 2019szu, 2019ujb, 2019unb, 2019xaq, 2019zbv, 2019zeu, 2020abjc, 2020adkm, 2020ank, 2020exj, 2020jii, 2020onb, 2020qef, 2020qlb, 2020rmv, 2020uew, 2020xga, 2020xgd, 2020znr, 2020zzb, 2021bnw, 2021een, 2021ejo, 2021ek, 2021fpl, 2021gtr, 2021hpc, 2021hpx, 2021mkr, 2021nxq, 2021txk, 2021vuw, 2021xfu, 2021yrp, 2021zcl, CSS160710, DES14S2qri, DES14X2byo, DES14X3taz, DES15E2mlf, DES15X1noe, DES15X3hm, DES16C2aix, DES16C2nm, DES16C3dmp, DES16C3ggu, DES17X1amf, DES17X1blv, iPTF13ajg, iPTF13dcc, iPTF13ehe, iPTF16bad, iPTF16eh, LSQ12dlf, LSQ14an, LSQ14bdq, LSQ14fxj, LSQ14mo, OGLE15qz, OGLE15sd, OGLE15xx, OGLE16dmu, PS110awh, PS110bzj, PS110ky, PS110pm, PS111afv, PS111aib, PS111ap, PS111bam, PS111bdn, PS112bqf, PS112cil, PS113or, PS114bj, PS15cjz, PTF09atu, PTF09cnd, PTF10aagc, PTF10bfz, PTF10nmn, PTF10uhf, PTF10vqv, PTF12dam, PTF12mxx, SCP06F6, SNLS06D4eu, SNLS07D2bv, SSS120810.

\end{document}